\documentclass{aa}  
\usepackage{graphicx}
\usepackage{txfonts}
\usepackage{xcolor}
\usepackage{courier}
\usepackage{gensymb}
\usepackage{soul}

\usepackage[utf8]{inputenc}
\DeclareUnicodeCharacter{2212}{-}

\usepackage{hyperref}
\hypersetup{colorlinks,allcolors=blue}

\begin{document} 
    \title{Evaluating star formation rates at $z=5$}
   \authorrunning{D. Ismail et al.}
    
   \author{D. Ismail\inst{1}
          \and
          K. Kraljic\inst{1}
          \and
          M. B\'ethermin\inst{1}
          \and
          A. U. Kapoor\inst{2}
          \and
          F. Renaud\inst{1}
          \and 
          C. Accard\inst{1}
          \and 
          J. Freundlich\inst{1} 
          \and
          Y. Dubois\inst{3}
          \and 
          S. Han\inst{3}
          \and 
          J.K. Jang\inst{4}
          \and
          S. Jeon\inst{4}
          \and 
          T. Kimm\inst{4}
          \and
          J. Rhee\inst{3,5}
          \and
          S. Yi\inst{4}
          }

   \institute{Observatoire Astronomique de Strasbourg, Universit\'e de Strasbourg, CNRS, UMR 7550, 67000 Strasbourg, France \\
            \email{diana.ismail@astro.unistra.fr}
            \and 
            Sterrenkundig Observatorium, Universiteit Gent, Krijgslaan 281 S9, 9000 Gent, Belgium 
            \and 
            Institut d’Astrophysique de Paris, Sorbonne Universit\'e, CNRS, UMR 7095, 98 bis bd Arago, 75014 Paris, France \
            \and
            Department of Astronomy and Yonsei University Observatory, Yonsei University, Seoul 03722, Republic of Korea
            \and 
            Korea Astronomy and Space Science Institute, 776, Daedeokdae-ro, Yuseong-gu, Daejeon 34055, Republic of Korea }

   \date{Received 9 January 2026; accepted 1 April 2026}

    \abstract
    {Inferring the star formation rates (SFR) in high redshift galaxies remains challenging, owing to observational limitations or uncertainties in calibration methods that link luminosities to SFRs. We utilize two state-of-the-art hydrodynamical simulations \textsc{NewHorizon} and \textsc{NewCluster}, post-processed with the radiative transfer code \textsc{Skirt}, to investigate the systematic uncertainties and biases in the inferred SFRs for $z=5$ galaxies {---} an epoch where galaxies build-up their stellar mass. We create synthetic observables for widely-used tracers: H$\alpha$ nebular line, [\textsc{Cii}] 158 $\mu$m fine-structure line, total infrared (IR) continuum luminosity, and hybrid (IR + UV). We find that H$\alpha$-inferred SFRs, time-averaged over 10 Myr, are sensitive to the choice of calibration and exhibit substantial scatter driven by dust attenuation, viewing angle, and dust-to-metal ratio. Adopting a steeper attenuation curve reduces this scatter significantly but does not fully eliminate systematic uncertainties. IR continuum-based SFRs trace intrinsic SFRs time-averaged over 100 Myr timescales when a well-sampled continuum emission between restframe 8 and 1000 $\mu$m is available and underestimate them with typical approaches when IR data are limited. Nevertheless, IR SFRs display a considerable scatter, largely due to UV photon leakage and strong variations in the star formation history. When UV data are available, hybrid (IR + UV) SFRs provide a more robust estimate, reducing scatter compared to IR-based SFRs while avoiding explicit attenuation corrections. Finally, we derive a [\textsc{Cii}]–SFR relation finding a steeper relation than previous studies, however with significant scatter linked to gas density and metallicity. Overall, IR-, hybrid-, and [\textsc{Cii}]-based tracers remain more robust than H$\alpha$ against variations in optical depth.}
    

\maketitle

%
\section{Introduction}
The star formation rate (SFR) is a fundamental quantity to understand galaxy evolution, as it provides a measure of how galaxies grow and evolve by converting gas into stars. It has been established that the cosmic star formation rate density increased rapidly in the first $1$ Gyr of evolution, reaching a peak around $z \sim 2-4$, before declining toward the present \citep{madau2014, zavala2021}. At redshifts $4<z<6$, the Universe is in a critical transition period, when these so-called "teenage galaxies" are still building up their stellar mass, making this epoch important to bridge the transition between primordial and more mature galaxies. However, accurately estimating the SFRs remains challenging due to dust attenuation, observational limits, and uncertainties in calibrations derived from the local Universe \citep[e.g.,][]{buat2014, shivaei2015, figueira2022}.

The most commonly used and direct tracer for SFR is the rest-frame ultraviolet (UV) continuum at wavelengths in the range $1500 \, \AA < \lambda < 2800 \, \AA$. Emitted by recently formed massive stars (O and B type), the UV luminosity traces the SFR over timescales $\sim 100$ Myr \citep[e.g.,][]{salim2007}. Although the UV continuum can be observed to very high redshifts \citep{Bouwens2009}, it presents several challenges as a diagnostic tool. At these wavelengths, a significant fraction of UV photons are reprocessed by dust (more than $40 \%$), particularly in heavily obscured star-forming galaxies. Moreover, the UV luminosity-to-SFR conversion shows a dependence on the metallicity \citep{madau2014}.

The H$\alpha$ nebular emission line, arising from ionized gas in star-forming regions, is another popular SFR tracer that until recently was only possible to observe up to $z \sim 2$. Thanks to the James Webb Space Telescope (JWST), and its NIRSpec instrument \citep{ferruit2022}, it is now possible to observe this emission line at $z > 3$. This nebular line probes the young stellar population, tracing timescales of $< 20$ Myr of massive stars, with $\rm M_* > 10 \, M_\odot$, which provides an instantaneous measurement of the SFR. Similarly to UV emission, H$\alpha$ is also affected by dust attenuation, however, it is argued to be the most reliable SFR estimator when the Balmer decrement is measured with high precision. By measuring the observed ratio of two Balmer line fluxes (typically H$\alpha$/H$\beta$) and comparing it to the intrinsic ratio, that is theoretically known, any deviation allows for a precise calculation of the dust optical depth along the line of sight \citep{hirashita2003, groves2012}.

While both UV and H$\alpha$ are affected by attenuation, the infrared (IR) radiation provides an indirect yet powerful diagnostic for SFR that traces the reprocessed stellar radiation absorbed by dust. At restframe wavelengths $\sim 8 - 1000 \, \mu$m, the emission is dominated by dust grains that are primarily heated by the young stellar population, with additional heating from other components (e.g., older stellar populations), tracing SFRs at timescales comparable to that of UV emission ($\sim 100$ Myr). At $z \sim 5$, restframe IR is shifted into the observed submillimeter (submm) to millimeter (mm) wavelengths, making it possible to observe with ground-based instruments such as the Atacama Millimeter/submillimeter Array (ALMA) and the NOrthern Extended Millimeter Array (NOEMA). In general, the IR luminosity can be a very robust SFR estimator, however, some surveys lack sufficient sampling of the IR spectral energy distribution (SED), particularly around the peak of dust emission, which could lead to biases in IR luminosity estimates due to necessary approximations and assumptions. 

Since UV light is dust attenuated and re-radiated in the IR continuum, while also a certain amount of UV light escapes the galaxy, it has been suggested that a hybrid SFR could be used to measure a bolometric SFR through the addition of the monochromatic FUV ($1500 \AA$) and the total IR ($8 - 1000 \, \mu$m) luminosities \citep{madau2014}. This approach has been widely used for UV-selected samples in both low redshift \citep[$z < 1$; e.g.,][]{freundlich2019} and high redshift \citep[e.g.,][]{bethermin2020} studies.

More recently, the singly-ionized carbon, [\textsc{Cii}] at 158 $\mu$m, has become a widely used tracer of star formation activity, especially at high-$z$ \citep[e.g.,][]{matthee2019, fudamoto2020, Mitsuhashi2024}. [\textsc{Cii}] luminosity is found to exhibit a tight correlation with the SFR \citep[e.g.,][]{stacey2010, delooze2014, herrera-camus2015}, sometimes making it a more reliable SFR tracer than the dust continuum, particularly when observations are limited. The fine-structure transition originates from both \textsc{Hii} regions and photodissociation regions \citep[PDRs;][]{vallini2013}, however it is also argued that it originates primarily from PDRs \citep[e.g.,][]{stacey2010}. As one of the brightest far-IR emission lines \citep{stacey2010}, [\textsc{Cii}] is observable with ALMA band 7 at $z \sim 5$.

The UV continuum, the H$\alpha$ nebular line, the IR continuum, and the [\textsc{Cii}] fine-structure line are key tracers of star formation in current high-$z$ studies. Despite their prevalence, the calibrations and underlying uncertainties remain poorly constrained. Moreover, observational studies at high-$z$ are inherently biased toward the most massive and luminous systems due to sensitivity limits, leaving the properties of lower-mass systems unexplored and calibrations limited to high mass systems. To overcome observational limitations, we leverage two state-of-the-art hydrodynamical simulations: \textsc{NewHorizon} \citep{dubois2021NH}, representing average-density field environment, and \textsc{NewCluster} \citep{han2025b}, capturing the formation and evolution of galaxies in a dense cluster environment. The simulated galaxies are post-processed with the 3D radiative transfer code \textsc{Skirt} \citep{camps2020baes} to generate synthetic SEDs from the UV to millimeter wavelengths. We aim to systematically quantify the accuracy and scatter of SFR estimation at $z \sim 5$ using the nebular H$\alpha$ emission line, the total IR continuum, the hybrid (IR + UV) approach, and the [\textsc{Cii}] $158 \,\mu$m fine structure line, providing insights into the interpretation of current and future observational data. Additionally, with having access to intrinsic properties, we also investigate the impact of the physical properties on SFR diagnostics. 

This paper is organized as follows. In Sect. \ref{section:sample}, we provide a brief description of the simulations used and the sample selection. In Sect. \ref{section:methods}, we describe the post-processing methodology we follow using \textsc{Skirt} radiative transfer code and the reconstruction of observables. In Sect. \ref{section:results}, we compare the inferred SFRs with the intrinsic ones across tracers. In Sect. \ref{section:discussion}, we explore the impact of dust-to-metal ratios on SFRs, H$\alpha$ attenuation curve, relation between \textsc{[Cii]} with $L_{\rm IR}$, and realistic IR luminosity estimates. The results are then summarized in Section \ref{section:conclusion}. 

\section{Sample: Hydrodynamical simulations}\label{section:sample}
\subsection{NewHorizon and NewCluster simulations}\label{section:sims}

In this work, we use zoom-in cosmological \textsc{NewHorizon}\footnote{\href{https://new.horizon-simulation.org/}{https://new.horizon-simulation.org/}} simulation presented in \cite{dubois2021NH}, here we only provide a brief summary. The \textsc{NewHorizon} simulation is a sub-volume of (16 Mpc)$^3$ embedded within the large-scale parent Horizon-AGN cosmological simulation \citep{dubois2014, kaviraj2017}. \textsc{NewHorizon} is run with the adaptive mesh refinement RAMSES code \citep{Teyssier2002} using cosmological parameters based on the 7-yr Wilkinson Microwave Anisotropy Probe (WMAP-7) data with a $\Lambda$CDM cosmology \citep[$\Omega_m$ = 0.272, $\Omega_\Lambda$ = 0.728, $H_0$ = 70.4 $\rm km\,s^{-1} \,Mpc^{-1}$;][]{komatsu2011wmap7}. The mass resolutions of dark matter and stellar particles are $1.2\times10^6 \, M_\odot$ and $1.3\times10^4 \, M_\odot$, respectively, while the spatial resolution reaches 34 pc in the densest regions. \textsc{NewHorizon} accounts for gas heating from a uniform UV radiation field \citep{haardt1996}, with a simple self-shielding approximation applied in optically thick regions \citep{rosdahl2012}. Radiative cooling is modeled through collisional ionization, excitation, recombination, bremsstrahlung, and Compton effect down $\sim 10^4$ K, and additional metal-line cooling of gas down to 0.1 K \citep{sutherland&dopita93,rosen&bregman1995}. Star formation is modeled in gas denser than 10 H cm$^{-3}$ following a Schmidt relation with a non-uniform star formation efficiency (SFE) per free-fall time \citep{federrath2012, kimm2017}. Feedback from Type II supernovae is included using a mechanical scheme \citep{kimm2015}. The simulation also follows the growth and evolution of massive black holes and the associated feedback from active galactic nuclei \citep[AGN; ][]{dubois2012} via the release of mass, momentum, and energy at low Eddington rates \citep[][]{dubois2012}, and of thermal energy at higher rates \citep[][]{teyssiere2011}.

We also use the \textsc{NewCluster}\footnote{\href{https://gemsimulation.com/}{https://gemsimulation.com/}} \citep{han2025b} simulation, a zoom-in cosmological simulation following the formation and evolution of a galaxy cluster and therefore serving as a massive halo counterpart to the average-density field captured by \textsc{NewHorizon}. 
\textsc{NewCluster} shares many of its subgrid prescriptions with \textsc{NewHorizon}, hence we highlight only their differences and refer the reader to \citet{han2025b} for its detailed description. \textsc{NewCluster}, run with the OpenMP version of \textsc{Ramses} \citep{Teyssier2002, Han2025a}, is a zoom-in region targeting a cluster of virial mass $M_{\rm vir} \sim 4.7 \times 10^{14} , M_\odot$ with a radius of $3.5 \, R_{\rm vir}$ ($\sim 17.7$ Mpc/$h$) at $z=0$. The mass resolutions of dark matter and stellar particles are $1.3\times10^6 \, M_\odot$ and $2\times10^4 \, M_\odot$, respectively, with a maximum spatial resolution of 68 pc reached in the densest regions. Star formation is modeled in gas denser than 5 H cm$^{-3}$ following the prescription adopted in \textsc{NewHorizon}. In addition to feedback from Type II supernovae, \textsc{NewCluster} also includes Type Ia, both implemented following mechanical scheme of \citep{kimm2014}, and stellar winds. Chemical enrichment for ten tracked elements is computed using the Starburst99 code \citep{leitherer1999,leitherer2014} with yields from stellar winds assuming the Geneva stellar wind model \citep{schaller1992,maeder&meynet2000}, and tabulated yields from \citet{koyabashi2006} and \citet{iwamoto1999} for Type II and Type Ia supernovae, respectively. Unlike \textsc{NewHorizon}, \textsc{NewCluster} also includes dust formation, growth, destruction, and size evolution \citep[see][]{byun2025NCdust}, following the model of \cite{dubois2024}.

Both simulations adopt a Chabrier IMF for stellar particles, which is consistently used throughout the radiative transfer modeling and SFR conversions.

\subsection{Sample Selection}\label{section:sample-selection}
For this study, we select the $1\%$ most massive sources from each simulation at $z = 5$ with $ M_* > 10^{9} \, M_\odot$ for \textsc{NewCluster} and $ M_* > 10^{8} \, M_\odot$ for \textsc{NewHorizon}. This selection allows us to investigate the most massive sources from different environments (cluster and field galaxies), that could shed light on the various biases originating from the intrinsic properties of galaxies. We note that by extending our analysis to \textsc{NewCluster} galaxies, it allows us to access higher masses that are currently observed but not reached in \textsc{NewHorizon} due to its smaller volume. 
 
Stellar particles and gas cells are extracted within a cubic region extending to ten effective radii (R$_{1/2}$; stellar half-mass radius). Each galaxy is visually inspected to verify whether the selection captures the entire galaxy and whether it is a merger. The latter are removed from the sample and for the former, we re-adjust the extraction box size that reaches $20 - 30 \, R_{1/2}$ for a few galaxies. With this selection, we obtain a total of 97 galaxies from both simulations, covering a wide range in the parameter space of physical properties. The stellar mass varies between $10^8$ and $10^{10}$ M$_\odot$, the gas mass varies between $10^{7.1}$ and $10^{9.8}$ M$_\odot$, and the stellar half-mass radius varies between 0.15 and 3.4 kpc. These sources have SFRs averaged over 100 Myr that vary between 0.2 and 58 $\rm M_\odot \, yr^{-1}$ and instantaneous SFRs averaged over 10 Myr varies between 0.1 and 126 $\rm M_\odot \, yr^{-1}$. The properties of these galaxies are summarized in Table \ref{tab:simulation-properties}.

\begin{table}[t!]
\centering
\caption{Physical properties of the \textsc{NewHorizon} and \textsc{NewCluster} selected galaxies.}
\label{tab:simulation-properties}
\begin{tabular}{lcc}
\hline \hline 
 & \textsc{NewHorizon} & \textsc{NewCluster} \\
\hline
Number of sources                  & 40 & 57 \\
$\rm M_{*}$ {[}M$_\odot${]}  & $10^8 - 10^{8.9}$ & $10^9 - 10^{10}$ \\
$^\dagger \rm M_{gas}$ {[}M$_\odot${]}      & $10^{7.1} - 10^{8.7}$ & $10^{8.6} - 10^{9.8}$   \\
SFR$_{100}$ [M$_\odot \, \rm yr^{-1}$] & $0.2 - 3$ & $2 - 58$ \\
SFR$_{10}$ [M$_\odot \, \rm yr^{-1}$] & $0.1 - 20 $  & $0.4 - 126$  \\
$\rm R_{1/2}$ [kpc]                & $0.15 - 1.04$ & $0.39 - 3.38$ \\
\hline
\end{tabular}
\vspace{0.1cm}
\parbox{0.9\linewidth}{\footnotesize $^\dagger$ The ranges correspond to the gas mass enclosed within the galaxy's stellar half-mass radius.}
\end{table}

\section{Methods} \label{section:methods}
\subsection{Post-processing with SKIRT}\label{section:SKIRT}
We post-process our sample galaxies with the open-source 3D Monte Carlo radiative transfer code \textsc{Skirt}\footnote{We use version 9 of \textsc{Skirt}; \href{https://github.com/SKIRT/SKIRT9}{https://github.com/SKIRT/SKIRT9}} \citep{baes2015, camps2020baes}, important to reproduce multi-wavelength (from UV to mm) observables affected by the absorption, scattering and re-emission by dust grains. Below, we describe the \textsc{Skirt} configuration used to perform the radiative transfer analysis of our sample. 

To prepare the inputs for \textsc{Skirt}, we extract gas cells and star particles from the \textsc{NewHorizon} and \textsc{NewCluster} simulations at the chosen snapshot (see Sect. \ref{section:sample-selection}). The coordinate system is rotated to align with the angular momentum vector of the stellar particles, so that the viewing angle corresponds to face-on configuration. For each galaxy, stellar particles (primary sources) are treated based on their age: the evolved stellar component with ages above 10 Myr and the star-forming regions with ages $\leq$ 10 Myr. The SEDs of evolved star particles are modeled using the \citet{bruzual2003charlot} library with a \citet{chabrier2003} initial mass function (IMF). This model requires the stellar mass at birth, age, and metallicity, all of which are available from the \textsc{NewHorizon} and \textsc{NewCluster} snapshot data. In contrast, star-forming regions are modeled using the \texttt{TODDLERS} \citep{kapoor2023, kapoor2024} template library, which includes thermal dust emission originating from starburst regions, while incorporating \textsc{Hii} regions and PDRs. The \texttt{TODDLERS} library requires five parameters: mass of the gas cloud, system's age, metallicity of the gas, SFE, and birth-cloud density. 

We choose the smoothed particle data import for both old stars and star-forming regions, which requires the spatial positions and the smoothing length that we fix to 100 pc\footnote{By increasing the smoothing length to 500 pc, we do not find any impact on the IR and [\textsc{Cii}] luminosities. An increase of $\sim 0.12$ dex is found for H$\alpha$ luminosities.}. The smoothing length is chosen to be larger than the spatial resolution (34 pc in \textsc{NewHorizon} and 68 pc in \textsc{NewCluster}), which ensures that at least one stellar particle and one gas cell are in the vicinity of each other. For star-forming regions, we also import the velocities of the young stellar particles and estimate the velocity dispersion from the gas cells within 100 pc of each young stellar particle.

In \textsc{NewCluster}, dust evolution is followed on the fly, including both formation and destruction processes. However, we do not use this information in order to ensure a fair comparison with NewHorizon, which does not track dust evolution. In this case, the dust distribution is estimated in \textsc{Skirt} using gas cells as a transfer medium. Gas cells are imported as cuboidal cell data that requires information about their positions, metallicities ($Z_{\rm gas}$), masses ($M_{\rm gas}$), and temperature ($T_{\rm gas}$) ---all available from snapshot data. We estimate the SFE (= $M_* / M_{\rm gas}$; $M_*$ being the stellar mass at birth) and birth cloud density using gas cells and stellar particles within 100 pc. For each galaxy, we adopt a global SFE corresponding to the median value across star-forming regions, as the SFE distribution shows little variation, while the density is allowed to vary for individual star-forming regions. This approach captures the typical efficiency of star formation per galaxy while retaining local variations in the physical conditions of the birth clouds. We limit the dust distribution to regions where the gas temperature is below $10^6$ K since dust grains are destroyed due to thermal sputtering around the hotter gas \citep[e.g.,][]{draine1979salpeter, hirashita2015}. The dust mass, $M_{\rm dust}$, is then calculated as follows: 
\begin{equation}\label{eq:dust-mass-skirt}
M_{\rm dust} = 
    \begin{cases}
    \, f_{\rm dust} \, Z_{\rm gas} \, M_{\rm gas} \, & \text{if } T_{\rm gas} < 10^6 \, \rm K \\
    \, 0 & \text{otherwise}
    \end{cases},
\end{equation}

\noindent where $f_{\rm dust}$ is the dust-to-metal (DTM) ratio, or in other words, the amount of metals locked up in dust grains. We use $f_{\rm dust} = 16\%$ following the relation derived by \citet{vogelsberger2022b} between the DTM ratio and the redshift by comparing the simulated UV luminosity functions to observational ones. The relation is given in the form $f_{\rm dust}(z) = 0.9 \, (z/2)^{-1.92}$ for $z\geq 2$. Given the substantial uncertainty in this quantity and the lack of a clear consensus on its value, we examine the effects of varying the DTM ratio in Sect. \ref{sec:impact-of-dtm}.

A \textsc{Themis} dust mix \citep{jones2017themis} is adopted---a model based on the physical conditions surrounding the dust grains, varying between the diffuse and dense ISM, making it a global model to use. This model requires setting a distribution of grain size bins for the silicate and hydrocarbon populations, which we set to 15 bins for each population.  Stochastic heating of dust grains is taken into account, as assuming local thermodynamic equilibrium (LTE) mainly affects small grains emitting in the restframe mid-IR. While an LTE assumption has minimal impact on the far-IR/submm emission, it still results in an underestimation of the total IR luminosity \citep[e.g.,][]{vijayan2022}.

The wavelength grid is discretized over 100 logarithmic grid points between 1 and $1000 \, \mu$m in restframe. We adopt separate spatial cell grouping scheme, which provides higher accuracy. Random density samples of 500 are adopted---responsible for estimating the mass in each spatial cell. We choose a grid refinement with a minimum of 6 and a maximum of 12, and a maximum dust fraction in each cell of $10^{-6}$. We choose a grid at the galaxy's center; in this case we use the center of the stellar mass with a grid size depending on the extent of each source. We place four detectors (or instruments) assuming four different inclinations of the galaxy's disk ranging from face-on to edge-on, in order to infer the impact of the viewing angle on derived SEDs.

\begin{figure}
    \centering
    \includegraphics[width=\linewidth]{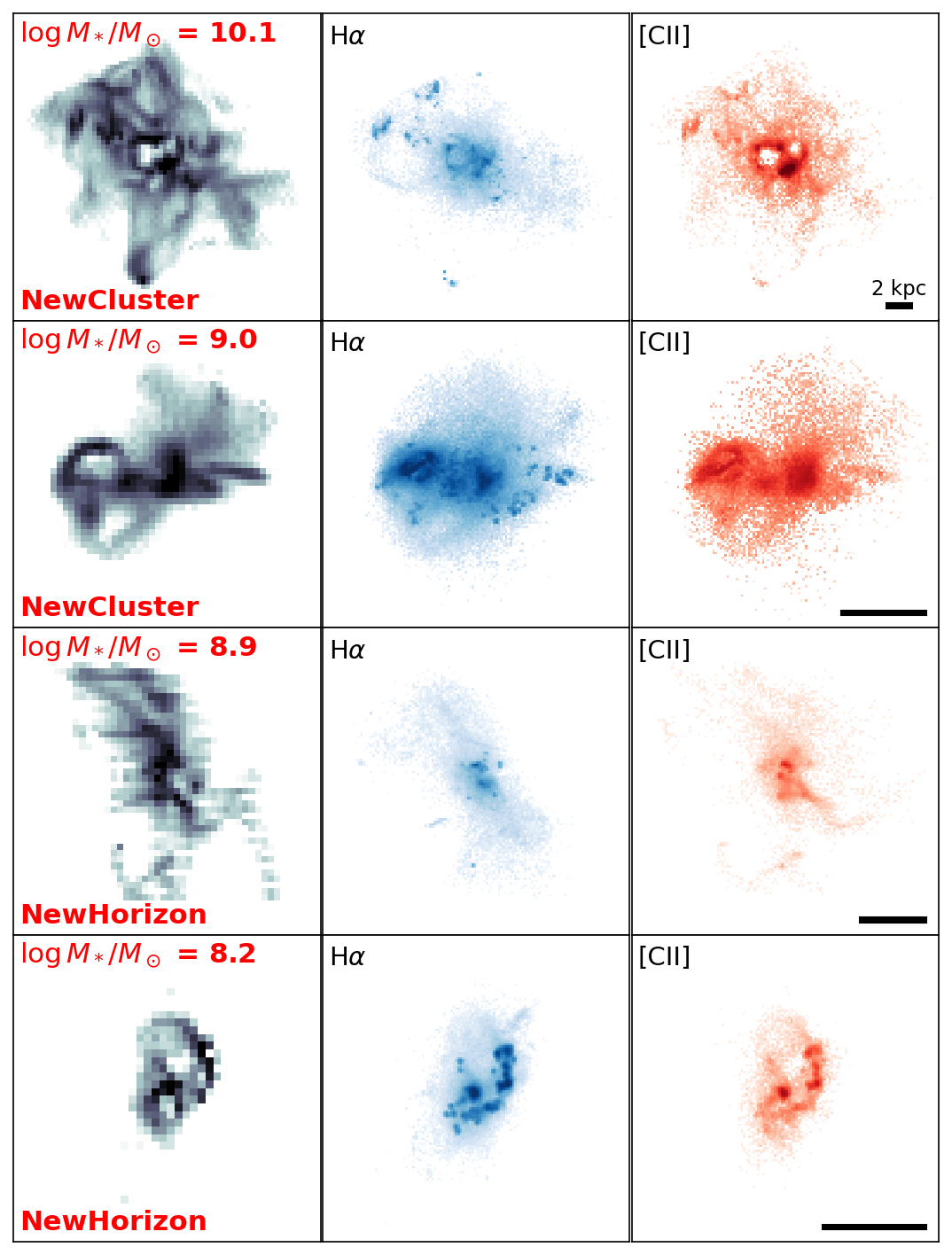}
    \caption{Projection of the gas mass (left) and post-processed \textsc{Skirt} output of the H$\alpha$ (middle) and [\textsc{Cii}] (right) emission of the most and least massive \textsc{NewCluster} (top two rows) and \textsc{NewHorizon} (bottom two rows) galaxies in a face-on configuration. The bar shows the physical scale of each source in kpc. }
    \label{fig:maps}
\end{figure}

The \textsc{Skirt} radiative transfer calculations are performed using $5 \times 10^7$ photon packets for both \textsc{NewHorizon} and \textsc{NewCluster}. We adopt resolutions based on the instruments' capabilities, while also aiming for a reasonable binning of the output SEDs enough to resolve the emission lines. In the optical/NIR bands, we adopt $R = 2700$ compatible with the best resolving power of JWST NIRSpec IFU, and $R = 10000$ at far-IR wavelengths, which corresponds to $\sim 30$ km/s velocity resolutions at $\sim 300$ GHz, comparable to typical resolutions of observed lines for high-$z$ galaxies \citep[e.g.,][]{bethermin2020}. The number of photon packets is chosen to ensure a low Poisson noise at the adopted resolution while keeping the computational cost per source tractable (see Appendix \ref{apdx:photon_packets}). We iterate over secondary emission for self-consistent calculation with the code converging when the total absorbed dust luminosity falls below 1\% of the total absorbed stellar luminosity or when it changes by less than 3\% compared to the previous iteration. \citet{camps2018} demonstrated that neglecting self-absorption can underestimate submm luminosities by a factor of 2.5, particularly for compact star-forming galaxies. We also account for the cosmic microwave background (CMB) effects by adopting the correction provided in the \textsc{Skirt} code. At $z=5$, the CMB temperatures, $T_{\rm CMB} = T_{\rm CMB}^{z=0} (1+z) = 16 \, \rm K$, become comparable to dust temperatures, thus affecting the far-IR luminosities \citep{dacunha2013}. Lastly, we use the flat $\Lambda$CDM cosmological parameters adopted in the \textsc{NewHorizon} and \textsc{NewCluster} simulations (see Sect. \ref{section:sims}). 

\begin{figure*}[t]
    \includegraphics[width=\textwidth]{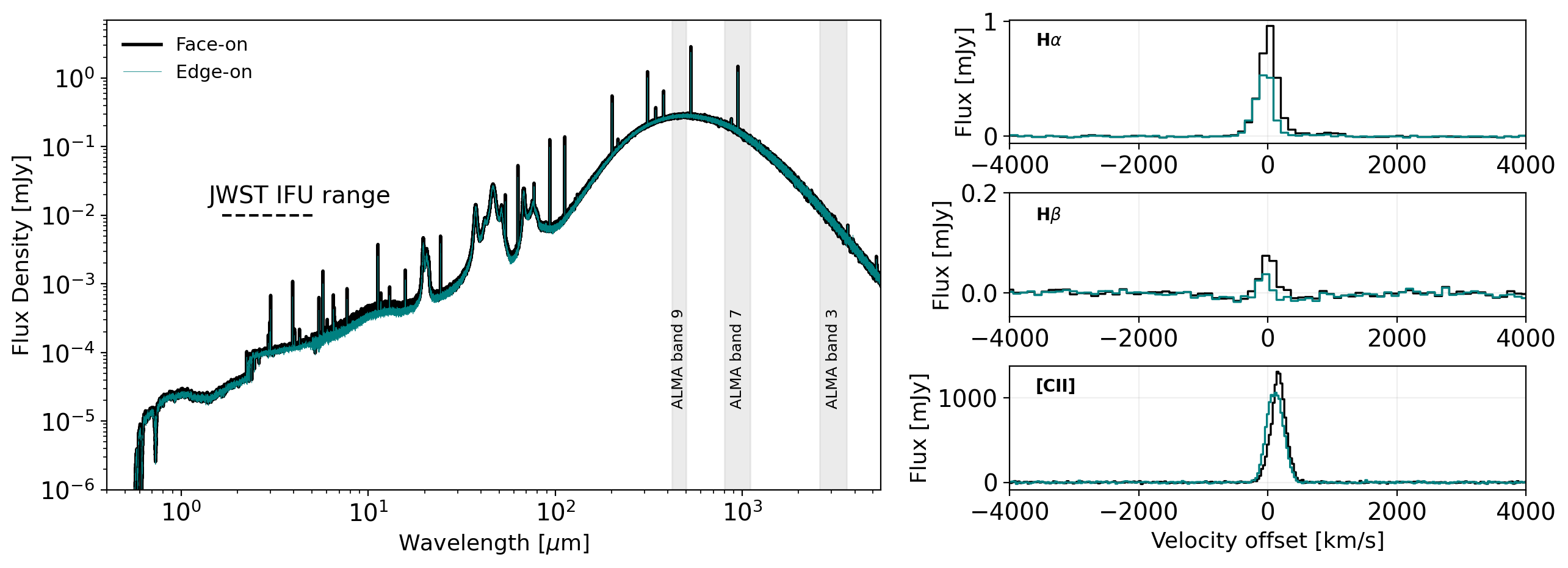}
    \caption{Integrated spectral energy distribution of one representative \textsc{NewCluster} galaxy as produced by the \textsc{Skirt} radiative transfer simulations assuming a dust-to-metal ratio of 16\%. \textit{Left:} Full SED in the observed frame, shown for both the face-on (black) and edge-on (teal) configurations, spanning wavelengths from the ultra-violet to the millimeter. Shaded vertical regions show the coverage of ALMA bands 3, 7, and 9, and the dashed line demonstrates the JWST IFU range. \textit{Right:} Examples of continuum-subtracted emission lines extracted from the same SED.}
    \label{fig:sed}
\end{figure*}

Figure \ref{fig:maps} shows the most and least massive sources in both simulations, demonstrating the gas mass distribution and the post-processed emission of H$\alpha$ and \textsc{[Cii]} at $158 \, \mu$m lines. This figure highlights that while both tracers follow the morphology of the gas, the H$\alpha$ emission exhibits peaks of emission tracing the dense ionized regions of star formation. In contrast, the \textsc{[Cii]} emission is more diffuse which traces the total gas reservoir and exhibits a few bright peaks compared to H$\alpha$. We also demonstrate the observed integrated spectral energy distribution (SED) in Fig. \ref{fig:sed} (left) produced by \textsc{Skirt} for one of the \textsc{NewCluster} sources in both the face-on (black) and edge-on (teal) configurations assuming a DTM = 16\%. This illustrates the continuum and the resolved emission lines across the electromagnetic spectrum considered in this work, from $\lambda \sim 10^{-1}$ to $5\times10^{3}\,\mu$m. The two configurations show the impact of inclination on the observed SEDs, where shorter wavelengths are more affected by dust attenuation (given the small DTM assumed), and little to no impact on IR wavelengths. To the right, we show zoom-ins on the continuum-subtracted emission lines in both viewing angles that are directly used in this study. The increased dust column-densities in the edge-on configuration demonstrate the effect of dust on shorter wavelength emission lines H$\alpha$ and H$\beta$ compared to the far-IR [\textsc{Cii}] emission at $158 \, \mu$m. 

\subsection{Luminosities and SFR conversions}\label{section:sfr}
From the SEDs produced by \textsc{Skirt}, we use the line-free continuum\footnote{To extract the line-free continuum, we use the \textsc{Scipy.signal.medfilt} function with a kernel size of 99 to smooth out the SED. While it is not a standard technique for far-IR continuum extraction, median filtering has been employed to isolate broad spectral features from narrower line emission \citep[e.g.,][]{arun2025}. We verified its reliability by comparison with a manual line-masking approach.} emission in the IR range between restframe 8 and 1000 $\mu$m to estimate the total IR luminosity. We also estimate the line luminosities for H$\alpha$, H$\beta$, and [\textsc{Cii}] at $158 \, \mu$m at different inclinations using the continuum-free fluxes. In \textsc{Skirt}, we adopt spectral resolutions that mimic those of JWST-IFU and ALMA where the spectral lines are resolved with spectral channel widths of $\rm \sim 100 \, km \, s^{-1}$ for both H$\alpha$ and H$\beta$, and $\rm \sim 30 \, km \, s^{-1}$ for the [\textsc{Cii}] line (see Sect. \ref{section:SKIRT}). We perform an automatic fitting procedure, using the curve\_fit function from Scipy, modelling each spectral line with a single Gaussian to estimate its flux (and luminosity) at each inclination. Because this routine returns a fit regardless of whether an emission line is present or not, we apply a signal-to-noise ratio (SNR) $> 5$ threshold to identify reliable detections. The SNR is defined as the ratio between the maximum line flux and the noise level, where the noise is estimated using the median absolute deviation of the spectrum excluding the emission line, measured over the velocity range 2500 and 4000 $\rm km \, s^{-1}$. 

We finally convert luminosities into SFRs. Starting with H$\alpha$, one of the most widely used conversions between its luminosity ($L_{\rm H\alpha}$) and SFR was derived by \citet{kennicutt1983} and calibrated for a \citet{chabrier2003} IMF by \citet{murphy2011}: 
\begin{equation}\label{eq:halpha-murphy11}
    \rm SFR_{H\alpha} \, [M_\odot yr^{-1}] = 5.37 \times 10^{-42} \, \textit{L}_{H\alpha} \, [erg \, s^{-1}]
\end{equation}

\noindent where $L_{\rm H\alpha}$ is the attenuation-corrected luminosity. More recently, \citet{reddy2022} derived a conversion factor for high-$z$ galaxies based on the BPASS population synthesis models with lower stellar metallicities ($Z_* = 0.001$) that includes stellar binaries assuming a \citet{chabrier2003} IMF with a maximal mass of 100 M$_\odot$:
\begin{equation}\label{eq:halpha-reddy22}
    \rm SFR_{H\alpha} \, [M_\odot yr^{-1}] = 2.14 \times 10^{-42}  \, \textit{L}_{H\alpha} \, [erg \, s^{-1}],
\end{equation}

To correct for attenuation, we use the relation between the intrinsic (F$_{\rm \lambda, int}$) and observed (F$_{\rm \lambda, obs}$) fluxes that is given in the form: 
\begin{equation}
    \rm F_{\lambda, int} = F_{\lambda, obs} \times 10^{0.4 \, A(\lambda)} , 
\end{equation}

\noindent where $ A(\lambda) = k(\lambda) \, E(B-V)$ is the attenuation at $\lambda$, $k(\lambda)$ is the attenuation curve value at $\lambda$, and $E(B-V)$ is the color excess between the $B$ and $V$ bands. In this case, the Balmer decrement is used to estimate the attenuation by comparing the observed line ratio of H$\alpha$ to H$\beta$ with the theoretical one. The H$\alpha$ attenuation, $A \rm _{H\alpha}$, can be summarized as follows: 
\begin{equation}\label{eq:attenuation}
    A_{\rm H\alpha} =  \frac{2.5 \, k_{\rm H\alpha} }{k_{\rm H\beta} - k_{\rm H\alpha}} \, \log \left( \frac{(\rm F_{H\alpha}  / F_{H\beta} )_{obs}}{(\rm F_{H\alpha} / F_{H\beta} )_{int}} \right), 
\end{equation}

\noindent where $\rm (F_{H\alpha} / F_{H\beta})_{obs}$ and $\rm (F_{H\alpha} / F_{H\beta})_{int}$ are the observed (attenuated) and intrinsic (dust-free) line ratios, respectively. We use the intrinsic ratio of 2.86, assuming Case B recombination with electron temperature $T = 10^4$ K and electron density $n_e = 100$ cm$^{-3}$ \citep{osterbrock2006}. $k_{\rm H\alpha} = 2.53$ and $k_{\rm H\beta} = 3.61$ represent the attenuation curves at $\rm H\alpha$ and $\rm H\beta$ wavelengths, respectively, adopting the \citet{calzetti2000} attenuation law for local starbursts.

In addition to the attenuation correction applied to the observed H$\alpha$ flux, this line is contaminated by the [\textsc{Nii}] doublet at restframe wavelengths 6548 and 6583 $\AA$. Several studies have investigated the impact of [\textsc{Nii}] emission on the total flux of H$\alpha$, particularly its evolution with redshift \citep[e.g.,][]{kewley2013,shapley2015, kashino2017}. Recent works suggest that the contribution from [\textsc{Nii}] becomes negligible at $z \gtrsim 4$ \citep[e.g.,][]{sanders2023, cameron2023} due to sub-solar metallicities, however other studies indicate that a small correction might be necessary with $96\%$ of the total flux dominated by H$\alpha$ \citep{sandles2024}. In this study, we adopt the latter. 

The IR luminosity is most commonly converted into SFR using the \citet{kennicutt1998} conversion modified for a \citet{chabrier2003} IMF:
\begin{equation}\label{eq:sfr-lir}
    \rm SFR_{IR} \, [M_\odot yr^{-1}] = 1.09 \times 10^{-10} \, \textit{L}_{IR} \, [\textit{L}_\odot], 
\end{equation}

\noindent where $L_{\rm IR}$ is the total IR luminosity integrated between restframe 8 and 1000 $\mu$m. While this conversion is not the most recent, it remains widely used for high-$z$ galaxies. At these redshifts, observational constraints typically provide access to only sparse continuum measurements \citep[e.g., ALPINE-\textsc{[Cii]} survey;][]{lefevre2020, fudamoto2020}. In this work, however, we compute $L_{\rm IR}$ using the full IR continuum emission from the \textsc{Skirt} outputs, representing an idealized scenario in which the dust emission is fully sampled. In Sect. \ref{section:ir-lum-realistic}, we discuss the implications for real observations and whether measuring the peak of dust emission at these redshifts (e.g., with ALMA Band 9) would be sufficient. 

\begin{figure}[t!]
    \centering
    \includegraphics[width=0.85\linewidth]{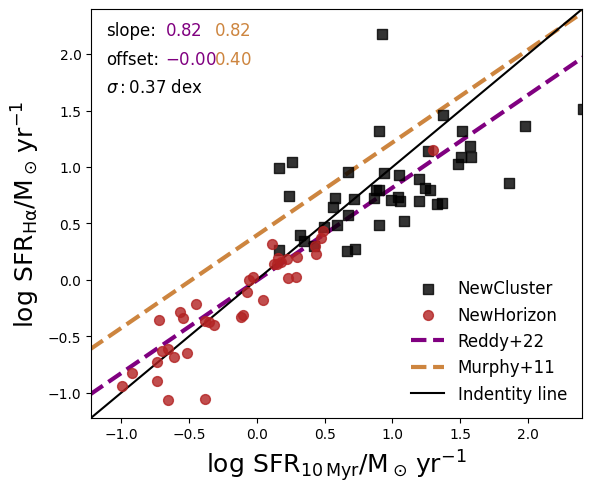}
    \caption{H$\alpha$-derived SFRs using Eq. \ref{eq:halpha-reddy22} versus SFR$_{\rm 10\,Myr}$ for \textsc{NewCluster} (black squares) and \textsc{NewHorizon} (red circles) simulated galaxies in a face-on configuration. The identity line is shown in black, the purple and orange dotted lines show the best-fit of the H$\alpha$-derived SFRs using the conversion of \citet{reddy2022} and \citet{murphy2011}, respectively (see Sect. \ref{section:halpha-sfr} for details). In the top left corner, the slopes and offsets are shown with the same color code as the lines, and the scatter $\sigma$ of the inferred SFRs.}
    \label{fig:sfr-halpha}
\end{figure}

Studies have also defined the total SFR which is the addition of the IR and far-UV SFRs \citep[e.g.,][]{bethermin2020, Schaerer2020}. This hybrid approach mitigates the dust attenuation that biases UV-based SFRs and missed IR contributions from less-obscured regions. The total SFR is defined as: 
\begin{equation}\label{eq:sfr-hybrid}
    \rm SFR_{IR + UV} \, [\textit{M}_\odot \, \rm yr^{-1}] = SFR_{IR} + SFR_{UV},
\end{equation}

\noindent where $\rm SFR_{IR} = \kappa_{IR} \, \textit{L}_{IR}$ is the total IR-inferred SFR as defined in Eq. \ref{eq:sfr-lir}. $\rm SFR_{UV} = \kappa_{UV} \, \textit{L}_{UV}$ where $L_{\rm UV}$ is the monochromatic luminosity at $1500 \, \AA$ and $\kappa_{\rm UV} = 1.47 \times 10^{-10} \, M_\odot \, \rm yr^{-1} \, \textit{L}_\odot^{-1}$ the conversion factor from luminosity to SFR assuming a Charbier IMF \citep{madau2014}. We note that we do not pursue a purely UV-based SFR analysis in this work since its interpretation is strongly affected by dust attenuation, whose impact has been extensively studied \citep[see][and references therein]{madau2014} and remains challenging to constrain.

Numerous studies have also explored the relation between the \textsc{[Cii]} luminosity, $L_{\textsc{[Cii]}}$, and the SFR as it probes the cooling of neutral and ionized gas heated by young starts \citep[e.g.,][]{stacey2010, delooze2014, herrera-camus2015, olsen2017,lagache2018, harikane2020, Schaerer2020, accard2025}. The relation can be expressed by a power-law: 
\begin{equation}
    \log \, (L_{[\rm \textsc{Cii}]}/L_\odot) = \alpha + \gamma \, \log \, (\rm SFR/M_\odot yr^{-1}), 
\end{equation}

\noindent where $\alpha$ and $\gamma$ are calibration coefficients that depend on the considered samples. \citet{delooze2014} investigated the reliability of the [\textsc{Cii}] line as a SFR tracer using a diverse sample of galaxies with various types. They found a wide range of calibration coefficients between metal-poor dwarf galaxies, local starbursts, ULIRGs, and high redshift \textst{($z < 2$)} galaxies. \citet{lagache2018} extended this analysis to $4<z<8$ using semi-analytical simulations coupled with the photoionization code \texttt{CLOUDY}, while assuming that the [\textsc{Cii}] line originates only from PDRs. They found a relation similar to the one found in \citet{delooze2014} for starburst galaxies and reported a small evolution of the $L_{[\rm \textsc{Cii}]} - \rm SFR$ relation with redshift for $L_{[\rm \textsc{Cii}]} > 10^7 \, L_\odot$. In this study, we calibrate the coefficients to our sample, which we discuss in Sect. \ref{section:cii-sfr}. 

\begin{figure}[t!]
    \centering
    \includegraphics[width=0.85\linewidth]{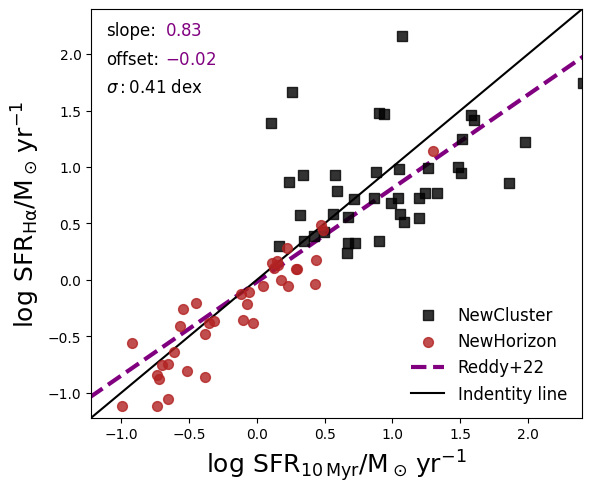}
    \caption{Same as Fig. \ref{fig:sfr-halpha} for an edge-on viewing angle.}
    \label{fig:sfr-halpha-edgeon}
\end{figure}

\section{Inferred SFRs}\label{section:results}
In this section, we derive the SFRs of each tracer: H$\alpha$ in Sect. \ref{section:halpha-sfr}, IR continuum in Sect. \ref{section:ir-sfr}, hybrid (IR + UV) in Sect. \ref{section:hybrid-sfr} and compare them to $\rm SFR_{true}$---the time-averaged SFR estimated from the simulation over 10 or 100 Myr. We note that when estimating $\rm SFR_{true}$, we correct stellar particle masses with ages $> 5 \, \rm Myr$ for mass loss due to mass ejection during feedback processes: 31\% in \textsc{NewHorizon} and 40\% in \textsc{NewCluster}. For each of the three tracers, we perform a power-law, finding the slope and offset based on the adopted conversions in Sect. \ref{section:sfr}. In Sect. \ref{section:cii-sfr}, we calibrate the $L_{\rm \textsc{[Cii]}} - \rm SFR$ coefficients to our sample of galaxies. The scatter of each tracer is estimated as the standard deviation of the residuals $= \log (\rm SFR_{tracer} / SFR_{true})$.

\subsection{H$\alpha$ SFR}\label{section:halpha-sfr}
In Fig. \ref{fig:sfr-halpha}, the Balmer-corrected H$\alpha$ SFRs as a function of SFR$_{\rm 10 \, Myr}$ are shown for a face-on configuration of \textsc{NewHorizon} and \textsc{NewCluster} simulated galaxies. A power-law fit to the simulation data using the \citet{reddy2022} conversion (Eq. \ref{eq:halpha-reddy22}) yields a slope of ($0.83 \pm 0.05$) and an insignificant offset of 0.005 dex. This means, we overestimate SFRs with intrinsic $\rm SFR_{10 \, Myr}$ below $5 \, M_\odot \, \rm yr^{-1}$ and overestimate larger values up to a factor of $\sim 1.5$. In contrast, the \citet{murphy2011} conversion (Eq. \ref{eq:halpha-murphy11}) yields the same slope, however, SFRs are systematically overestimated of $\rm SFR_{10 \, Myr} < 300 \, M_\odot \, \rm yr^{-1}$ with an offset of 0.4 dex. Based on these results, we adopt the \citet{reddy2022} conversion for the rest of the study, for which we find a scatter of 0.37 dex. 

In Fig. \ref{fig:sfr-halpha-edgeon}, we inspect the effect of inclination on the derived SFRs since geometrical configurations have a direct effect on attenuation especially at shorter wavelengths. For an edge-on configuration, we find that the inferred SFRs reproduce SFR$_{\rm 10 \, Myr}$ similar to the face-on configuration with a slope of ($0.83 \pm 0.06$) and a small offset of -0.02 dex. However, we find an increase in scatter up to 0.41 dex due to an increase in column densities along the line of sight. The scatter varies between different inclinations, reaching $\sim 0.5$ dex for an inclination of $60\degree$ (see Fig. \ref{fig:sfr-tracers} for effect of different inclinations). We note that H$\alpha$ is detected in most sources across inclinations, while H$\beta$ detections vary across inclinations, thus varying the number of sources for which we estimate a SFR, and might slightly vary the scatter.

A clear difference is observed between the two simulations: \textsc{NewHorizon} sources exhibit a lower scatter of $\sim 0.2$ dex, while \textsc{NewCluster} ones show a broader distribution with a scatter of $\sim 0.5$ dex and a small number of outliers. This trend continues to hold at different inclinations (see Fig. \ref{fig:sfr-halpha-edgeon} and \ref{fig:sfr-tracers}). We have inspected the line profiles and spectral fits of the \textsc{NewCluster} sources contributing to the larger scatter seen (i.e., those with residuals $> 0.7$ dex), and do not find any failure originating from the methodology of spectral fitting. These differences can originate from several reasons. \textsc{NewCluster} galaxies are generally more massive and with higher gas masses as compared to \textsc{NewHorizon} ones, leading to higher dust masses at a fixed DTM (see Eq. \ref{eq:dust-mass-skirt}), and larger dust optical depths. Comparison of lower-mass \textsc{NewCluster} galaxies with \textsc{NewHorizon} counterparts reveals similar behavior, suggesting that the scatter is  primarily driven by underlying physical processes rather than simulation-specific conditions.

While on average a single attenuation curve has shown to produce adequate results for $z = 5$ galaxies from the low- to the high-mass end, increased dust optical depths, dust distributions and dust geometry make attenuation curves non-trivial \citep[e.g.,][]{Scicluna2015, lin2021}. This highlights that optical tracers at these redshifts are highly susceptible to uncertainties introduced by complex dust physics, particularly for high mass systems. We further explore the attenuation curves in Sect. \ref{section:attenuation}.

\subsection{IR SFR}\label{section:ir-sfr}
\begin{figure}[t!]
    \centering
    \includegraphics[width=0.85\linewidth]{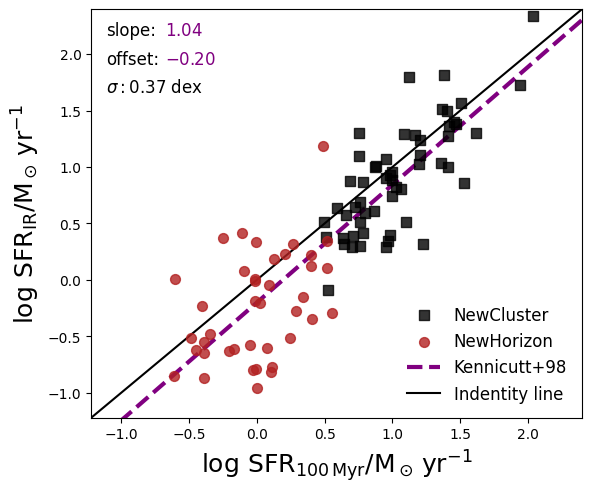}
    \caption{Same as Fig. \ref{fig:sfr-halpha}, for the IR-inferred SFRs versus time-averaged SFR over 100 Myr. The SFRs are derived using \citet{kennicutt1998} conversion (Eq. \ref{eq:sfr-lir}) and the best fit is the purple dashed line.}
    \label{fig:sfr-ir}
\end{figure}
The inferred SFRs from the total IR luminosity using the \citet{kennicutt1998} conversion reproduces the SFRs traced by the simulations over 100 Myr timescales with a slope close to unity ($1.09 \pm 0.06$) and an offset of -0.19 dex. In contrast to $\rm SFR_{H\alpha}$, we systematically underestimate $\rm SFR_{100 \, Myr}$ below $\sim 30 -50 \, M_\odot \, \rm yr^{-1}$ and overestimate larger SFRs by a factor of $\sim 1.5$. 

We find a scatter of 0.37 dex whose origin is generally due to differences in stochastic heating of small dust grains \citep{draine2001}. As expected, we do not see variations in the resulting SFRs (and a constant dispersion) when changing the galaxies' inclination (Fig. \ref{fig:sfr-tracers}). However, we can see a difference in the scatter between \textsc{NewHorizon} (0.41 dex) and \textsc{NewCluster} galaxies (0.28 dex), which could be attributed to several factors such as galaxy mass, star formation histories (SFH), and dust content. 

\begin{figure}[t!]
    \centering
    \includegraphics[width=0.85\linewidth]{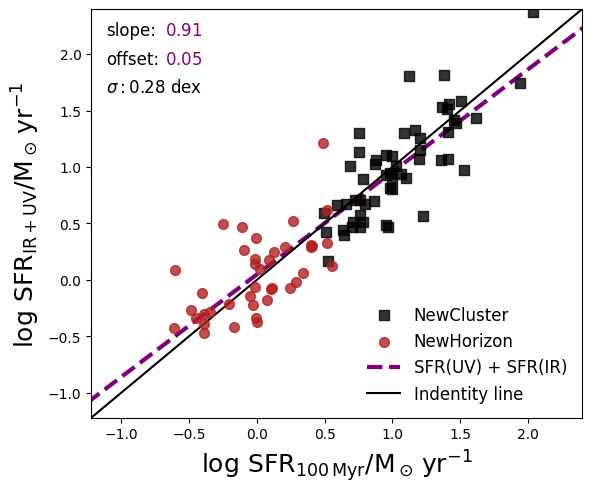}
    \caption{Same as Fig. \ref{fig:sfr-halpha}, for the Hybrid-inferred SFRs (IR + UV) versus time-averaged SFR over 100 Myr. The SFRs are derived using the conversion in Eq. \ref{eq:sfr-hybrid} and the best fit is the purple dashed line.}
    \label{fig:sfr-hybrid}
\end{figure}

As mentioned in Sect. \ref{section:halpha-sfr}, \textsc{NewHorizon} galaxies have lower dust content compared to \textsc{NewCluster} ones. Systems with lower dust masses tend to have a higher leakage of UV photons and thus lower IR luminosities \citep{buat2007}, where it is argued that SFRs become questionable when $L_{\rm IR} < 10^{9.5} \, L_\odot$ \citep{rieke2009}. We estimate the monochromatic far-UV luminosities for a face-on configuration using the rest-frame $1500 \, \AA$ flux densities and infer the IR-to-UV ratio. We find that this ratio varies homogeneously across both samples (see Fig. \ref{fig:uv-lir-scatter}), which explains some of the increased scatter observed for \textsc{NewHorizon} galaxies. We discuss this further in Sect. \ref{section:hybrid-sfr}. 

Studies have also shown that low-mass galaxies (at $M_* < 10^{8.5} \, M_\odot$) experience more stochastic, bursty star formation compared to their higher mass counterparts \citep[e.g.,][]{ciesla2024, perry2025, alba2025}. Both the \textsc{NewHorizon} and \textsc{NewCluster} simulations capture this bursty star formation \citep[see details in ][]{dubois2021NH, han2025b}. Indeed, the SFH of our galaxies shows evidence of bursty star formation over timescales of $\sim 10$ Myr (see Appendix \ref{apdx:sfh} and Fig. \ref{fig:sfr-scattered}). This means that, while the IR conversion to SFR offers a good estimate of SFRs time-averaged over 100 Myr, it fails to do so when galaxies undergo bursty episodes of star formation leading to increased uncertainties, particularly for low mass \textsc{NewHorizon} galaxies ($M_* < 10^9 \, M_\odot$), that is in agreement with the literature.

\subsection{Hybrid SFR (IR + UV)}\label{section:hybrid-sfr}
Figure \ref{fig:sfr-hybrid} shows the hybrid SFR estimated using Eq. \ref{eq:sfr-hybrid} versus $\rm SFR_{100 \, Myr}$ for our sources. We find a slope very close to unity ($0.93 \pm 0.04$) and an offset of 0.1 dex. As expected from the previous discussion on $\rm SFR_{IR}$, the dispersion significantly decreases to 0.27 dex. 

\begin{figure*}[ht!]
    \centering
    \includegraphics[width=\linewidth]{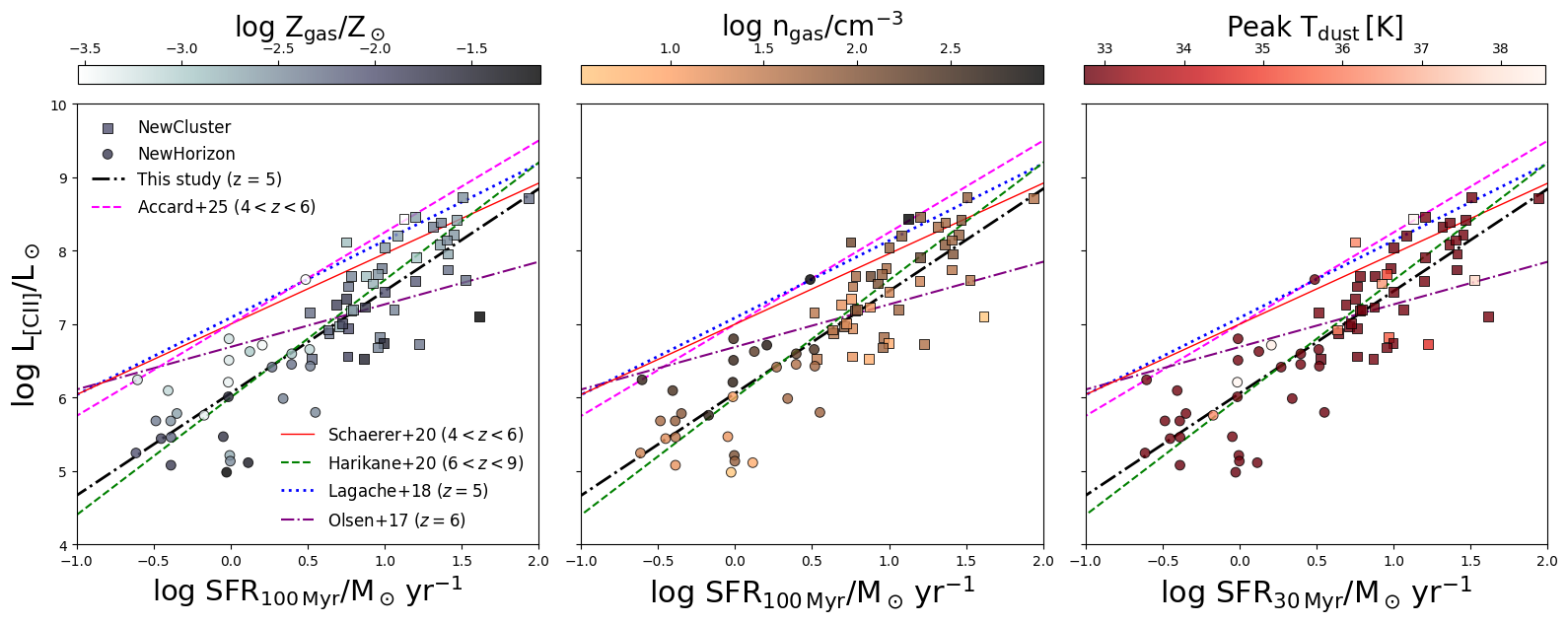}
    \caption{[\textsc{Cii}] luminosity as a function of SFR$_{\rm 100 \,Myr}$ of \textsc{NewHorizon} and \textsc{NewCluster} simulated sources shown in circles and squares, respectively. The data points are color-coded based on the median of gas metallicities (\textit{left}), the median of gas densities (\textit{middle}), and the peak temperature (\textit{right}) in a given source (see Sect. \ref{section:cii-sfr} for details). The black dashed-dotted line shows the $\rm L_{[\textsc{Cii}]} - SFR$ relation we obtain in this study. The purple dashed-dotted, blue dotted, red, green dashed, and magenta dashed lines represent $\rm L_{[\textsc{Cii}]} - SFR$ relations by  \citet[$z=6$]{olsen2017}, \citet[$z = 5$]{lagache2018}, \citet[$4 < z < 6$]{Schaerer2020}, \citet[$6 < z < 9$]{harikane2020}, and \citet[][$4 < z < 6$]{accard2025}, respectively.}
    \label{fig:Lcii-sfr}
\end{figure*} 

The scatter significantly decreases for the less massive \textsc{NewHorizon} sources, dropping from 0.41 dex for IR-inferred SFRs (see Fig. \ref{fig:sfr-ir}) to 0.3 dex, however only a slight decrease in scatter is observed for \textsc{NewCluster} sources (from 0.28 to 0.25 dex). The lower gas content ($\propto$ dust content) in these systems leads to significant UV photon leakage (see Fig. \ref{fig:uv-lir-scatter}) which makes the IR luminosity an incomplete tracer of the total star formation. By accounting for the unattenuated UV component, the hybrid tracer recovers the escaped radiation, providing a more robust SFR time-averaged over 100 Myr particularly for the low-mass regimes ($M_* < 10^9 \, M_\odot$). Additionally, the offset decreases from 0.1 to 0.06 dex when rotating the galaxies to an edge-on configuration (Fig. \ref{fig:sfr-tracers}). This improvement is likely due to the increased optical depth along the line of sight, which decreases the sensitivity to UV luminosity variations that may contribute to the observed offset.

Despite these improvements, a scatter of 0.27 dex remains. This suggests that while the hybrid SFR corrects for underestimated IR luminosities due to UV-photon leakage, it cannot eliminate uncertainties arising from SFH stochasticity and the stochastic heating of small dust grains---both of which contribute to the scatter of IR luminosities.

\subsection{[\textsc{Cii}] SFR}\label{section:cii-sfr}
In Fig. \ref{fig:Lcii-sfr}, we plot $L_{[\rm \textsc{Cii}]}$ as a function of $\rm SFR_{100 \, Myr}$. We perform a linear regression fit to our sample of simulated sources, deriving the following relation: 
\begin{equation}
       \log \, (L_{[\rm \textsc{Cii}]}/L_\odot) = (6.06 \pm 0.09) + (1.39 \pm 0.10) \log \, \rm (SFR/M_\odot yr^{-1}). 
\end{equation}

The slope we find is generally steeper than most values derived in the literature. \citet{Schaerer2020} derived the relation for massive ($M_* > 10^{10} \, M_\odot$) $z \sim 5$ ALMA-observed sources from the ALPINE \citep{lefevre2020, fudamoto2020} sample of galaxies, finding a slope close to unity ($0.96 \pm 0.09$).  Our values are, however, closer to the findings of \citet{accard2025} where the reported slope $\sim 1.25 \pm 0.25$ using resolved measurements of ALMA-observed galaxies at $z \sim 5$. \citet[][]{harikane2020} find an even steeper $\rm slope = 1.6$ for a sample of $z= 6-9$ ALMA-observed galaxies. \citet{lagache2018} derive a relation that varies with redshift, where the $\rm slope = 1.05$ at $z=5$ for a sample of semi-analytical simulated galaxies with $L_{\textsc{[Cii]}} > 10^7 \, L_\odot$. Interestingly, some of our sources, particularly at similar $L_{\textsc{[Cii]}}$, lie on the relation found in \citet{lagache2018}. \citet{olsen2017} found a significantly shallower slope of $0.58 \pm 0.11$ at $z=6$ for a sample of 30 main-sequence galaxies from the \textsc{mufasa} cosmological simulation \citep[][]{dave2016}, post-processed with \texttt{CLOUDY}. Their inferred \textsc{[Cii]} luminosities are generally fainter ($6 \times 10^6 < L_{[\textsc{Cii}]} < 6 \times 10^7 \, L_\odot$) than those in \citet{lagache2018}. By limiting our sample to the respective luminosities of the mentioned samples, we find slopes of $0.99 \pm 0.18$ and $0.37 \pm 0.09$ {---} both of which agree with \citet{lagache2018} and \citet{olsen2017}. This highlights that the slope of the $ L_{\rm [\textsc{Cii}]} - \rm SFR$ relation may depend on the luminosity range probed. 

Inferred SFRs show a uniform behavior across both samples. We find that they reproduce $\rm SFR_{100 \, Myr}$ with a scatter of 0.35 dex at all inclinations (see Fig. \ref{fig:sfr-tracers}).  Given that [\textsc{Cii}] is generally optically thin and is unaffected by dust attenuation, we do not expect significant variations in the integrated luminosities or the corresponding inferred SFRs. \citet{delooze2014} demonstrated that while a correlation exists between the [\textsc{Cii}] emission and SFR, this relation is dispersed where they find a scatter of $\sim 0.41$ dex for observed galaxies. Recent results of the resolved \textsc{[Cii]} - SFR relation \citep{accard2025} of ALMA-observed galaxies at $z\sim5$ show a smaller intrinsic of $0.19$ dex, suggesting that properly accounting for measurements and their uncertainties can significantly decrease the scatter. It is also shown that the relation is highly dependent on gas density, metallicity, and dust temperature \citep[see also][]{herrera-camus2015, liang2024}. 

\begin{figure*}[ht!]
    \centering
    \includegraphics[width=\linewidth]{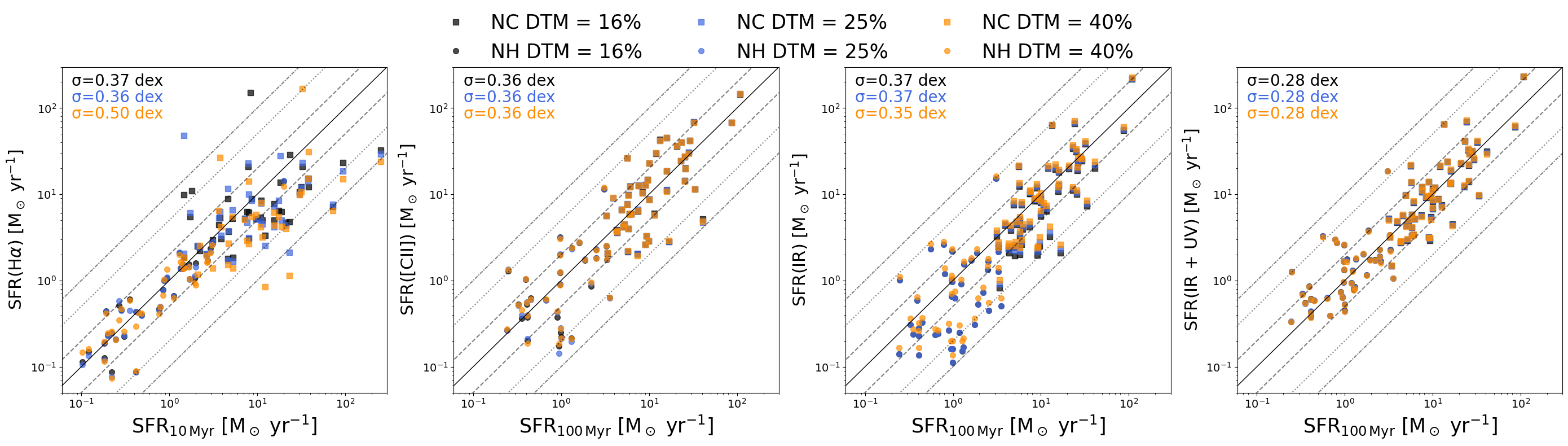}
    \caption{DTM effect on SFR estimations of H$\alpha$ versus SFR$_{\rm 10 \, Myr}$, [\textsc{Cii}] at $158 \, \mu$m versus SFR$_{\rm 100 \, Myr}$, IR versus SFR$_{\rm \, 100 Myr}$, and Hybrid (UV+IR) versus SFR$_{\rm \, 100 Myr}$ (\textit{left to right}), for \textsc{NewCluster} (squares) and \textsc{NewHorizon} (circles) simulated galaxies. The DTM ratios considered are $16\%$ (\textit{black}), $25\%$ (\textit{blue}), and $40\%$ (\textit{orange}). The solid grey line is the identity line, while the dashed, dotted and dash-dot lines indicate offsets of $\pm 0.3$, $\pm 0.7$, and $\pm 1$ dex, respectively. On the top left corner of each plot, the dispersion of the entire sample is shown. }
    \label{fig:sfr-dtm}
\end{figure*}

\begin{figure}[t!]
    \centering
    \includegraphics[width=\linewidth]{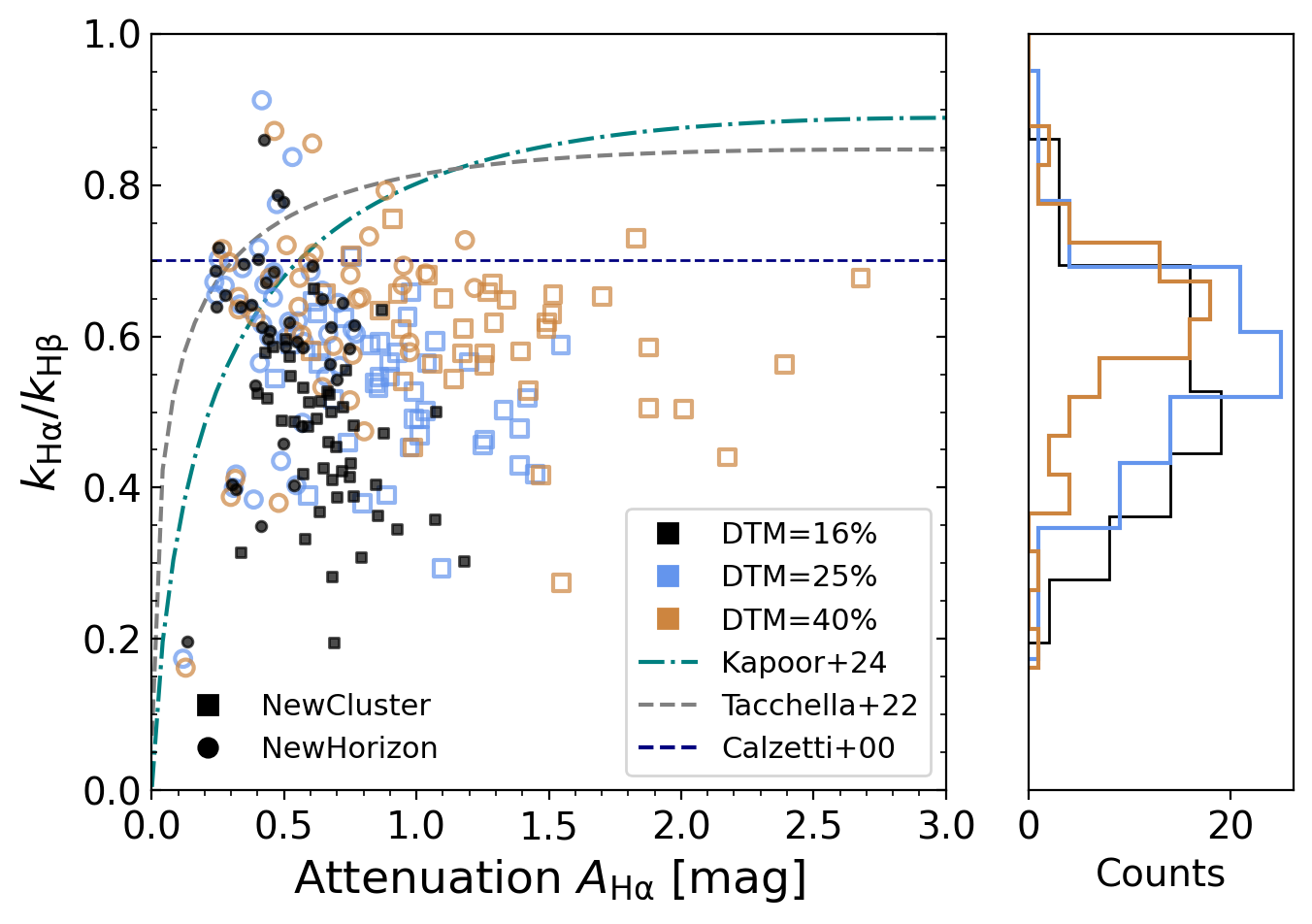}
    \caption{\textit{Left:} Ratio of the H$\alpha$ and H$\beta$ attenuation laws derived versus the H$\alpha$ attenuation, for \textsc{NewHorizon} (circles) and \textsc{NewCluster} (squares) simulated galaxies. The black, blue, and orange colors represent the DTM ratios of 16\%, 25\%, and 40\%, respectively. The ratio derived for simulated galaxies from \citet{tacchella2022} (grey dashed line) and \citet{kapoor2024} (teal dashed-dotted line), and that derived by \citet{calzetti2000} (blue dashed line) for local starbursts, are shown. \textit{Right:} Distribution of the derived attenuation law ratios for our sources at different DTM ratios. }
    \label{fig:attenuation}
\end{figure}

While our data are unaffected by observational uncertainties, we further look into the dependence of this correlation on the physical parameters of the ISM. In \textsc{Skirt}, the [\textsc{Cii}] line is produced using the \texttt{TODDLERS} template for star-forming regions. To this end, for each galaxy we compute the median gas density and the median metallicity of star-forming regions with stellar ages $<10$ Myr (see Sect. \ref{section:SKIRT} for our detailed configuration in \textsc{Skirt}). In the left and middle panels, we show the $ L_{[\rm \textsc{Cii}]} - \rm SFR$ color-coded by the gas metallicity and density, respectively. It is clear that galaxies with lower metallicity lie above the relation and higher metallicity lie below, while the gas density shows the opposite of the trend found for the metallicity.

We also asses the dependence on the dust temperature (right panel). For simplicity, we estimate the dust temperature by assuming Wien's law\footnote{At a fixed dust emissivity $\beta$, the correction applied to the Wien peak temperature to derive, for example, the modified blackbody temperature results in a constant rescaling \citep{liang2019}.} ($\rm \lambda_{peak} \, T = 2.989 \times 10^3 \, \mu m \, K$). However, most of the sources' peak temperatures are $\rm \sim 33 \, K$ making the sample relatively homogeneous. Only a few sources have higher temperatures, reaching $\sim 40 \rm \, K$, and over this limited 10 K range, a temperature dependence is not evident.

\section{Discussion}\label{section:discussion}
\subsection{Impact of DTM ratio on SFRs}\label{sec:impact-of-dtm}
The DTM has been studied in depth, particularly in the local Universe \citep[e.g.,][]{remy-ruyer2014, devis2019}. At higher redshifts, studies suggest that the DTM ratio has a dependence on both the gas metallicity \citep[e.g.,][]{Li2019} and redshift \citep[e.g.,][]{inoue2003}. \citet{popping2017} found a varying DTM with gas metallicity, stellar mass, and redshift, with DTM < $18 \%$ at $z = 5$ for all metallicities and stellar masses, which is consistent with the relation of \citet{vogelsberger2022b}. A ratio similar to the Galactic one is also adopted throughout the literature \citep[$\sim 40\%$][]{liang2019}. In this section we explore higher DTM ratios, particularly 25\% and 40\% (in addition to the previously adopted value of $16 \%$), to understand its impact on the inferred-SFRs. The results are shown in Fig. \ref{fig:sfr-dtm} for all four tracers. 

As expected, the scatter increases substantially for the H$\alpha$-derived SFRs (from 0.37 to 0.5 dex). At these wavelengths, the attenuation strongly affects the ability to recover the intrinsic SFRs, and as previously stated, larger dust reservoirs (i.e., higher optical depths for a fixed galaxy size) introduce more scatter due to the complexity of attenuation curves \citep[e.g.,][]{Scicluna2015, lin2021}. Additionally, the total sample decreases due to H$\beta$ non-detections for DTM = 40\% with a total of 71 sources, compared to 85 sources for a DTM of 16\%, further increasing the scatter. In contrast, [\textsc{Cii}], IR continuum, and hybrid (UV+IR) SFRs show little-to-no variation when changing the DTM. Nevertheless, we find a modest increase in the IR-inferred SFRs, by a factor of 1.13, when adopting a DTM of 40\%. This trend reflects the higher dust masses at larger DTM values, which enhance the far-IR peak of the SED and thus increase the total IR luminosities. In addition, the scatter in the IR-inferred SFRs decreases slightly, from 0.37 to 0.35 dex, which is driven by the increased dust column densities limiting the leakage of UV photons.

For the remainder of the study, we adopt $\rm DTM = 16 \%$ unless stated otherwise. 

\begin{figure*}[t]
    \centering
    \includegraphics[width=0.85\linewidth]{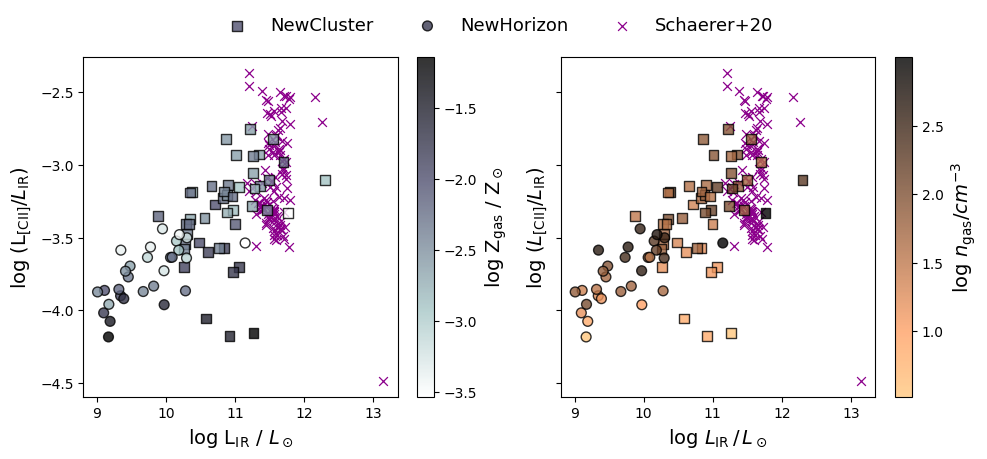}
    \caption{[\textsc{Cii}]/IR luminosity ratios versus IR luminosity of \textsc{NewCluster} (squares) and \textsc{NewHorizon} (circles). We color code our sources based on the median gas metallicity (left) and median gas density (right) of each galaxy. In purple, the observed ALPINE $z \sim 4-6$ sources \citep{Schaerer2020} are plotted for comparison.}
    \label{fig:Lcii-Lir}
\end{figure*}

\subsection{H$\alpha$ attenuation}\label{section:attenuation}
The H$\alpha$-derived SFRs are strongly dependent on the assumed attenuation law. While the \citet{calzetti2000} laws have been widely used for high-$z$ studies, particularly the calibrations made for local starbursts, we are still limited in our understanding of the ISM at these redshifts. Here, we explore the attenuation curve for our simulated sources from both \textsc{NewHorizon} and \textsc{NewCluster} simulations. 

Using the \texttt{TODDLERS} library, we process the sources with \textsc{Skirt} while excluding dust (\textit{includeDust="false"}) following the same criteria as outlined in Sect. \ref{section:SKIRT}. This allows us to recover the intrinsic H$\alpha$ line flux. Combined with the attenuated line fluxes, we estimate the H$\alpha$ attenuation $A_{\rm H\alpha} = -2.5 \log (F_{\rm obs} / F_{\rm int})$. We also estimate the ratio of H$\alpha$ and H$\beta$ attenuation curves, $k_{\rm H\alpha} / k_{\rm H\beta}$, using Eq. \ref{eq:attenuation}. In Fig. \ref{fig:attenuation}, the derived attenuation curve ratios for a varying DTM are shown as compared to some of the attenuation laws in the literature.

We find that our sources exhibit a large scatter and systematically lie below the relations derived in both \citet{tacchella2022} and \citet{kapoor2024} for local simulated sources and post-processed with \textsc{Skirt}. We note that their curves were derived from spatially resolved maps, that we do not pursue in this study. As expected, the attenuation ($A_{\rm H\alpha}$) increases with DTM. Across both \textsc{NewHorizon} and \textsc{NewCluster} sources, the resulting $k_{\rm H\alpha} / k_{\rm H\beta}$ ratios span broad ranges: for DTM = 16\% the values range from $\sim 0.4-0.7$, for DTM = 25\% from $\sim 0.5-0.7$, and for DTM = 40\% from $\sim 0.55-0.72$ (see histogram in Fig. \ref{fig:attenuation}). These values are generally in agreement with earlier high redshift studies \citep[$ z < 3$; e.g.,][]{buat2012, kriek2013, reddy2015} and more recent ones with JWST \citep{cooper2025olivia}, who argue that attenuation curves are steeper (i.e., lower $k_{\rm H\alpha} / k_{\rm H\beta}$ values) at these redshifts as compared to the widely used curve with a ratio $\sim 0.7$ found by \citet{calzetti2000} for local starbursts. Additionally, the attenuation curves flatten with DTM as expected due to an increase in the optical depth \citep{narayanan2018}. Sources from both simulations exhibit differences in the attenuation curve where \textsc{NewHorizon} galaxies exhibit shallower curves (i.e., higher $k_{\rm H\alpha} / k_{\rm H\beta}$ values) and vice versa, which are attributed to their physical conditions \citep[e.g.,][]{shivaei2020, shivaei2025}. By using an average value of the attenuation curve ratio of 0.58, we re-estimate the H$\alpha$-corrected SFRs (see Fig. \ref{fig:new-attenuation-sfr}), and find a significant decrease in the scatter reaching 0.16 dex at SFR $< 5 \rm \, M_\odot \, yr^{-1}$ however underestimates the SFR by a factor of $\sim 2$ at larger SFRs. These results highlight the importance of obtaining a more diverse library of attenuation curves that could better constrain the corrections applied to the attenuation at high-$z$.

\subsection{Do we find a [\textsc{Cii}]-deficit?}\label{section:lcii-deficit}
In the local Universe, the \textsc{[Cii]} deficit is defined as a systematic decrease of the $L_{\rm [\textsc{Cii}]}/L_{\rm IR}$ ratio with increasing $L_{\rm IR}$, observed for luminous and ultra-luminous infrared galaxies \citep[e.g.,][]{munoz2016, diaz-santos2017}. In contrast, at high-$z$, it is referred to describe galaxies with $L_{\rm [\textsc{Cii}]}/L_{\rm IR} < 10^{-3}$ for IR-bright sources ($\gtrsim 10^{12} \, L_\odot$), even in the absence of a decreasing trend. This deficit is attributed to different factors such as increased dust column densities, reduced photoelectric heating, or [\textsc{Cii}] saturation at high gas temperature \citep[e.g.,][]{casey2014, Goicoechea2015, munoz2016}.  

\begin{figure}[t!]
    \centering
    \includegraphics[width=\linewidth]{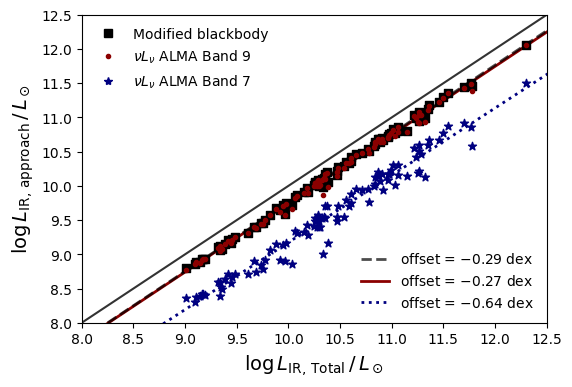}
    \caption{IR luminosities derived using different approaches versus the IR luminosity derived in Sect. \ref{eq:sfr-lir} for our entire sample. The black squares show $L_{\rm IR}$ derived using a modified blackbody, the red circles are $L_{\rm IR}$ derived from the monochromatic luminosity from ALMA band 9, and blue stars those derived from the monochromatic luminosity from ALMA band 7. The black line is the identity line, whereas the dashed black, solid red, and dotted blue, represent the best-fit of each inferred $L_{\rm IR}$.}
    \label{fig:lir-realistic}
\end{figure}

In Fig. \ref{fig:Lcii-Lir}, we plot the $L_{\rm [\textsc{Cii}]} / L_{\rm IR}$ ratio estimated for the \textsc{NewHorizon} and \textsc{NewCluster} simulated galaxies as a function of the IR luminosity. The $L_{\rm [\textsc{Cii}]} / L_{\rm IR}$ ratio varies between $\sim 10^{-4}$ and $\sim 2\times 10^{-3}$ for the entire sample. Within our luminosity ranges considered, particularly within the $10^{11} - 10^{12} \, L_\odot$ range, the ratio is comparable to $z \sim 5$ sources. For comparison, we plot the ALPINE-[\textsc{Cii}] survey sources \citep{Schaerer2020} that align with the most luminous \textsc{NewCluster} sources, even though they occupy a narrow range in the $L_{\rm [\textsc{Cii}]} / L_{\rm IR} - L_{\rm IR}$ space. While ALPINE sources are more massive than \textsc{NewCluster} galaxies, we are able to reproduce on average the observed ratios of bright \textsc{[Cii]} emitters. In general, we find that galaxies from both simulations display an increasing trend in the $L_{\rm [\textsc{Cii}]} / L_{\rm IR} - L_{\rm IR}$ plane. However, some sources deviate from the general trend, showing exceptionally low ratios. To explore the deviation from the general trend, we investigate the ratio's dependence on both gas metallicity and gas density, represented by color-maps in Fig. \ref{fig:Lcii-Lir}. In the left panel, a clear deviation is observed for some sources with increased metallicities, where the $L_{\rm [\textsc{Cii}]} / L_{\rm IR}$ ratio decreases with increasing $L_{\rm IR}$ particularly for the \textsc{NewCluster} sources. In the right panel, the gas density shows a similar trend, where a [\textsc{Cii}]-deficit appears to follow sources with lower gas density. At this stage, it is difficult to draw conclusions from a few data points. As reported previously, we do not find a difference in the [\textsc{Cii}] luminosities and a modest increase of 13\% in IR luminosities while varying the DTM ratios, which means that increased dust column densities would not have a significant effect of the \textsc{[Cii]}-IR luminosity ratios of our sample.

\subsection{$L_{\rm IR}$ with realistic IR sampling}\label{section:ir-lum-realistic}
In Sect. \ref{section:ir-sfr}, we presented an ideal-case scenario of estimating $L_{\rm IR}$ assuming that we are able to sample the IR SED between restframe 8 and 1000 $\mu$m. However, this is not a realistic coverage for high-$z$ galaxies whose current sampling is limited to a few measurements in the far-IR with instruments like ALMA and, when available, mid-IR from JWST-MIRI. Particularly, some of the [\textsc{Cii}] surveys at $z > 4$ are limited to one photometric measurement in the IR, which has motivated approaches based on stacking \citep[e.g.,][]{bethermin2020} or deriving dust temperatures from a single continuum measurement combined with priors or scaling relations \citep{sommovigo2022}. \citet{bethermin2020} also derive the ratio of monochromatic continuum luminosity to the total IR luminosity. Here, we investigate two approaches: 1) luminosities inferred from a modified blackbody and 2) monochromatic luminosities ($\nu L_\nu$) at restframe frequency $\nu$.

To understand whether a modified blackbody (MBB) could estimate a reliable $L_{\rm IR}$, we assume we are able to sample the peak of dust emission in addition to the photometric data from ALMA band 7 where the [\textsc{Cii}] line is observed. For galaxies at $z=5$, the peak of dust emission is roughly in the observed wavelength range $\lambda \sim 0.4 - 0.5$ mm which is covered by ALMA Band 9. From \textsc{Skirt}, we obtain the broadband fluxes\footnote{ \textsc{Skirt} computes broadband fluxes by convolving them with the built-in transmission curves for each band. See more: \href{https://skirt.ugent.be/skirt9/class_broad_band.html}{skirt.ugent.be/skirt9/class$\_$broad$\_$band.html}.} of ALMA bands 7 and 9 with central wavelengths at 450 and 940 $\mu$m, respectively (see Fig. \ref{fig:sed}). The data is fit with a single-temperature MBB in its optically thin form and sampled using \textsc{emcee} \citep{emcee}. We follow the method described in \citet{ismail2023} which also accounts for the CMB effects \citep{dacunha2013}, however, we fix the value of the dust emissivity index to $\beta = 2.2$ following the findings in \citet{ismail2023, boogaard2025, algera2025b}. In the case of monochromatic luminosities, we estimate $\nu L_\nu$ for both ALMA bands 7 and 9 that correspond to restframe wavelengths $\lambda_{\rm Band \, 9}^{\rm rest} = 75 \, \mu$m and $\lambda_{\rm Band \, 7}^{\rm rest} \sim 157 \, \mu$m at $z=5$, respectively.

In Fig. \ref{fig:lir-realistic}, we compare IR luminosities inferred from the different approaches to the total $L_{\rm IR}$ derived from a fully sampled continuum in Sect. \ref{section:ir-sfr}. We also fit a power-law to assess the relation between them. We find that all three cases, the MBB and the monochromatic luminosities, systematically underestimate the total IR luminosity, however, the resulting slopes are $\sim 1$. The MBB fit underestimate the total luminosity by a factor of $\sim 2$ ($-0.29 \pm 0.06$ dex). The is expected since the MBB models the far-IR emission of large thermal dust grains but fails to capture the contribution from small grains and PAHs at wavelengths $\lambda_{\rm rest} < 50 \, \mu$m. Interestingly, the monochromatic luminosity at ALMA band 9 performs similarly to the MBB with an offset of $-0.27 \pm 0.08$ dex. This arises because Band 9 closely samples the peak of the dust SED of our sample. Given that our galaxies exhibit a relatively uniform peak dust temperature of $\sim 33$ K (see Sect. \ref{section:cii-sfr}), the monochromatic measurement at restframe $75 \, \mu$m captures the bulk of the energy emission with little scatter, which would not be the case if the dust temperature varies. The monochromatic luminosity in Band 7, on the other hand, underestimates the total $L_{\rm IR}$ by a factor of $\sim 4.4$ ($\-0.64 \pm 0.19$ dex) that is smaller the the factor found in \citet[][$\sim 6.5$]{bethermin2020}, however, their SEDs exhibit dust temperatures $> 40$ K. While the correlation stays tight following a nearly linear slope ($0.98 \pm 0.02$), the scatter is larger since the emission is sensitive to the dust mass and dust grain properties at these wavelengths, assuming a constant dust temperature. 

Nevertheless, the linearity of these relations provide a framework to correct for the total $L_{\rm IR}$ for $z=5$ galaxy samples where current observations are limited in photometric measurements.

\section{Summary and Conclusions}\label{section:conclusion}
In this study, we analyze simulated galaxies at $z=5$ drawn from two hydrodynamical cosmological zoom-in simulations, \textsc{NewHorizon} and \textsc{NewCluster}, which are post-processed with the 3D radiative transfer code \textsc{Skirt}. We derive SFRs using observables relevant at high-$z$: the H$\alpha$ nebular emission line, the [\textsc{Cii}] 158 $\mu$m fine-structure line, the FUV and IR continua. These observables are generated at spectral resolutions comparable to those of current facilities, namely JWST and ALMA. We further examine how the inferred SFRs depend on the physical properties of the interstellar medium, as well as on intrinsic factors such as the viewing angle and the dust-to-metal ratio. Our main findings are summarized below.

\begin{itemize}
    \item The H$\alpha$–inferred SFRs, corrected via the Balmer decrement, reproduce the intrinsic SFRs with timescales of 10 Myr timescales using the \citet{reddy2022} conversion, while it is overestimated when assuming the \citet{kennicutt1983} conversion. Nevertheless, the inferred SFRs exhibit the largest dispersion among all tracers, with a scatter varying between $0.3$ to $0.5$ dex for the entire sample, due to a dependence on both the viewing angle and the assumed dust-to-metal ratio. For more massive sources ($M_* > 10^9 \, M_\odot$), the scatter is significantly larger ($\sim 0.5$ dex) than for the lower-mass galaxies ($\sim 0.2$ dex), primarily owing to their larger dust reservoirs and correspondingly higher dust optical depths, which increase the complexity of attenuation corrections. We infer an effective attenuation ratio of $k_{\rm H\alpha} / k_{\rm H\beta}$ spanning $0.4-0.7$ across different dust-to-metal ratio assumptions, generally steeper than the canonical ratio of $0.7$ usually assumed. Adopting a value of 0.58 reduces the scatter at low SFRs but leads to an underestimation of the inferred SFRs by a factor of $\sim 2$ at high intrinsic SFRs.

    \item  The IR-inferred SFRs reproduce intrinsic SFRs time-averaged over 100 Myr with a slope close to unity and a small offset of -0.19 underestimating intrinsically low SFRs. These SFRs exhibit a significant scatter of 0.37 dex that is due to bursty SFHs and to UV-photon leakage which becomes significant in galaxies with lower gas content. Under realistic sampling of the IR energy distribution, we derive correction factors to recover the IR luminosities from limited data. We find that $L_{\rm IR}$ is underestimated by a factor of $\sim2$ when using a MBB and monochromatic luminosities at the peak of the SED (ALMA Band 9 in this case). Measurements along the RJ tail with ALMA band 7 exhibit a larger underestimation by a factor of $\sim 4.4$. 
    
    \item The hybrid SFR ($\rm UV + IR$) shows a reduced scatter of 0.27 dex compared to IR-based SFRs. The inferred SFRs reproduce intrinsic SFRs over 100 Myr timescales, with a slope close to unity and a small offset of 0.1 dex. This hybrid estimator mitigates the impact of UV-photon leakage that contributes to the scatter in IR continuum-based SFRs, while simultaneously avoiding the need for uncertain dust attenuation corrections to the UV emission, making it a robust SFR tracer for UV-selected samples particularly for galaxies with lower gas content.

    \item The [\textsc{Cii}]-inferred SFRs show a consistent, yet relatively high scatter of 0.36 dex around the identity line. We find the [\textsc{Cii}] to be independent of the viewing angle and DTM ratio, making it more robust as a SFR tracer than the former optical line. The $L_{\textsc{[Cii]}} - \rm SFR$ relation for our sample of galaxies exhibits a steep slope of $\sim 1.4$ and a high scatter. We demonstrate that the steepness of this relation is sensitive to the range of luminosities assumed for the fit, and the scatter to be dependent on both gas metallicity and gas density, which is in agreement with literature studies. We also show that our most massive galaxies from \textsc{NewCluster} reproduce the $L_{\textsc{[Cii]}}/L_{\rm IR}$ ratios of ALMA-observed sources from the ALPINE survey. A general increasing trend is observed between $L_{\textsc{[Cii]}}/L_{\rm IR}$ and $L_{\rm IR}$, however, a [\textsc{Cii}]-deficit is not very clear with the current galaxy sample.

\end{itemize}

In summary, this work provides quantitative benchmarks for widely-used observables at $z = 5$ across galaxies spanning a wide range of stellar masses ($8 < \log M_*/M_\odot < 10$). In the future, we aim at further exploring these tracers in more detail, particularly emission lines, while also integrating non-uniform dust distribution and properties that are modeled on-the-fly in the \textsc{NewCluster} simulation. 

\begin{acknowledgements}
We thank the anonymous referee for the insightful comments that helped improve the manuscript.
S.K.Y. acknowledges support from the Korean National Research Foundation (RS-2025-00514475 and RS-2022-NR070872). 
This work was granted access to the HPC resources of KISTI under the allocations KSC-2021-CRE-0486, KSC-2022-CRE-0088, KSC-2022-CRE-0344, KSC-2022-CRE-0409, KSC-2023-CRE-0343, KSC-2024-CHA-0009, and KSC-2025-CRE-0031 and of GENCI under the allocation A0150414625 and A0180416216. The large data transfer was supported by KREONET which is managed and operated by KISTI. 
This work was granted access to the HPC resources of CINES under the allocations c2016047637, A0020407637, and A0070402192 by Genci, KSC-2017-G2-0003, KSC-2020-CRE-0055, and KSC-2020-CRE-0280 by KISTI, and as a "Grand Challenge" project granted by GENCI on the AMD Rome extension of the Joliot Curie supercomputer at TGCC. 
D.I. is grateful to Alfred Gulum for his unwavering enthusiasm, even when the scatter was not.
T.K. is supported by the National Research Foundation of Korea (RS-2022-NR070872 and RS-2025-00516961) and the Yonsei Fellowship, funded by Lee Youn Jae.
A.U.K. acknowledges support from the Belgian Federal Science Policy Office (BELSPO) via the ESA-PRODEX programme.
C.A. acknowledges that this work of the Interdisciplinary Thematic Institute IRMIA++, as part of the ITI 2021-2028 program of the University of Strasbourg, CNRS and Inserm, was supported by IdEx Unistra (ANR-10-IDEX-0002), and by SFRI-STRAT’US project (ANR-20-SFRI-0012) under the framework of the French Investments for the Future Program.
\end{acknowledgements}

\bibliographystyle{aa}
\bibliography{biblio}

@ARTICLE{dubois2024,
       author = {{Dubois}, Yohan and {Rodr{\'\i}guez Montero}, Francisco and {Guerra}, Corentin and {Trebitsch}, Maxime and {Han}, San and {Beckmann}, Ricarda and {Yi}, Sukyoung K. and {Lewis}, Joseph and {Jang}, J.~K.},
        title = "{Galaxies with grains: unraveling dust evolution and extinction curves with hydrodynamical simulations}",
      journal = {\aap},
         year = 2024,
        month = jul,
       volume = {687},
          eid = {A240},
        pages = {A240},
          doi = {10.1051/0004-6361/202449784},
archivePrefix = {arXiv},
       eprint = {2402.18515},
 primaryClass = {astro-ph.GA},
       adsurl = {https://ui.adsabs.harvard.edu/abs/2024A&A...687A.240D}
}

@ARTICLE{dave2016,
       author = {{Dav{\'e}}, Romeel and {Thompson}, Robert and {Hopkins}, Philip F.},
        title = "{MUFASA: galaxy formation simulations with meshless hydrodynamics}",
      journal = {\mnras},
         year = 2016,
        month = nov,
       volume = {462},
       number = {3},
        pages = {3265-3284},
          doi = {10.1093/mnras/stw1862},
archivePrefix = {arXiv},
       eprint = {1604.01418},
 primaryClass = {astro-ph.GA},
       adsurl = {https://ui.adsabs.harvard.edu/abs/2016MNRAS.462.3265D}
}

@ARTICLE{arun2025,
       author = {{Arun}, R.},
        title = "{When the Wall Fell: Study of Polycyclic Aromatic Hydrocarbons in T Chamaeleontis Using JWST}",
      journal = {\aj},
         year = 2025,
        month = oct,
       volume = {170},
       number = {4},
          eid = {196},
        pages = {196},
          doi = {10.3847/1538-3881/adf637},
archivePrefix = {arXiv},
       eprint = {2507.21639},
 primaryClass = {astro-ph.SR},
       adsurl = {https://ui.adsabs.harvard.edu/abs/2025AJ....170..196A}
}

@ARTICLE{maeder&meynet2000,
       author = {{Maeder}, A. and {Meynet}, G.},
        title = "{Stellar evolution with rotation. VI. The Eddington and Omega -limits, the rotational mass loss for OB and LBV stars}",
      journal = {\aap},
         year = 2000,
        month = sep,
       volume = {361},
        pages = {159-166},
          doi = {10.48550/arXiv.astro-ph/0006405},
archivePrefix = {arXiv},
       eprint = {astro-ph/0006405},
 primaryClass = {astro-ph},
       adsurl = {https://ui.adsabs.harvard.edu/abs/2000A&A...361..159M}
}

@ARTICLE{schaller1992,
       author = {{Schaller}, G. and {Schaerer}, D. and {Meynet}, G. and {Maeder}, A.},
        title = "{New Grids of Stellar Models from 0.8-SOLAR-MASS to 120-SOLAR-MASSES at Z=0.020 and Z=0.001}",
      journal = {\aaps},
         year = 1992,
        month = dec,
       volume = {96},
        pages = {269},
       adsurl = {https://ui.adsabs.harvard.edu/abs/1992A&AS...96..269S}
}

@ARTICLE{leitherer1999,
       author = {{Leitherer}, Claus and {Schaerer}, Daniel and {Goldader}, Jeffrey D. and {Delgado}, Rosa M. Gonz{\'a}lez and {Robert}, Carmelle and {Kune}, Denis Foo and {de Mello}, Du{\'\i}lia F. and {Devost}, Daniel and {Heckman}, Timothy M.},
        title = "{Starburst99: Synthesis Models for Galaxies with Active Star Formation}",
      journal = {\apjs},
         year = 1999,
        month = jul,
       volume = {123},
       number = {1},
        pages = {3-40},
          doi = {10.1086/313233},
archivePrefix = {arXiv},
       eprint = {astro-ph/9902334},
 primaryClass = {astro-ph},
       adsurl = {https://ui.adsabs.harvard.edu/abs/1999ApJS..123....3L}
}

@ARTICLE{leitherer2014,
       author = {{Leitherer}, Claus and {Ekstr{\"o}m}, Sylvia and {Meynet}, Georges and {Schaerer}, Daniel and {Agienko}, Katerina B. and {Levesque}, Emily M.},
        title = "{The Effects of Stellar Rotation. II. A Comprehensive Set of Starburst99 Models}",
      journal = {\apjs},
         year = 2014,
        month = may,
       volume = {212},
       number = {1},
          eid = {14},
        pages = {14},
          doi = {10.1088/0067-0049/212/1/14},
archivePrefix = {arXiv},
       eprint = {1403.5444},
 primaryClass = {astro-ph.GA},
       adsurl = {https://ui.adsabs.harvard.edu/abs/2014ApJS..212...14L}
}

@ARTICLE{teyssiere2011,
   author = {{Teyssier}, R. and {Moore}, B. and {Martizzi}, D. and {Dubois}, Y. and 
	{Mayer}, L.},
    title = "{Mass distribution in galaxy clusters: the role of Active Galactic Nuclei feedback}",
  journal = {\mnras},
archivePrefix = "arXiv",
   eprint = {1003.4744},
     year = 2011,
    month = jun,
   volume = 414,
    pages = {195-208},
      doi = {10.1111/j.1365-2966.2011.18399.x},
   adsurl = {http://cdsads.u-strasbg.fr/abs/2011MNRAS.414..195T}
}

@ARTICLE{rosen&bregman1995,
       author = {{Rosen}, Alexander and {Bregman}, Joel N.},
        title = "{Global Models of the Interstellar Medium in Disk Galaxies}",
      journal = {\apj},
         year = 1995,
        month = feb,
       volume = {440},
        pages = {634},
          doi = {10.1086/175303},
       adsurl = {https://ui.adsabs.harvard.edu/abs/1995ApJ...440..634R}
}

@article{sutherland&dopita93,
	Adsurl = {http://adsabs.harvard.edu/abs/1993ApJS...88..253S},
	Author = {{Sutherland}, R.~S. and {Dopita}, M.~A.},
	Doi = {10.1086/191823},
	Journal = {\apjs},
	Month = sep,
	Pages = {253-327},
	Title = {{Cooling functions for low-density astrophysical plasmas}},
	Volume = 88,
	Year = 1993}

@ARTICLE{dubois2021NH,
       author = {{Dubois}, Yohan and {Beckmann}, Ricarda and {Bournaud}, Fr{\'e}d{\'e}ric and {Choi}, Hoseung and {Devriendt}, Julien and {Jackson}, Ryan and {Kaviraj}, Sugata and {Kimm}, Taysun and {Kraljic}, Katarina and {Laigle}, Clotilde and {Martin}, Garreth and {Park}, Min-Jung and {Peirani}, S{\'e}bastien and {Pichon}, Christophe and {Volonteri}, Marta and {Yi}, Sukyoung K.},
        title = "{Introducing the NEWHORIZON simulation: Galaxy properties with resolved internal dynamics across cosmic time}",
      journal = {\aap},
         year = 2021,
        month = jul,
       volume = {651},
          eid = {A109},
        pages = {A109},
          doi = {10.1051/0004-6361/202039429},
archivePrefix = {arXiv},
       eprint = {2009.10578},
 primaryClass = {astro-ph.GA},
       adsurl = {https://ui.adsabs.harvard.edu/abs/2021A&A...651A.109D}
}

@ARTICLE{dubois2014,
       author = {{Dubois}, Y. and {Pichon}, C. and {Welker}, C. and {Le Borgne}, D. and {Devriendt}, J. and {Laigle}, C. and {Codis}, S. and {Pogosyan}, D. and {Arnouts}, S. and {Benabed}, K. and {Bertin}, E. and {Blaizot}, J. and {Bouchet}, F. and {Cardoso}, J. -F. and {Colombi}, S. and {de Lapparent}, V. and {Desjacques}, V. and {Gavazzi}, R. and {Kassin}, S. and {Kimm}, T. and {McCracken}, H. and {Milliard}, B. and {Peirani}, S. and {Prunet}, S. and {Rouberol}, S. and {Silk}, J. and {Slyz}, A. and {Sousbie}, T. and {Teyssier}, R. and {Tresse}, L. and {Treyer}, M. and {Vibert}, D. and {Volonteri}, M.},
        title = "{Dancing in the dark: galactic properties trace spin swings along the cosmic web}",
      journal = {\mnras},
         year = 2014,
        month = oct,
       volume = {444},
       number = {2},
        pages = {1453-1468},
          doi = {10.1093/mnras/stu1227},
archivePrefix = {arXiv},
       eprint = {1402.1165},
 primaryClass = {astro-ph.CO},
       adsurl = {https://ui.adsabs.harvard.edu/abs/2014MNRAS.444.1453D}
}

@ARTICLE{kaviraj2017,
       author = {{Kaviraj}, S. and {Laigle}, C. and {Kimm}, T. and {Devriendt}, J.~E.~G. and {Dubois}, Y. and {Pichon}, C. and {Slyz}, A. and {Chisari}, E. and {Peirani}, S.},
        title = "{The Horizon-AGN simulation: evolution of galaxy properties over cosmic time}",
      journal = {\mnras},
         year = 2017,
        month = jun,
       volume = {467},
       number = {4},
        pages = {4739-4752},
          doi = {10.1093/mnras/stx126},
archivePrefix = {arXiv},
       eprint = {1605.09379},
 primaryClass = {astro-ph.GA},
       adsurl = {https://ui.adsabs.harvard.edu/abs/2017MNRAS.467.4739K}
}

@ARTICLE{baes2015,
       author = {{Baes}, M. and {Camps}, P.},
        title = "{SKIRT: The design of a suite of input models for Monte Carlo radiative transfer simulations}",
      journal = {Astronomy and Computing},
         year = 2015,
        month = sep,
       volume = {12},
        pages = {33-44},
          doi = {10.1016/j.ascom.2015.05.006},
archivePrefix = {arXiv},
       eprint = {1505.07708},
 primaryClass = {astro-ph.IM},
       adsurl = {https://ui.adsabs.harvard.edu/abs/2015A&C....12...33B}
}

@ARTICLE{camps2020baes,
       author = {{Camps}, P. and {Baes}, M.},
        title = "{SKIRT 9: Redesigning an advanced dust radiative transfer code to allow kinematics, line transfer and polarization by aligned dust grains}",
      journal = {Astronomy and Computing},
         year = 2020,
        month = apr,
       volume = {31},
          eid = {100381},
        pages = {100381},
          doi = {10.1016/j.ascom.2020.100381},
archivePrefix = {arXiv},
       eprint = {2003.00721},
 primaryClass = {astro-ph.GA},
       adsurl = {https://ui.adsabs.harvard.edu/abs/2020A&C....3100381C}
}

@ARTICLE{komatsu2011wmap7,
       author = {{Komatsu}, E. and {Smith}, K.~M. and {Dunkley}, J. and {Bennett}, C.~L. and {Gold}, B. and {Hinshaw}, G. and {Jarosik}, N. and {Larson}, D. and {Nolta}, M.~R. and {Page}, L. and {Spergel}, D.~N. and {Halpern}, M. and {Hill}, R.~S. and {Kogut}, A. and {Limon}, M. and {Meyer}, S.~S. and {Odegard}, N. and {Tucker}, G.~S. and {Weiland}, J.~L. and {Wollack}, E. and {Wright}, E.~L.},
        title = "{Seven-year Wilkinson Microwave Anisotropy Probe (WMAP) Observations: Cosmological Interpretation}",
      journal = {\apjs},
         year = 2011,
        month = feb,
       volume = {192},
       number = {2},
          eid = {18},
        pages = {18},
          doi = {10.1088/0067-0049/192/2/18},
archivePrefix = {arXiv},
       eprint = {1001.4538},
 primaryClass = {astro-ph.CO},
       adsurl = {https://ui.adsabs.harvard.edu/abs/2011ApJS..192...18K}
}

@ARTICLE{bruzual2003charlot,
       author = {{Bruzual}, G. and {Charlot}, S.},
        title = "{Stellar population synthesis at the resolution of 2003}",
      journal = {\mnras},
         year = 2003,
        month = oct,
       volume = {344},
       number = {4},
        pages = {1000-1028},
          doi = {10.1046/j.1365-8711.2003.06897.x},
archivePrefix = {arXiv},
       eprint = {astro-ph/0309134},
 primaryClass = {astro-ph},
       adsurl = {https://ui.adsabs.harvard.edu/abs/2003MNRAS.344.1000B}
}

@ARTICLE{chabrier2003,
       author = {{Chabrier}, Gilles},
        title = "{Galactic Stellar and Substellar Initial Mass Function}",
      journal = {\pasp},
         year = 2003,
        month = jul,
       volume = {115},
       number = {809},
        pages = {763-795},
          doi = {10.1086/376392},
archivePrefix = {arXiv},
       eprint = {astro-ph/0304382},
 primaryClass = {astro-ph},
       adsurl = {https://ui.adsabs.harvard.edu/abs/2003PASP..115..763C}
}

@ARTICLE{dacunha2013,
       author = {{da Cunha}, Elisabete and {Groves}, Brent and {Walter}, Fabian and {Decarli}, Roberto and {Weiss}, Axel and {Bertoldi}, Frank and {Carilli}, Chris and {Daddi}, Emanuele and {Elbaz}, David and {Ivison}, Rob and {Maiolino}, Roberto and {Riechers}, Dominik and {Rix}, Hans-Walter and {Sargent}, Mark and {Smail}, Ian},
        title = "{On the Effect of the Cosmic Microwave Background in High-redshift (Sub-)millimeter Observations}",
      journal = {\apj},
         year = 2013,
        month = mar,
       volume = {766},
       number = {1},
          eid = {13},
        pages = {13},
          doi = {10.1088/0004-637X/766/1/13},
archivePrefix = {arXiv},
       eprint = {1302.0844},
 primaryClass = {astro-ph.CO},
       adsurl = {https://ui.adsabs.harvard.edu/abs/2013ApJ...766...13D}
}

@ARTICLE{draine1979salpeter,
       author = {{Draine}, B.~T. and {Salpeter}, E.~E.},
        title = "{Destruction mechanisms for interstellar dust.}",
      journal = {\apj},
         year = 1979,
        month = jul,
       volume = {231},
        pages = {438-455},
          doi = {10.1086/157206},
       adsurl = {https://ui.adsabs.harvard.edu/abs/1979ApJ...231..438D}
}

@ARTICLE{hirashita2015,
       author = {{Hirashita}, Hiroyuki and {Nozawa}, Takaya and {Villaume}, Alexa and {Srinivasan}, Sundar},
        title = "{Dust processing in elliptical galaxies}",
      journal = {\mnras},
         year = 2015,
        month = dec,
       volume = {454},
       number = {2},
        pages = {1620-1633},
          doi = {10.1093/mnras/stv2095},
archivePrefix = {arXiv},
       eprint = {1509.03978},
 primaryClass = {astro-ph.GA},
       adsurl = {https://ui.adsabs.harvard.edu/abs/2015MNRAS.454.1620H}
}

@ARTICLE{jones2017themis,
       author = {{Jones}, A.~P. and {K{\"o}hler}, M. and {Ysard}, N. and {Bocchio}, M. and {Verstraete}, L.},
        title = "{The global dust modelling framework THEMIS}",
      journal = {\aap},
         year = 2017,
        month = jun,
       volume = {602},
          eid = {A46},
        pages = {A46},
          doi = {10.1051/0004-6361/201630225},
archivePrefix = {arXiv},
       eprint = {1703.00775},
 primaryClass = {astro-ph.GA},
       adsurl = {https://ui.adsabs.harvard.edu/abs/2017A&A...602A..46J}
}

@ARTICLE{vijayan2022,
       author = {{Vijayan}, Aswin P. and {Wilkins}, Stephen M. and {Lovell}, Christopher C. and {Thomas}, Peter A. and {Camps}, Peter and {Baes}, Maarten and {Trayford}, James and {Kuusisto}, Jussi and {Roper}, William J.},
        title = "{First Light And Reionisation Epoch Simulations (FLARES) - III. The properties of massive dusty galaxies at cosmic dawn}",
      journal = {\mnras},
         year = 2022,
        month = apr,
       volume = {511},
       number = {4},
        pages = {4999-5017},
          doi = {10.1093/mnras/stac338},
archivePrefix = {arXiv},
       eprint = {2108.00830},
 primaryClass = {astro-ph.GA},
       adsurl = {https://ui.adsabs.harvard.edu/abs/2022MNRAS.511.4999V}
}

@ARTICLE{kimm2015,
       author = {{Kimm}, Taysun and {Cen}, Renyue and {Devriendt}, Julien and {Dubois}, Yohan and {Slyz}, Adrianne},
        title = "{Towards simulating star formation in turbulent high-z galaxies with mechanical supernova feedback}",
      journal = {\mnras},
         year = 2015,
        month = aug,
       volume = {451},
       number = {3},
        pages = {2900-2921},
          doi = {10.1093/mnras/stv1211},
archivePrefix = {arXiv},
       eprint = {1501.05655},
 primaryClass = {astro-ph.GA},
       adsurl = {https://ui.adsabs.harvard.edu/abs/2015MNRAS.451.2900K}
}

@ARTICLE{dubois2012,
       author = {{Dubois}, Yohan and {Devriendt}, Julien and {Slyz}, Adrianne and {Teyssier}, Romain},
        title = "{Self-regulated growth of supermassive black holes by a dual jet-heating active galactic nucleus feedback mechanism: methods, tests and implications for cosmological simulations}",
      journal = {\mnras},
         year = 2012,
        month = mar,
       volume = {420},
       number = {3},
        pages = {2662-2683},
          doi = {10.1111/j.1365-2966.2011.20236.x},
archivePrefix = {arXiv},
       eprint = {1108.0110},
 primaryClass = {astro-ph.CO},
       adsurl = {https://ui.adsabs.harvard.edu/abs/2012MNRAS.420.2662D}
}

@ARTICLE{haardt1996,
       author = {{Haardt}, Francesco and {Madau}, Piero},
        title = "{Radiative Transfer in a Clumpy Universe. II. The Ultraviolet Extragalactic Background}",
      journal = {\apj},
         year = 1996,
        month = apr,
       volume = {461},
        pages = {20},
          doi = {10.1086/177035},
archivePrefix = {arXiv},
       eprint = {astro-ph/9509093},
 primaryClass = {astro-ph},
       adsurl = {https://ui.adsabs.harvard.edu/abs/1996ApJ...461...20H}
}

@ARTICLE{rosdahl2012,
       author = {{Rosdahl}, J. and {Blaizot}, J.},
        title = "{Extended Ly{\ensuremath{\alpha}} emission from cold accretion streams}",
      journal = {\mnras},
         year = 2012,
        month = jun,
       volume = {423},
       number = {1},
        pages = {344-366},
          doi = {10.1111/j.1365-2966.2012.20883.x},
archivePrefix = {arXiv},
       eprint = {1112.4408},
 primaryClass = {astro-ph.CO},
       adsurl = {https://ui.adsabs.harvard.edu/abs/2012MNRAS.423..344R}
}

@ARTICLE{camps2018,
       author = {{Camps}, Peter and {Tr{\v{c}}ka}, Ana and {Trayford}, James and {Baes}, Maarten and {Theuns}, Tom and {Crain}, Robert A. and {McAlpine}, Stuart and {Schaller}, Matthieu and {Schaye}, Joop},
        title = "{Data Release of UV to Submillimeter Broadband Fluxes for Simulated Galaxies from the EAGLE Project}",
      journal = {\apjs},
         year = 2018,
        month = feb,
       volume = {234},
       number = {2},
          eid = {20},
        pages = {20},
          doi = {10.3847/1538-4365/aaa24c},
archivePrefix = {arXiv},
       eprint = {1712.05583},
 primaryClass = {astro-ph.GA},
       adsurl = {https://ui.adsabs.harvard.edu/abs/2018ApJS..234...20C}
}

@ARTICLE{vogelsberger2022b,
       author = {{Vogelsberger}, Mark and {Nelson}, Dylan and {Pillepich}, Annalisa and {Shen}, Xuejian and {Marinacci}, Federico and {Springel}, Volker and {Pakmor}, R{\"u}diger and {Tacchella}, Sandro and {Weinberger}, Rainer and {Torrey}, Paul and {Hernquist}, Lars},
        title = "{High-redshift JWST predictions from IllustrisTNG: dust modelling and galaxy luminosity functions}",
      journal = {\mnras},
         year = 2020,
        month = mar,
       volume = {492},
       number = {4},
        pages = {5167-5201},
          doi = {10.1093/mnras/staa137},
archivePrefix = {arXiv},
       eprint = {1904.07238},
 primaryClass = {astro-ph.GA},
       adsurl = {https://ui.adsabs.harvard.edu/abs/2020MNRAS.492.5167V}
}

@ARTICLE{kapoor2023,
       author = {{Kapoor}, Anand Utsav and {Baes}, Maarten and {van der Wel}, Arjen and {Gebek}, Andrea and {Camps}, Peter and {Nersesian}, Angelos and {Meidt}, Sharon E. and {Smith}, Aaron and {Vicens}, Sebastien and {D'Eugenio}, Francesco and {Martorano}, Marco and {Barrientos}, Daniela and {Sartorio}, Nina Sanches},
        title = "{TODDLERS: a new UV-mm emission library for star-forming regions - I. Integration with SKIRT and public release}",
      journal = {\mnras},
         year = 2023,
        month = dec,
       volume = {526},
       number = {3},
        pages = {3871-3901},
          doi = {10.1093/mnras/stad2977},
archivePrefix = {arXiv},
       eprint = {2310.00388},
 primaryClass = {astro-ph.GA},
       adsurl = {https://ui.adsabs.harvard.edu/abs/2023MNRAS.526.3871K}
}

@ARTICLE{kapoor2024,
       author = {{Kapoor}, Anand Utsav and {Baes}, Maarten and {van der Wel}, Arjen and {Gebek}, Andrea and {Camps}, Peter and {Smith}, Aaron and {Boquien}, M{\'e}d{\'e}ric and {Andreadis}, Nick and {Vicens}, Sebastien},
        title = "{TODDLERS: A new UV-millimeter emission library for star-forming regions: II. Star-formation rate indicators using Auriga zoom simulations}",
      journal = {\aap},
         year = 2024,
        month = dec,
       volume = {692},
          eid = {A79},
        pages = {A79},
          doi = {10.1051/0004-6361/202451207},
archivePrefix = {arXiv},
       eprint = {2410.01067},
 primaryClass = {astro-ph.GA},
       adsurl = {https://ui.adsabs.harvard.edu/abs/2024A&A...692A..79K}
}

@ARTICLE{inoue2003,
       author = {{Inoue}, Akio K.},
        title = "{Evolution of Dust-to-Metal Ratio in Galaxies}",
      journal = {\pasj},
         year = 2003,
        month = oct,
       volume = {55},
        pages = {901-909},
          doi = {10.1093/pasj/55.5.901},
archivePrefix = {arXiv},
       eprint = {astro-ph/0308204},
 primaryClass = {astro-ph},
       adsurl = {https://ui.adsabs.harvard.edu/abs/2003PASJ...55..901I}
}

@BOOK{osterbrock2006,
       author = {{Osterbrock}, Donald E. and {Ferland}, Gary J.},
        title = "{Astrophysics of gaseous nebulae and active galactic nuclei}",
         year = 2006,
       adsurl = {https://ui.adsabs.harvard.edu/abs/2006agna.book.....O}
}

@ARTICLE{popping2017,
       author = {{Popping}, Gerg{\"o} and {Puglisi}, Annagrazia and {Norman}, Colin A.},
        title = "{Dissecting the IRX-{\ensuremath{\beta}} dust attenuation relation: exploring the physical origin of observed variations in galaxies}",
      journal = {\mnras},
         year = 2017,
        month = dec,
       volume = {472},
       number = {2},
        pages = {2315-2333},
          doi = {10.1093/mnras/stx2202},
archivePrefix = {arXiv},
       eprint = {1706.06587},
 primaryClass = {astro-ph.GA},
       adsurl = {https://ui.adsabs.harvard.edu/abs/2017MNRAS.472.2315P}
}

@ARTICLE{madau2014,
       author = {{Madau}, Piero and {Dickinson}, Mark},
        title = "{Cosmic Star-Formation History}",
      journal = {\araa},
         year = 2014,
        month = aug,
       volume = {52},
        pages = {415-486},
          doi = {10.1146/annurev-astro-081811-125615},
archivePrefix = {arXiv},
       eprint = {1403.0007},
 primaryClass = {astro-ph.CO},
       adsurl = {https://ui.adsabs.harvard.edu/abs/2014ARA&A..52..415M}
}

@ARTICLE{Bouwens2009,
       author = {{Bouwens}, R.~J. and {Illingworth}, G.~D. and {Franx}, M. and {Chary}, R. -R. and {Meurer}, G.~R. and {Conselice}, C.~J. and {Ford}, H. and {Giavalisco}, M. and {van Dokkum}, P.},
        title = "{UV Continuum Slope and Dust Obscuration from z \raisebox{-0.5ex}\textasciitilde 6 to z \raisebox{-0.5ex}\textasciitilde 2: The Star Formation Rate Density at High Redshift}",
      journal = {\apj},
         year = 2009,
        month = nov,
       volume = {705},
       number = {1},
        pages = {936-961},
          doi = {10.1088/0004-637X/705/1/936},
archivePrefix = {arXiv},
       eprint = {0909.4074},
 primaryClass = {astro-ph.CO},
       adsurl = {https://ui.adsabs.harvard.edu/abs/2009ApJ...705..936B}
}

@ARTICLE{buat2014,
       author = {{Buat}, V. and {Heinis}, S. and {Boquien}, M. and {Burgarella}, D. and {Charmandaris}, V. and {Boissier}, S. and {Boselli}, A. and {Le Borgne}, D. and {Morrison}, G.},
        title = "{Ultraviolet to infrared emission of z > 1 galaxies: Can we derive reliable star formation rates and stellar masses?}",
      journal = {\aap},
         year = 2014,
        month = jan,
       volume = {561},
          eid = {A39},
        pages = {A39},
          doi = {10.1051/0004-6361/201322081},
archivePrefix = {arXiv},
       eprint = {1310.7712},
 primaryClass = {astro-ph.CO},
       adsurl = {https://ui.adsabs.harvard.edu/abs/2014A&A...561A..39B}
}

@ARTICLE{shivaei2015,
       author = {{Shivaei}, Irene and {Reddy}, Naveen A. and {Steidel}, Charles C. and {Shapley}, Alice E.},
        title = "{Investigating H{\ensuremath{\alpha}}, UV, and IR Star-formation Rate Diagnostics for a Large Sample of z {\ensuremath{\sim}} 2 Galaxies}",
      journal = {\apj},
         year = 2015,
        month = may,
       volume = {804},
       number = {2},
          eid = {149},
        pages = {149},
          doi = {10.1088/0004-637X/804/2/149},
archivePrefix = {arXiv},
       eprint = {1503.03929},
 primaryClass = {astro-ph.GA},
       adsurl = {https://ui.adsabs.harvard.edu/abs/2015ApJ...804..149S}
}

@ARTICLE{figueira2022,
       author = {{Figueira}, M. and {Pollo}, A. and {Ma{\l}ek}, K. and {Buat}, V. and {Boquien}, M. and {Pistis}, F. and {Cassar{\`a}}, L.~P. and {Vergani}, D. and {Hamed}, M. and {Salim}, S.},
        title = "{SFR estimations from z = 0 to z = 0.9. A comparison of SFR calibrators for star-forming galaxies}",
      journal = {\aap},
         year = 2022,
        month = nov,
       volume = {667},
          eid = {A29},
        pages = {A29},
          doi = {10.1051/0004-6361/202141701},
archivePrefix = {arXiv},
       eprint = {2209.04390},
 primaryClass = {astro-ph.GA},
       adsurl = {https://ui.adsabs.harvard.edu/abs/2022A&A...667A..29F}
}

@ARTICLE{salim2007,
       author = {{Salim}, Samir and {Rich}, R. Michael and {Charlot}, St{\'e}phane and {Brinchmann}, Jarle and {Johnson}, Benjamin D. and {Schiminovich}, David and {Seibert}, Mark and {Mallery}, Ryan and {Heckman}, Timothy M. and {Forster}, Karl and {Friedman}, Peter G. and {Martin}, D. Christopher and {Morrissey}, Patrick and {Neff}, Susan G. and {Small}, Todd and {Wyder}, Ted K. and {Bianchi}, Luciana and {Donas}, Jos{\'e} and {Lee}, Young-Wook and {Madore}, Barry F. and {Milliard}, Bruno and {Szalay}, Alex S. and {Welsh}, Barry Y. and {Yi}, Sukyoung K.},
        title = "{UV Star Formation Rates in the Local Universe}",
      journal = {\apjs},
         year = 2007,
        month = dec,
       volume = {173},
       number = {2},
        pages = {267-292},
          doi = {10.1086/519218},
archivePrefix = {arXiv},
       eprint = {0704.3611},
 primaryClass = {astro-ph},
       adsurl = {https://ui.adsabs.harvard.edu/abs/2007ApJS..173..267S}
}

@ARTICLE{fudamoto2020,
       author = {{Fudamoto}, Y. and {Oesch}, P.~A. and {Faisst}, A. and {B{\'e}thermin}, M. and {Ginolfi}, M. and {Khusanova}, Y. and {Loiacono}, F. and {Le F{\`e}vre}, O. and {Capak}, P. and {Schaerer}, D. and {Silverman}, J.~D. and {Cassata}, P. and {Yan}, L. and {Amorin}, R. and {Bardelli}, S. and {Boquien}, M. and {Cimatti}, A. and {Dessauges-Zavadsky}, M. and {Fujimoto}, S. and {Gruppioni}, C. and {Hathi}, N.~P. and {Ibar}, E. and {Jones}, G.~C. and {Koekemoer}, A.~M. and {Lagache}, G. and {Lemaux}, B.~C. and {Maiolino}, R. and {Narayanan}, D. and {Pozzi}, F. and {Riechers}, D.~A. and {Rodighiero}, G. and {Talia}, M. and {Toft}, S. and {Vallini}, L. and {Vergani}, D. and {Zamorani}, G. and {Zucca}, E.},
        title = "{The ALPINE-ALMA [CII] survey. Dust attenuation properties and obscured star formation at z {\ensuremath{\sim}} 4.4-5.8}",
      journal = {\aap},
         year = 2020,
        month = nov,
       volume = {643},
          eid = {A4},
        pages = {A4},
          doi = {10.1051/0004-6361/202038163},
archivePrefix = {arXiv},
       eprint = {2004.10760},
 primaryClass = {astro-ph.GA},
       adsurl = {https://ui.adsabs.harvard.edu/abs/2020A&A...643A...4F}
}

@ARTICLE{delooze2014,
       author = {{De Looze}, Ilse and {Cormier}, Diane and {Lebouteiller}, Vianney and {Madden}, Suzanne and {Baes}, Maarten and {Bendo}, George J. and {Boquien}, M{\'e}d{\'e}ric and {Boselli}, Alessandro and {Clements}, David L. and {Cortese}, Luca and {Cooray}, Asantha and {Galametz}, Maud and {Galliano}, Fr{\'e}d{\'e}ric and {Graci{\'a}-Carpio}, Javier and {Isaak}, Kate and {Karczewski}, Oskar {\L}. and {Parkin}, Tara J. and {Pellegrini}, Eric W. and {R{\'e}my-Ruyer}, Aur{\'e}lie and {Spinoglio}, Luigi and {Smith}, Matthew W.~L. and {Sturm}, Eckhard},
        title = "{The applicability of far-infrared fine-structure lines as star formation rate tracers over wide ranges of metallicities and galaxy types}",
      journal = {\aap},
         year = 2014,
        month = aug,
       volume = {568},
          eid = {A62},
        pages = {A62},
          doi = {10.1051/0004-6361/201322489},
archivePrefix = {arXiv},
       eprint = {1402.4075},
 primaryClass = {astro-ph.GA},
       adsurl = {https://ui.adsabs.harvard.edu/abs/2014A&A...568A..62D}
}

@ARTICLE{herrera-camus2015,
       author = {{Herrera-Camus}, R. and {Bolatto}, A.~D. and {Wolfire}, M.~G. and {Smith}, J.~D. and {Croxall}, K.~V. and {Kennicutt}, R.~C. and {Calzetti}, D. and {Helou}, G. and {Walter}, F. and {Leroy}, A.~K. and {Draine}, B. and {Brandl}, B.~R. and {Armus}, L. and {Sandstrom}, K.~M. and {Dale}, D.~A. and {Aniano}, G. and {Meidt}, S.~E. and {Boquien}, M. and {Hunt}, L.~K. and {Galametz}, M. and {Tabatabaei}, F.~S. and {Murphy}, E.~J. and {Appleton}, P. and {Roussel}, H. and {Engelbracht}, C. and {Beirao}, P.},
        title = "{[C II] 158 {\ensuremath{\mu}}m Emission as a Star Formation Tracer}",
      journal = {\apj},
         year = 2015,
        month = feb,
       volume = {800},
       number = {1},
          eid = {1},
        pages = {1},
          doi = {10.1088/0004-637X/800/1/1},
archivePrefix = {arXiv},
       eprint = {1409.7123},
 primaryClass = {astro-ph.GA},
       adsurl = {https://ui.adsabs.harvard.edu/abs/2015ApJ...800....1H}
}

@ARTICLE{Mitsuhashi2024,
       author = {{Mitsuhashi}, Ikki and {Tadaki}, Ken-ichi and {Ikeda}, Ryota and {Herrera-Camus}, Rodrigo and {Aravena}, Manuel and {De Looze}, Ilse and {F{\"o}rster Schreiber}, Natascha M. and {Gonz{\'a}lez-L{\'o}pez}, Jorge and {Spilker}, Justin and {Assef}, Roberto J. and {Bouwens}, Rychard and {Barcos-Munoz}, Loreto and {Birkin}, Jack and {Bowler}, Rebecca A.~A. and {Calistro Rivera}, Gabriela and {Davies}, Rebecca and {Da Cunha}, Elisabete and {D{\'\i}az-Santos}, Tanio and {Ferrara}, Andrea and {Fisher}, Deanne B. and {Lee}, Lilian L. and {Li}, Juno and {Lutz}, Dieter and {Rela{\~n}o}, Monica and {Naab}, Thorsten and {Palla}, Marco and {Posses}, Ana and {Solimano}, Manuel and {Tacconi}, Linda and {{\"U}bler}, Hannah and {van der Giessen}, Stefan and {Veilleux}, Sylvain},
        title = "{The ALMA-CRISTAL survey: Widespread dust-obscured star formation in typical star-forming galaxies at z = 4{\textendash}6}",
      journal = {\aap},
         year = 2024,
        month = oct,
       volume = {690},
          eid = {A197},
        pages = {A197},
          doi = {10.1051/0004-6361/202348782},
archivePrefix = {arXiv},
       eprint = {2311.17671},
 primaryClass = {astro-ph.GA},
       adsurl = {https://ui.adsabs.harvard.edu/abs/2024A&A...690A.197M}
}

@ARTICLE{kennicutt1998,
       author = {{Kennicutt}, Jr., Robert C.},
        title = "{Star Formation in Galaxies Along the Hubble Sequence}",
      journal = {\araa},
         year = 1998,
        month = jan,
       volume = {36},
        pages = {189-232},
          doi = {10.1146/annurev.astro.36.1.189},
archivePrefix = {arXiv},
       eprint = {astro-ph/9807187},
 primaryClass = {astro-ph},
       adsurl = {https://ui.adsabs.harvard.edu/abs/1998ARA&A..36..189K}
}

@ARTICLE{groves2012,
       author = {{Groves}, Brent and {Brinchmann}, Jarle and {Walcher}, Carl Jakob},
        title = "{The Balmer decrement of Sloan Digital Sky Survey galaxies}",
      journal = {\mnras},
         year = 2012,
        month = jan,
       volume = {419},
       number = {2},
        pages = {1402-1412},
          doi = {10.1111/j.1365-2966.2011.19796.x},
archivePrefix = {arXiv},
       eprint = {1109.2597},
 primaryClass = {astro-ph.CO},
       adsurl = {https://ui.adsabs.harvard.edu/abs/2012MNRAS.419.1402G}
}

@ARTICLE{ferruit2022,
       author = {{Ferruit}, P. and {Jakobsen}, P. and {Giardino}, G. and {Rawle}, T. and {Alves de Oliveira}, C. and {Arribas}, S. and {Beck}, T.~L. and {Birkmann}, S. and {B{\"o}ker}, T. and {Bunker}, A.~J. and {Charlot}, S. and {de Marchi}, G. and {Franx}, M. and {Henry}, A. and {Karakla}, D. and {Kassin}, S.~A. and {Kumari}, N. and {L{\'o}pez-Caniego}, M. and {L{\"u}tzgendorf}, N. and {Maiolino}, R. and {Manjavacas}, E. and {Marston}, A. and {Moseley}, S.~H. and {Muzerolle}, J. and {Pirzkal}, N. and {Rauscher}, B. and {Rix}, H. -W. and {Sabbi}, E. and {Sirianni}, M. and {te Plate}, M. and {Valenti}, J. and {Willott}, C.~J. and {Zeidler}, P.},
        title = "{The Near-Infrared Spectrograph (NIRSpec) on the James Webb Space Telescope. II. Multi-object spectroscopy (MOS)}",
      journal = {\aap},
         year = 2022,
        month = may,
       volume = {661},
          eid = {A81},
        pages = {A81},
          doi = {10.1051/0004-6361/202142673},
archivePrefix = {arXiv},
       eprint = {2202.03306},
 primaryClass = {astro-ph.IM},
       adsurl = {https://ui.adsabs.harvard.edu/abs/2022A&A...661A..81F}
}

@ARTICLE{matthee2019,
       author = {{Matthee}, J. and {Sobral}, D. and {Boogaard}, L.~A. and {R{\"o}ttgering}, H. and {Vallini}, L. and {Ferrara}, A. and {Paulino-Afonso}, A. and {Boone}, F. and {Schaerer}, D. and {Mobasher}, B.},
        title = "{Resolved UV and [C II] Structures of Luminous Galaxies within the Epoch of Reionization}",
      journal = {\apj},
         year = 2019,
        month = aug,
       volume = {881},
       number = {2},
          eid = {124},
        pages = {124},
          doi = {10.3847/1538-4357/ab2f81},
archivePrefix = {arXiv},
       eprint = {1903.08171},
 primaryClass = {astro-ph.GA},
       adsurl = {https://ui.adsabs.harvard.edu/abs/2019ApJ...881..124M}
}

@ARTICLE{stacey2010,
       author = {{Stacey}, G.~J. and {Hailey-Dunsheath}, S. and {Ferkinhoff}, C. and {Nikola}, T. and {Parshley}, S.~C. and {Benford}, D.~J. and {Staguhn}, J.~G. and {Fiolet}, N.},
        title = "{A 158 {\ensuremath{\mu}}m [C II] Line Survey of Galaxies at z \raisebox{-0.5ex}\textasciitilde 1-2: An Indicator of Star Formation in the Early Universe}",
      journal = {\apj},
         year = 2010,
        month = dec,
       volume = {724},
       number = {2},
        pages = {957-974},
          doi = {10.1088/0004-637X/724/2/957},
archivePrefix = {arXiv},
       eprint = {1009.4216},
 primaryClass = {astro-ph.CO},
       adsurl = {https://ui.adsabs.harvard.edu/abs/2010ApJ...724..957S}
}

@ARTICLE{vallini2013,
       author = {{Vallini}, Livia and {Gallerani}, Simona and {Ferrara}, Andrea and {Baek}, Sunghye},
        title = "{Far-infrared line emission from high-redshift galaxies}",
      journal = {\mnras},
         year = 2013,
        month = aug,
       volume = {433},
       number = {2},
        pages = {1567-1572},
          doi = {10.1093/mnras/stt828},
archivePrefix = {arXiv},
       eprint = {1305.2202},
 primaryClass = {astro-ph.CO},
       adsurl = {https://ui.adsabs.harvard.edu/abs/2013MNRAS.433.1567V}
}

@ARTICLE{hirashita2003,
       author = {{Hirashita}, H. and {Buat}, V. and {Inoue}, A.~K.},
        title = "{Star formation rate in galaxies from UV, IR, and H{\ensuremath{\alpha}}  estimators}",
      journal = {\aap},
         year = 2003,
        month = oct,
       volume = {410},
        pages = {83-100},
          doi = {10.1051/0004-6361:20031144},
archivePrefix = {arXiv},
       eprint = {astro-ph/0308531},
 primaryClass = {astro-ph},
       adsurl = {https://ui.adsabs.harvard.edu/abs/2003A&A...410...83H}
}

@ARTICLE{Li2019,
       author = {{Li}, Qi and {Narayanan}, Desika and {Dav{\'e}}, Romeel},
        title = "{The dust-to-gas and dust-to-metal ratio in galaxies from z = 0 to 6}",
      journal = {\mnras},
         year = 2019,
        month = nov,
       volume = {490},
       number = {1},
        pages = {1425-1436},
          doi = {10.1093/mnras/stz2684},
archivePrefix = {arXiv},
       eprint = {1906.09277},
 primaryClass = {astro-ph.GA},
       adsurl = {https://ui.adsabs.harvard.edu/abs/2019MNRAS.490.1425L}
}

@ARTICLE{remy-ruyer2014,
       author = {{R{\'e}my-Ruyer}, A. and {Madden}, S.~C. and {Galliano}, F. and {Galametz}, M. and {Takeuchi}, T.~T. and {Asano}, R.~S. and {Zhukovska}, S. and {Lebouteiller}, V. and {Cormier}, D. and {Jones}, A. and {Bocchio}, M. and {Baes}, M. and {Bendo}, G.~J. and {Boquien}, M. and {Boselli}, A. and {DeLooze}, I. and {Doublier-Pritchard}, V. and {Hughes}, T. and {Karczewski}, O. {\L}. and {Spinoglio}, L.},
        title = "{Gas-to-dust mass ratios in local galaxies over a 2 dex metallicity range}",
      journal = {\aap},
         year = 2014,
        month = mar,
       volume = {563},
          eid = {A31},
        pages = {A31},
          doi = {10.1051/0004-6361/201322803},
archivePrefix = {arXiv},
       eprint = {1312.3442},
 primaryClass = {astro-ph.GA},
       adsurl = {https://ui.adsabs.harvard.edu/abs/2014A&A...563A..31R}
}

@ARTICLE{kimm2017,
       author = {{Kimm}, Taysun and {Katz}, Harley and {Haehnelt}, Martin and {Rosdahl}, Joakim and {Devriendt}, Julien and {Slyz}, Adrianne},
        title = "{Feedback-regulated star formation and escape of LyC photons from mini-haloes during reionization}",
      journal = {\mnras},
         year = 2017,
        month = apr,
       volume = {466},
       number = {4},
        pages = {4826-4846},
          doi = {10.1093/mnras/stx052},
archivePrefix = {arXiv},
       eprint = {1608.04762},
 primaryClass = {astro-ph.GA},
       adsurl = {https://ui.adsabs.harvard.edu/abs/2017MNRAS.466.4826K}
}

@ARTICLE{devis2019,
       author = {{De Vis}, P. and {Jones}, A. and {Viaene}, S. and {Casasola}, V. and {Clark}, C.~J.~R. and {Baes}, M. and {Bianchi}, S. and {Cassara}, L.~P. and {Davies}, J.~I. and {De Looze}, I. and {Galametz}, M. and {Galliano}, F. and {Lianou}, S. and {Madden}, S. and {Manilla-Robles}, A. and {Mosenkov}, A.~V. and {Nersesian}, A. and {Roychowdhury}, S. and {Xilouris}, E.~M. and {Ysard}, N.},
        title = "{A systematic metallicity study of DustPedia galaxies reveals evolution in the dust-to-metal ratios}",
      journal = {\aap},
         year = 2019,
        month = mar,
       volume = {623},
          eid = {A5},
        pages = {A5},
          doi = {10.1051/0004-6361/201834444},
archivePrefix = {arXiv},
       eprint = {1901.09040},
 primaryClass = {astro-ph.GA},
       adsurl = {https://ui.adsabs.harvard.edu/abs/2019A&A...623A...5D}
}

@ARTICLE{han2025b,
       author = {{Han}, San and {Yi}, Sukyoung K. and {Dubois}, Yohan and {Rhee}, Jinsu and {Jeon}, Seyoung and {Jang}, J.~K. and {Byun}, Gyeong-Hwan and {Cadiou}, Corentin and {Kim}, Juhan and {Kimm}, Taysun and {Pichon}, Christophe},
        title = "{Introducing NewCluster: the first half of the history of a high-resolution cluster simulation}",
      journal = {arXiv e-prints},
         year = 2025,
        month = jul,
          eid = {arXiv:2507.06301},
        pages = {arXiv:2507.06301},
          doi = {10.48550/arXiv.2507.06301},
archivePrefix = {arXiv},
       eprint = {2507.06301},
 primaryClass = {astro-ph.GA},
       adsurl = {https://ui.adsabs.harvard.edu/abs/2025arXiv250706301H}
}

@ARTICLE{han2025a,
       author = {{Han}, San and {Dubois}, Yohan and {Lee}, Jaehyun and {Kim}, Juhan and {Cadiou}, Corentin and {Yi}, Sukyoung K.},
        title = "{RAMSES-yOMP: Performance Optimizations for the Astrophysical Hydrodynamic Simulation Code RAMSES}",
      journal = {\apj},
         year = 2025,
        month = jan,
       volume = {978},
       number = {1},
          eid = {96},
        pages = {96},
          doi = {10.3847/1538-4357/ad98f4},
archivePrefix = {arXiv},
       eprint = {2411.14631},
 primaryClass = {astro-ph.IM},
       adsurl = {https://ui.adsabs.harvard.edu/abs/2025ApJ...978...96H}
}

@ARTICLE{Teyssier2002,
       author = {{Teyssier}, R.},
        title = "{Cosmological hydrodynamics with adaptive mesh refinement. A new high resolution code called RAMSES}",
      journal = {\aap},
         year = 2002,
        month = apr,
       volume = {385},
        pages = {337-364},
          doi = {10.1051/0004-6361:20011817},
archivePrefix = {arXiv},
       eprint = {astro-ph/0111367},
 primaryClass = {astro-ph},
       adsurl = {https://ui.adsabs.harvard.edu/abs/2002A&A...385..337T}
}

@ARTICLE{federrath2012,
       author = {{Federrath}, Christoph and {Klessen}, Ralf S.},
        title = "{The Star Formation Rate of Turbulent Magnetized Clouds: Comparing Theory, Simulations, and Observations}",
      journal = {\apj},
         year = 2012,
        month = dec,
       volume = {761},
       number = {2},
          eid = {156},
        pages = {156},
          doi = {10.1088/0004-637X/761/2/156},
archivePrefix = {arXiv},
       eprint = {1209.2856},
 primaryClass = {astro-ph.SR},
       adsurl = {https://ui.adsabs.harvard.edu/abs/2012ApJ...761..156F}
}

@ARTICLE{koyabashi2006,
       author = {{Kobayashi}, Chiaki and {Umeda}, Hideyuki and {Nomoto}, Ken'ichi and {Tominaga}, Nozomu and {Ohkubo}, Takuya},
        title = "{Galactic Chemical Evolution: Carbon through Zinc}",
      journal = {\apj},
         year = 2006,
        month = dec,
       volume = {653},
       number = {2},
        pages = {1145-1171},
          doi = {10.1086/508914},
archivePrefix = {arXiv},
       eprint = {astro-ph/0608688},
 primaryClass = {astro-ph},
       adsurl = {https://ui.adsabs.harvard.edu/abs/2006ApJ...653.1145K}
}

@ARTICLE{iwamoto1999,
       author = {{Iwamoto}, Koichi and {Brachwitz}, Franziska and {Nomoto}, Ken'ICHI and {Kishimoto}, Nobuhiro and {Umeda}, Hideyuki and {Hix}, W. Raphael and {Thielemann}, Friedrich-Karl},
        title = "{Nucleosynthesis in Chandrasekhar Mass Models for Type IA Supernovae and Constraints on Progenitor Systems and Burning-Front Propagation}",
      journal = {\apjs},
         year = 1999,
        month = dec,
       volume = {125},
       number = {2},
        pages = {439-462},
          doi = {10.1086/313278},
archivePrefix = {arXiv},
       eprint = {astro-ph/0002337},
 primaryClass = {astro-ph},
       adsurl = {https://ui.adsabs.harvard.edu/abs/1999ApJS..125..439I}
}

@ARTICLE{kimm2014,
       author = {{Kimm}, Taysun and {Cen}, Renyue},
        title = "{Escape Fraction of Ionizing Photons during Reionization: Effects due to Supernova Feedback and Runaway OB Stars}",
      journal = {\apj},
         year = 2014,
        month = jun,
       volume = {788},
       number = {2},
          eid = {121},
        pages = {121},
          doi = {10.1088/0004-637X/788/2/121},
archivePrefix = {arXiv},
       eprint = {1405.0552},
 primaryClass = {astro-ph.GA},
       adsurl = {https://ui.adsabs.harvard.edu/abs/2014ApJ...788..121K}
}

@ARTICLE{reddy2022,
       author = {{Reddy}, Naveen A. and {Topping}, Michael W. and {Shapley}, Alice E. and {Steidel}, Charles C. and {Sanders}, Ryan L. and {Du}, Xinnan and {Coil}, Alison L. and {Mobasher}, Bahram and {Price}, Sedona H. and {Shivaei}, Irene},
        title = "{The Effects of Stellar Population and Gas Covering Fraction on the Emergent Ly{\ensuremath{\alpha}} Emission of High-redshift Galaxies}",
      journal = {\apj},
         year = 2022,
        month = feb,
       volume = {926},
       number = {1},
          eid = {31},
        pages = {31},
          doi = {10.3847/1538-4357/ac3b4c},
archivePrefix = {arXiv},
       eprint = {2108.05363},
 primaryClass = {astro-ph.GA},
       adsurl = {https://ui.adsabs.harvard.edu/abs/2022ApJ...926...31R}
}

@ARTICLE{kennicutt1983,
       author = {{Kennicutt}, Jr., R.~C.},
        title = "{The rate of star formation in normal disk galaxies.}",
      journal = {\apj},
         year = 1983,
        month = sep,
       volume = {272},
        pages = {54-67},
          doi = {10.1086/161261},
       adsurl = {https://ui.adsabs.harvard.edu/abs/1983ApJ...272...54K}
}

@ARTICLE{sanders2023,
       author = {{Sanders}, Ryan L. and {Shapley}, Alice E. and {Topping}, Michael W. and {Reddy}, Naveen A. and {Brammer}, Gabriel B.},
        title = "{Excitation and Ionization Properties of Star-forming Galaxies at z = 2.0-9.3 with JWST/NIRSpec}",
      journal = {\apj},
         year = 2023,
        month = sep,
       volume = {955},
       number = {1},
          eid = {54},
        pages = {54},
          doi = {10.3847/1538-4357/acedad},
archivePrefix = {arXiv},
       eprint = {2301.06696},
 primaryClass = {astro-ph.GA},
       adsurl = {https://ui.adsabs.harvard.edu/abs/2023ApJ...955...54S}
}

@ARTICLE{cameron2023,
       author = {{Cameron}, Alex J. and {Saxena}, Aayush and {Bunker}, Andrew J. and {D'Eugenio}, Francesco and {Carniani}, Stefano and {Maiolino}, Roberto and {Curtis-Lake}, Emma and {Ferruit}, Pierre and {Jakobsen}, Peter and {Arribas}, Santiago and {Bonaventura}, Nina and {Charlot}, Stephane and {Chevallard}, Jacopo and {Curti}, Mirko and {Looser}, Tobias J. and {Maseda}, Michael V. and {Rawle}, Tim and {Rodr{\'\i}guez Del Pino}, Bruno and {Smit}, Renske and {{\"U}bler}, Hannah and {Willott}, Chris and {Witstok}, Joris and {Egami}, Eiichi and {Eisenstein}, Daniel J. and {Johnson}, Benjamin D. and {Hainline}, Kevin and {Rieke}, Marcia and {Robertson}, Brant E. and {Stark}, Daniel P. and {Tacchella}, Sandro and {Williams}, Christina C. and {Willmer}, Christopher N.~A. and {Bhatawdekar}, Rachana and {Bowler}, Rebecca and {Boyett}, Kristan and {Circosta}, Chiara and {Helton}, Jakob M. and {Jones}, Gareth C. and {Kumari}, Nimisha and {Ji}, Zhiyuan and {Nelson}, Erica and {Parlanti}, Eleonora and {Sandles}, Lester and {Scholtz}, Jan and {Sun}, Fengwu},
        title = "{JADES: Probing interstellar medium conditions at z {\ensuremath{\sim}} 5.5-9.5 with ultra-deep JWST/NIRSpec spectroscopy}",
      journal = {\aap},
         year = 2023,
        month = sep,
       volume = {677},
          eid = {A115},
        pages = {A115},
          doi = {10.1051/0004-6361/202346107},
archivePrefix = {arXiv},
       eprint = {2302.04298},
 primaryClass = {astro-ph.GA},
       adsurl = {https://ui.adsabs.harvard.edu/abs/2023A&A...677A.115C}
}

@ARTICLE{sandles2024,
       author = {{Sandles}, Lester and {D'Eugenio}, Francesco and {Maiolino}, Roberto and {Looser}, Tobias J. and {Arribas}, Santiago and {Baker}, William M. and {Bonaventura}, Nina and {Bunker}, Andrew J. and {Cameron}, Alex J. and {Carniani}, Stefano and {Charlot}, Stephane and {Chevallard}, Jacopo and {Curti}, Mirko and {Curtis-Lake}, Emma and {de Graaff}, Anna and {Eisenstein}, Daniel J. and {Hainline}, Kevin and {Ji}, Zhiyuan and {Johnson}, Benjamin D. and {Jones}, Gareth C. and {Kumari}, Nimisha and {Nelson}, Erica and {Perna}, Michele and {Rawle}, Tim and {Rix}, Hans-Walter and {Robertson}, Brant and {Del Pino}, Bruno Rodr{\'\i}guez and {Scholtz}, Jan and {Shivaei}, Irene and {Smit}, Renske and {Sun}, Fengwu and {Tacchella}, Sandro and {{\"U}bler}, Hannah and {Williams}, Christina C. and {Willott}, Chris and {Witstok}, Joris},
        title = "{JADES: Balmer decrement measurements at redshifts 4 < z < 7}",
      journal = {\aap},
         year = 2024,
        month = nov,
       volume = {691},
          eid = {A305},
        pages = {A305},
          doi = {10.1051/0004-6361/202347119},
archivePrefix = {arXiv},
       eprint = {2306.03931},
 primaryClass = {astro-ph.GA},
       adsurl = {https://ui.adsabs.harvard.edu/abs/2024A&A...691A.305S}
}

@ARTICLE{shapley2015,
       author = {{Shapley}, Alice E. and {Reddy}, Naveen A. and {Kriek}, Mariska and {Freeman}, William R. and {Sanders}, Ryan L. and {Siana}, Brian and {Coil}, Alison L. and {Mobasher}, Bahram and {Shivaei}, Irene and {Price}, Sedona H. and {de Groot}, Laura},
        title = "{The MOSDEF Survey: Excitation Properties of z {\ensuremath{\sim}} 2.3 Star-forming Galaxies}",
      journal = {\apj},
         year = 2015,
        month = mar,
       volume = {801},
       number = {2},
          eid = {88},
        pages = {88},
          doi = {10.1088/0004-637X/801/2/88},
archivePrefix = {arXiv},
       eprint = {1409.7071},
 primaryClass = {astro-ph.GA},
       adsurl = {https://ui.adsabs.harvard.edu/abs/2015ApJ...801...88S}
}

@ARTICLE{kewley2013,
       author = {{Kewley}, Lisa J. and {Dopita}, Michael A. and {Leitherer}, Claus and {Dav{\'e}}, Romeel and {Yuan}, Tiantian and {Allen}, Mark and {Groves}, Brent and {Sutherland}, Ralph},
        title = "{Theoretical Evolution of Optical Strong Lines across Cosmic Time}",
      journal = {\apj},
         year = 2013,
        month = sep,
       volume = {774},
       number = {2},
          eid = {100},
        pages = {100},
          doi = {10.1088/0004-637X/774/2/100},
archivePrefix = {arXiv},
       eprint = {1307.0508},
 primaryClass = {astro-ph.CO},
       adsurl = {https://ui.adsabs.harvard.edu/abs/2013ApJ...774..100K}
}

@ARTICLE{kashino2017,
       author = {{Kashino}, D. and {Silverman}, J.~D. and {Sanders}, D. and {Kartaltepe}, J.~S. and {Daddi}, E. and {Renzini}, A. and {Valentino}, F. and {Rodighiero}, G. and {Juneau}, S. and {Kewley}, L.~J. and {Zahid}, H.~J. and {Arimoto}, N. and {Nagao}, T. and {Chu}, J. and {Sugiyama}, N. and {Civano}, F. and {Ilbert}, O. and {Kajisawa}, M. and {Le F{\`e}vre}, O. and {Maier}, C. and {Masters}, D. and {Miyaji}, T. and {Onodera}, M. and {Puglisi}, A. and {Taniguchi}, Y.},
        title = "{The FMOS-COSMOS Survey of Star-forming Galaxies at z {\ensuremath{\approx}} 1.6. IV. Excitation State and Chemical Enrichment of the Interstellar Medium}",
      journal = {\apj},
         year = 2017,
        month = jan,
       volume = {835},
       number = {1},
          eid = {88},
        pages = {88},
          doi = {10.3847/1538-4357/835/1/88},
archivePrefix = {arXiv},
       eprint = {1604.06802},
 primaryClass = {astro-ph.GA},
       adsurl = {https://ui.adsabs.harvard.edu/abs/2017ApJ...835...88K}
}

@ARTICLE{lefevre2020,
       author = {{Le F{\`e}vre}, O. and {B{\'e}thermin}, M. and {Faisst}, A. and {Jones}, G.~C. and {Capak}, P. and {Cassata}, P. and {Silverman}, J.~D. and {Schaerer}, D. and {Yan}, L. and {Amorin}, R. and {Bardelli}, S. and {Boquien}, M. and {Cimatti}, A. and {Dessauges-Zavadsky}, M. and {Giavalisco}, M. and {Hathi}, N.~P. and {Fudamoto}, Y. and {Fujimoto}, S. and {Ginolfi}, M. and {Gruppioni}, C. and {Hemmati}, S. and {Ibar}, E. and {Koekemoer}, A. and {Khusanova}, Y. and {Lagache}, G. and {Lemaux}, B.~C. and {Loiacono}, F. and {Maiolino}, R. and {Mancini}, C. and {Narayanan}, D. and {Morselli}, L. and {M{\'e}ndez-Hern{\`a}ndez}, Hugo and {Oesch}, P.~A. and {Pozzi}, F. and {Romano}, M. and {Riechers}, D. and {Scoville}, N. and {Talia}, M. and {Tasca}, L.~A.~M. and {Thomas}, R. and {Toft}, S. and {Vallini}, L. and {Vergani}, D. and {Walter}, F. and {Zamorani}, G. and {Zucca}, E.},
        title = "{The ALPINE-ALMA [CII] survey. Survey strategy, observations, and sample properties of 118 star-forming galaxies at 4 < z < 6}",
      journal = {\aap},
         year = 2020,
        month = nov,
       volume = {643},
          eid = {A1},
        pages = {A1},
          doi = {10.1051/0004-6361/201936965},
archivePrefix = {arXiv},
       eprint = {1910.09517},
 primaryClass = {astro-ph.CO},
       adsurl = {https://ui.adsabs.harvard.edu/abs/2020A&A...643A...1L}
}

@ARTICLE{lagache2018,
       author = {{Lagache}, G. and {Cousin}, M. and {Chatzikos}, M.},
        title = "{The [CII] 158 {\ensuremath{\mu}}m line emission in high-redshift galaxies}",
      journal = {\aap},
         year = 2018,
        month = jan,
       volume = {609},
          eid = {A130},
        pages = {A130},
          doi = {10.1051/0004-6361/201732019},
archivePrefix = {arXiv},
       eprint = {1711.00798},
 primaryClass = {astro-ph.GA},
       adsurl = {https://ui.adsabs.harvard.edu/abs/2018A&A...609A.130L}
}

@ARTICLE{diaz-santos2017,
       author = {{D{\'\i}az-Santos}, T. and {Armus}, L. and {Charmandaris}, V. and {Lu}, N. and {Stierwalt}, S. and {Stacey}, G. and {Malhotra}, S. and {van der Werf}, P.~P. and {Howell}, J.~H. and {Privon}, G.~C. and {Mazzarella}, J.~M. and {Goldsmith}, P.~F. and {Murphy}, E.~J. and {Barcos-Mu{\~n}oz}, L. and {Linden}, S.~T. and {Inami}, H. and {Larson}, K.~L. and {Evans}, A.~S. and {Appleton}, P. and {Iwasawa}, K. and {Lord}, S. and {Sanders}, D.~B. and {Surace}, J.~A.},
        title = "{A Herschel/PACS Far-infrared Line Emission Survey of Local Luminous Infrared Galaxies}",
      journal = {\apj},
         year = 2017,
        month = sep,
       volume = {846},
       number = {1},
          eid = {32},
        pages = {32},
          doi = {10.3847/1538-4357/aa81d7},
archivePrefix = {arXiv},
       eprint = {1705.04326},
 primaryClass = {astro-ph.GA},
       adsurl = {https://ui.adsabs.harvard.edu/abs/2017ApJ...846...32D}
}

@ARTICLE{munoz2016,
       author = {{Mu{\~n}oz}, Joseph A. and {Oh}, S. Peng},
        title = "{High-temperature saturation can produce the [C II] deficit in LIRGs and ULIRGs}",
      journal = {\mnras},
         year = 2016,
        month = dec,
       volume = {463},
       number = {2},
        pages = {2085-2091},
          doi = {10.1093/mnras/stw2102},
archivePrefix = {arXiv},
       eprint = {1510.00397},
 primaryClass = {astro-ph.GA},
       adsurl = {https://ui.adsabs.harvard.edu/abs/2016MNRAS.463.2085M}
}

@ARTICLE{Schaerer2020,
       author = {{Schaerer}, D. and {Ginolfi}, M. and {B{\'e}thermin}, M. and {Fudamoto}, Y. and {Oesch}, P.~A. and {Le F{\`e}vre}, O. and {Faisst}, A. and {Capak}, P. and {Cassata}, P. and {Silverman}, J.~D. and {Yan}, Lin and {Jones}, G.~C. and {Amorin}, R. and {Bardelli}, S. and {Boquien}, M. and {Cimatti}, A. and {Dessauges-Zavadsky}, M. and {Giavalisco}, M. and {Hathi}, N.~P. and {Fujimoto}, S. and {Ibar}, E. and {Koekemoer}, A. and {Lagache}, G. and {Lemaux}, B.~C. and {Loiacono}, F. and {Maiolino}, R. and {Narayanan}, D. and {Morselli}, L. and {M{\'e}ndez-Hern{\`a}ndez}, H. and {Pozzi}, F. and {Riechers}, D. and {Talia}, M. and {Toft}, S. and {Vallini}, L. and {Vergani}, D. and {Zamorani}, G. and {Zucca}, E.},
        title = "{The ALPINE-ALMA [C II] survey. Little to no evolution in the [C II]-SFR relation over the last 13 Gyr}",
      journal = {\aap},
         year = 2020,
        month = nov,
       volume = {643},
          eid = {A3},
        pages = {A3},
          doi = {10.1051/0004-6361/202037617},
archivePrefix = {arXiv},
       eprint = {2002.00979},
 primaryClass = {astro-ph.GA},
       adsurl = {https://ui.adsabs.harvard.edu/abs/2020A&A...643A...3S}
}

@ARTICLE{harikane2020,
       author = {{Harikane}, Yuichi and {Ouchi}, Masami and {Inoue}, Akio K. and {Matsuoka}, Yoshiki and {Tamura}, Yoichi and {Bakx}, Tom and {Fujimoto}, Seiji and {Moriwaki}, Kana and {Ono}, Yoshiaki and {Nagao}, Tohru and {Tadaki}, Ken-ichi and {Kojima}, Takashi and {Shibuya}, Takatoshi and {Egami}, Eiichi and {Ferrara}, Andrea and {Gallerani}, Simona and {Hashimoto}, Takuya and {Kohno}, Kotaro and {Matsuda}, Yuichi and {Matsuo}, Hiroshi and {Pallottini}, Andrea and {Sugahara}, Yuma and {Vallini}, Livia},
        title = "{Large Population of ALMA Galaxies at z > 6 with Very High [O III] 88 {\ensuremath{\mu}}m to [C II] 158 {\ensuremath{\mu}}m Flux Ratios: Evidence of Extremely High Ionization Parameter or PDR Deficit?}",
      journal = {\apj},
         year = 2020,
        month = jun,
       volume = {896},
       number = {2},
          eid = {93},
        pages = {93},
          doi = {10.3847/1538-4357/ab94bd},
archivePrefix = {arXiv},
       eprint = {1910.10927},
 primaryClass = {astro-ph.GA},
       adsurl = {https://ui.adsabs.harvard.edu/abs/2020ApJ...896...93H}
}

@ARTICLE{alba2025,
       author = {{Covelo-Paz}, Alba and {Giovinazzo}, Emma and {Oesch}, Pascal A. and {Meyer}, Romain A. and {Weibel}, Andrea and {Brammer}, Gabriel and {Fudamoto}, Yoshinobu and {Kerutt}, Josephine and {Lin}, Jamie and {Matharu}, Jasleen and {Naidu}, Rohan P. and {Velichko}, Anna and {Bollo}, Victoria and {Bouwens}, Rychard and {Chisholm}, John and {Illingworth}, Garth D. and {Kramarenko}, Ivan and {Magee}, Daniel and {Maseda}, Michael and {Matthee}, Jorryt and {Nelson}, Erica and {Reddy}, Naveen and {Schaerer}, Daniel and {Stefanon}, Mauro and {Xiao}, Mengyuan},
        title = "{An H{\ensuremath{\alpha}} view of galaxy buildup in the first 2 Gyr: Luminosity functions at z {\ensuremath{\sim}} 4{\ensuremath{-}}6.5 from NIRCam/grism spectroscopy}",
      journal = {\aap},
         year = 2025,
        month = feb,
       volume = {694},
          eid = {A178},
        pages = {A178},
          doi = {10.1051/0004-6361/202452363},
archivePrefix = {arXiv},
       eprint = {2409.17241},
 primaryClass = {astro-ph.GA},
       adsurl = {https://ui.adsabs.harvard.edu/abs/2025A&A...694A.178C}
}

@ARTICLE{lin2021,
       author = {{Lin}, Yen-Hsing and {Hirashita}, Hiroyuki and {Camps}, Peter and {Baes}, Maarten},
        title = "{Geometry effects on dust attenuation curves with different grain sources at high redshift}",
      journal = {\mnras},
         year = 2021,
        month = oct,
       volume = {507},
       number = {2},
        pages = {2755-2765},
          doi = {10.1093/mnras/stab2242},
archivePrefix = {arXiv},
       eprint = {2109.03072},
 primaryClass = {astro-ph.GA},
       adsurl = {https://ui.adsabs.harvard.edu/abs/2021MNRAS.507.2755L}
}

@ARTICLE{Scicluna2015,
       author = {{Scicluna}, P. and {Siebenmorgen}, R.},
        title = "{Extinction and dust properties in a clumpy medium}",
      journal = {\aap},
         year = 2015,
        month = dec,
       volume = {584},
          eid = {A108},
        pages = {A108},
          doi = {10.1051/0004-6361/201323149},
archivePrefix = {arXiv},
       eprint = {1510.06149},
 primaryClass = {astro-ph.SR},
       adsurl = {https://ui.adsabs.harvard.edu/abs/2015A&A...584A.108S}
}

@ARTICLE{ciesla2024,
       author = {{Ciesla}, L. and {Elbaz}, D. and {Ilbert}, O. and {Buat}, V. and {Magnelli}, B. and {Narayanan}, D. and {Daddi}, E. and {G{\'o}mez-Guijarro}, C. and {Arango-Toro}, R.},
        title = "{Identification of a transition from stochastic to secular star formation around z = 9 with JWST}",
      journal = {\aap},
         year = 2024,
        month = jun,
       volume = {686},
          eid = {A128},
        pages = {A128},
          doi = {10.1051/0004-6361/202348091},
archivePrefix = {arXiv},
       eprint = {2309.15720},
 primaryClass = {astro-ph.GA},
       adsurl = {https://ui.adsabs.harvard.edu/abs/2024A&A...686A.128C}
}

@ARTICLE{perry2025,
       author = {{Perry}, Marissa N. and {Taylor}, Anthony J. and {Ch{\'a}vez Ortiz}, {\'O}scar A. and {Finkelstein}, Steven L. and {C.~K. Leung}, Gene and {Bagley}, Micaela B. and {Fern{\'a}ndez}, Vital and {Arrabal Haro}, Pablo and {Chworowsky}, Katherine and {Cleri}, Nikko J. and {Dickinson}, Mark and {Ellis}, Richard S. and {Kartaltepe}, Jeyhan S. and {Koekemoer}, Anton M. and {Pacucci}, Fabio and {Papovich}, Casey and {Pirzkal}, Nor and {Tacchella}, Sandro},
        title = "{The Prevalence of Bursty Star Formation in Low-mass Galaxies at z = 1─7 from H{\ensuremath{\alpha}}-to-UV Diagnostics}",
      journal = {\apj},
         year = 2025,
        month = nov,
       volume = {994},
       number = {1},
          eid = {14},
        pages = {14},
          doi = {10.3847/1538-4357/ae102f},
archivePrefix = {arXiv},
       eprint = {2510.05388},
 primaryClass = {astro-ph.GA},
       adsurl = {https://ui.adsabs.harvard.edu/abs/2025ApJ...994...14P}
}

@ARTICLE{buat2007,
       author = {{Buat}, V. and {Takeuchi}, T.~T. and {Iglesias-P{\'a}ramo}, J. and {Xu}, C.~K. and {Burgarella}, D. and {Boselli}, A. and {Barlow}, T. and {Bianchi}, L. and {Donas}, J. and {Forster}, K. and {Friedman}, P.~G. and {Heckman}, T.~M. and {Lee}, Y.-W. and {Madore}, B.~F. and {Martin}, D.~C. and {Milliard}, B. and {Morissey}, P. and {Neff}, S. and {Rich}, M. and {Schiminovich}, D. and {Seibert}, M. and {Small}, T. and {Szalay}, A.~S. and {Welsh}, B. and {Wyder}, T. and {Yi}, S.~K.},
        title = "{The Local Universe as Seen in the Far-Infrared and Far-Ultraviolet: A Global Point of View of the Local Recent Star Formation}",
      journal = {\apjs},
         year = 2007,
        month = dec,
       volume = {173},
       number = {2},
        pages = {404-414},
          doi = {10.1086/516645},
archivePrefix = {arXiv},
       eprint = {astro-ph/0609738},
 primaryClass = {astro-ph},
       adsurl = {https://ui.adsabs.harvard.edu/abs/2007ApJS..173..404B}
}

@ARTICLE{rieke2009,
       author = {{Rieke}, G.~H. and {Alonso-Herrero}, A. and {Weiner}, B.~J. and {P{\'e}rez-Gonz{\'a}lez}, P.~G. and {Blaylock}, M. and {Donley}, J.~L. and {Marcillac}, D.},
        title = "{Determining Star Formation Rates for Infrared Galaxies}",
      journal = {\apj},
         year = 2009,
        month = feb,
       volume = {692},
       number = {1},
        pages = {556-573},
          doi = {10.1088/0004-637X/692/1/556},
archivePrefix = {arXiv},
       eprint = {0810.4150},
 primaryClass = {astro-ph},
       adsurl = {https://ui.adsabs.harvard.edu/abs/2009ApJ...692..556R}
}

@ARTICLE{bethermin2020,
       author = {{B{\'e}thermin}, M. and {Fudamoto}, Y. and {Ginolfi}, M. and {Loiacono}, F. and {Khusanova}, Y. and {Capak}, P.~L. and {Cassata}, P. and {Faisst}, A. and {Le F{\`e}vre}, O. and {Schaerer}, D. and {Silverman}, J.~D. and {Yan}, L. and {Amorin}, R. and {Bardelli}, S. and {Boquien}, M. and {Cimatti}, A. and {Davidzon}, I. and {Dessauges-Zavadsky}, M. and {Fujimoto}, S. and {Gruppioni}, C. and {Hathi}, N.~P. and {Ibar}, E. and {Jones}, G.~C. and {Koekemoer}, A.~M. and {Lagache}, G. and {Lemaux}, B.~C. and {Moreau}, C. and {Oesch}, P.~A. and {Pozzi}, F. and {Riechers}, D.~A. and {Talia}, M. and {Toft}, S. and {Vallini}, L. and {Vergani}, D. and {Zamorani}, G. and {Zucca}, E.},
        title = "{The ALPINE-ALMA [CII] survey: Data processing, catalogs, and statistical source properties}",
      journal = {\aap},
         year = 2020,
        month = nov,
       volume = {643},
          eid = {A2},
        pages = {A2},
          doi = {10.1051/0004-6361/202037649},
archivePrefix = {arXiv},
       eprint = {2002.00962},
 primaryClass = {astro-ph.GA},
       adsurl = {https://ui.adsabs.harvard.edu/abs/2020A&A...643A...2B}
}

@ARTICLE{sommovigo2022,
       author = {{Sommovigo}, L. and {Ferrara}, A. and {Carniani}, S. and {Pallottini}, A. and {Dayal}, P. and {Pizzati}, E. and {Ginolfi}, M. and {Markov}, V. and {Faisst}, A.},
        title = "{A new look at the infrared properties of z   5 galaxies}",
      journal = {\mnras},
         year = 2022,
        month = dec,
       volume = {517},
       number = {4},
        pages = {5930-5941},
          doi = {10.1093/mnras/stac2997},
archivePrefix = {arXiv},
       eprint = {2210.09312},
 primaryClass = {astro-ph.GA},
       adsurl = {https://ui.adsabs.harvard.edu/abs/2022MNRAS.517.5930S}
}

@ARTICLE{ismail2023,
       author = {{Ismail}, D. and {Beelen}, A. and {Buat}, V. and {Berta}, S. and {Cox}, P. and {Stanley}, F. and {Young}, A. and {Jin}, S. and {Neri}, R. and {Bakx}, T. and {Dannerbauer}, H. and {Butler}, K. and {Cooray}, A. and {Nanni}, A. and {Omont}, A. and {Serjeant}, S. and {van der Werf}, P. and {Vlahakis}, C. and {Wei{\ss}}, A. and {Yang}, C. and {Baker}, A.~J. and {Bendo}, G. and {Borsato}, E. and {Chartab}, N. and {Dye}, S. and {Eales}, S. and {Gavazzi}, R. and {Hughes}, D. and {Ivison}, R. and {Jones}, B.~M. and {Krips}, M. and {Lehnert}, M. and {Marchetti}, L. and {Messias}, H. and {Negrello}, M. and {Perez-Fournon}, I. and {Riechers}, D.~A. and {Urquhart}, S.},
        title = "{z-GAL: A NOEMA spectroscopic redshift survey of bright Herschel galaxies. II. Dust properties}",
      journal = {\aap},
         year = 2023,
        month = oct,
       volume = {678},
          eid = {A27},
        pages = {A27},
          doi = {10.1051/0004-6361/202346804},
archivePrefix = {arXiv},
       eprint = {2307.15747},
 primaryClass = {astro-ph.GA},
       adsurl = {https://ui.adsabs.harvard.edu/abs/2023A&A...678A..27I}
}

@ARTICLE{emcee,
       author = {{Foreman-Mackey}, Daniel and {Hogg}, David W. and {Lang}, Dustin and {Goodman}, Jonathan},
        title = "{emcee: The MCMC Hammer}",
      journal = {\pasp},
         year = 2013,
        month = mar,
       volume = {125},
       number = {925},
        pages = {306},
          doi = {10.1086/670067},
archivePrefix = {arXiv},
       eprint = {1202.3665},
 primaryClass = {astro-ph.IM},
       adsurl = {https://ui.adsabs.harvard.edu/abs/2013PASP..125..306F}
}

@ARTICLE{boogaard2025,
       author = {{Boogaard}, Leindert A. and {Walter}, Fabian and {Weiss}, Axel and {Colina}, Luis and {Hodge}, Jacqueline and {Bik}, Arjan and {Crespo G{\'o}mez}, Alejandro and {Daddi}, Emanuele and {Magdis}, Georgios E. and {Meyer}, Romain A. and {{\"O}stlin}, G{\"o}ran},
        title = "{Resolving the dusty star-forming galaxy GN20 at z=4.055 with NOEMA and JWST: A similar distribution of stars, gas and dust despite distinct apparent profiles}",
      journal = {arXiv e-prints},
         year = 2025,
        month = oct,
          eid = {arXiv:2510.17804},
        pages = {arXiv:2510.17804},
          doi = {10.48550/arXiv.2510.17804},
archivePrefix = {arXiv},
       eprint = {2510.17804},
 primaryClass = {astro-ph.GA},
       adsurl = {https://ui.adsabs.harvard.edu/abs/2025arXiv251017804B}
}

@ARTICLE{draine2001,
       author = {{Draine}, B.~T. and {Li}, Aigen},
        title = "{Infrared Emission from Interstellar Dust. I. Stochastic Heating of Small Grains}",
      journal = {\apj},
         year = 2001,
        month = apr,
       volume = {551},
       number = {2},
        pages = {807-824},
          doi = {10.1086/320227},
archivePrefix = {arXiv},
       eprint = {astro-ph/0011318},
 primaryClass = {astro-ph},
       adsurl = {https://ui.adsabs.harvard.edu/abs/2001ApJ...551..807D}
}

@ARTICLE{liang2024,
       author = {{Liang}, Lichen and {Feldmann}, Robert and {Murray}, Norman and {Narayanan}, Desika and {Hayward}, Christopher C. and {Angl{\'e}s-Alc{\'a}zar}, Daniel and {Bassini}, Luigi and {Richings}, Alexander J. and {Faucher-Gigu{\`e}re}, Claude-Andr{\'e} and {Chung}, Dongwoo T. and {Chan}, Jennifer Y.~H. and {Tolgay}, Do{\v{g}}a and {{\c{C}}atmabacak}, Onur and {Kere{\v{s}}}, Du{\v{s}}an and {Hopkins}, Philip F.},
        title = "{[C II] 158 {\ensuremath{\mu}}m emission as an indicator of galaxy star formation rate}",
      journal = {\mnras},
         year = 2024,
        month = feb,
       volume = {528},
       number = {1},
        pages = {499-541},
          doi = {10.1093/mnras/stad3792},
archivePrefix = {arXiv},
       eprint = {2301.04149},
 primaryClass = {astro-ph.GA},
       adsurl = {https://ui.adsabs.harvard.edu/abs/2024MNRAS.528..499L}
}

@ARTICLE{liang2019,
       author = {{Liang}, Lichen and {Feldmann}, Robert and {Kere{\v{s}}}, Du{\v{s}}an and {Scoville}, Nick Z. and {Hayward}, Christopher C. and {Faucher-Gigu{\`e}re}, Claude-Andr{\'e} and {Schreiber}, Corentin and {Ma}, Xiangcheng and {Hopkins}, Philip F. and {Quataert}, Eliot},
        title = "{On the dust temperatures of high-redshift galaxies}",
      journal = {\mnras},
         year = 2019,
        month = oct,
       volume = {489},
       number = {1},
        pages = {1397-1422},
          doi = {10.1093/mnras/stz2134},
archivePrefix = {arXiv},
       eprint = {1902.10727},
 primaryClass = {astro-ph.GA},
       adsurl = {https://ui.adsabs.harvard.edu/abs/2019MNRAS.489.1397L}
}

@ARTICLE{calzetti2000,
       author = {{Calzetti}, Daniela and {Armus}, Lee and {Bohlin}, Ralph C. and {Kinney}, Anne L. and {Koornneef}, Jan and {Storchi-Bergmann}, Thaisa},
        title = "{The Dust Content and Opacity of Actively Star-forming Galaxies}",
      journal = {\apj},
         year = 2000,
        month = apr,
       volume = {533},
       number = {2},
        pages = {682-695},
          doi = {10.1086/308692},
archivePrefix = {arXiv},
       eprint = {astro-ph/9911459},
 primaryClass = {astro-ph},
       adsurl = {https://ui.adsabs.harvard.edu/abs/2000ApJ...533..682C}
}

@ARTICLE{tacchella2022,
       author = {{Tacchella}, Sandro and {Smith}, Aaron and {Kannan}, Rahul and {Marinacci}, Federico and {Hernquist}, Lars and {Vogelsberger}, Mark and {Torrey}, Paul and {Sales}, Laura and {Li}, Hui},
        title = "{H {\ensuremath{\alpha}} emission in local galaxies: star formation, time variability, and the diffuse ionized gas}",
      journal = {\mnras},
         year = 2022,
        month = jun,
       volume = {513},
       number = {2},
        pages = {2904-2929},
          doi = {10.1093/mnras/stac818},
archivePrefix = {arXiv},
       eprint = {2112.00027},
 primaryClass = {astro-ph.GA},
       adsurl = {https://ui.adsabs.harvard.edu/abs/2022MNRAS.513.2904T}
}

@ARTICLE{olsen2017,
       author = {{Olsen}, Karen and {Greve}, Thomas R. and {Narayanan}, Desika and {Thompson}, Robert and {Dav{\'e}}, Romeel and {Niebla Rios}, Luis and {Stawinski}, Stephanie},
        title = "{S{\'I}GAME Simulations of the [CII], [OI], and [OIII] Line Emission from Star-forming Galaxies at z≃ 6}",
      journal = {\apj},
         year = 2017,
        month = sep,
       volume = {846},
       number = {2},
          eid = {105},
        pages = {105},
          doi = {10.3847/1538-4357/aa86b4},
archivePrefix = {arXiv},
       eprint = {1708.04936},
 primaryClass = {astro-ph.GA},
       adsurl = {https://ui.adsabs.harvard.edu/abs/2017ApJ...846..105O}
}

@ARTICLE{narayanan2018,
       author = {{Narayanan}, Desika and {Conroy}, Charlie and {Dav{\'e}}, Romeel and {Johnson}, Benjamin D. and {Popping}, Gerg{\"o}},
        title = "{A Theory for the Variation of Dust Attenuation Laws in Galaxies}",
      journal = {\apj},
         year = 2018,
        month = dec,
       volume = {869},
       number = {1},
          eid = {70},
        pages = {70},
          doi = {10.3847/1538-4357/aaed25},
archivePrefix = {arXiv},
       eprint = {1805.06905},
 primaryClass = {astro-ph.GA},
       adsurl = {https://ui.adsabs.harvard.edu/abs/2018ApJ...869...70N}
}

@ARTICLE{buat2012,
       author = {{Buat}, V. and {Noll}, S. and {Burgarella}, D. and {Giovannoli}, E. and {Charmandaris}, V. and {Pannella}, M. and {Hwang}, H.~S. and {Elbaz}, D. and {Dickinson}, M. and {Magdis}, G. and {Reddy}, N. and {Murphy}, E.~J.},
        title = "{GOODS-Herschel: dust attenuation properties of UV selected high redshift galaxies}",
      journal = {\aap},
         year = 2012,
        month = sep,
       volume = {545},
          eid = {A141},
        pages = {A141},
          doi = {10.1051/0004-6361/201219405},
archivePrefix = {arXiv},
       eprint = {1207.3528},
 primaryClass = {astro-ph.CO},
       adsurl = {https://ui.adsabs.harvard.edu/abs/2012A&A...545A.141B}
}

@ARTICLE{cooper2025olivia,
       author = {{Cooper}, Olivia R. and {Brammer}, Gabriel and {Heintz}, Kasper E. and {Toft}, Sune and {Casey}, Caitlin M. and {Setton}, David J. and {de Graaff}, Anna and {Boogaard}, Leindert and {Cleri}, Nikko J. and {Gillman}, Steven and {Gottumukkala}, Rashmi and {Greene}, Jenny E. and {Gullberg}, Bitten and {Hirschmann}, Michaela and {Hviding}, Raphael E. and {Lambrides}, Erini and {Leja}, Joel and {Long}, Arianna S. and {Manning}, Sinclaire M. and {Maseda}, Michael V. and {McConachie}, Ian and {McKinney}, Jed and {Narayanan}, Desika and {Price}, Sedona H. and {Strait}, Victoria and {Suess}, Katherine A. and {Weibel}, Andrea and {Williams}, Christina C.},
        title = "{RUBIES: JWST/NIRSpec Resolves Evolutionary Phases of Dusty Star-forming Galaxies at z {\ensuremath{\sim}} 2}",
      journal = {\apj},
         year = 2025,
        month = apr,
       volume = {982},
       number = {2},
          eid = {125},
        pages = {125},
          doi = {10.3847/1538-4357/adb8e1},
archivePrefix = {arXiv},
       eprint = {2410.08387},
 primaryClass = {astro-ph.GA},
       adsurl = {https://ui.adsabs.harvard.edu/abs/2025ApJ...982..125C}
}

@ARTICLE{kriek2013,
       author = {{Kriek}, Mariska and {Conroy}, Charlie},
        title = "{The Dust Attenuation Law in Distant Galaxies: Evidence for Variation with Spectral Type}",
      journal = {\apjl},
         year = 2013,
        month = sep,
       volume = {775},
       number = {1},
          eid = {L16},
        pages = {L16},
          doi = {10.1088/2041-8205/775/1/L16},
archivePrefix = {arXiv},
       eprint = {1308.1099},
 primaryClass = {astro-ph.CO},
       adsurl = {https://ui.adsabs.harvard.edu/abs/2013ApJ...775L..16K}
}

@ARTICLE{reddy2015,
       author = {{Reddy}, Naveen A. and {Kriek}, Mariska and {Shapley}, Alice E. and {Freeman}, William R. and {Siana}, Brian and {Coil}, Alison L. and {Mobasher}, Bahram and {Price}, Sedona H. and {Sanders}, Ryan L. and {Shivaei}, Irene},
        title = "{The MOSDEF Survey: Measurements of Balmer Decrements and the Dust Attenuation Curve at Redshifts z \raisebox{-0.5ex}\textasciitilde 1.4-2.6}",
      journal = {\apj},
         year = 2015,
        month = jun,
       volume = {806},
       number = {2},
          eid = {259},
        pages = {259},
          doi = {10.1088/0004-637X/806/2/259},
archivePrefix = {arXiv},
       eprint = {1504.02782},
 primaryClass = {astro-ph.GA},
       adsurl = {https://ui.adsabs.harvard.edu/abs/2015ApJ...806..259R}
}

@ARTICLE{shivaei2020,
       author = {{Shivaei}, Irene and {Reddy}, Naveen and {Rieke}, George and {Shapley}, Alice and {Kriek}, Mariska and {Battisti}, Andrew and {Mobasher}, Bahram and {Sanders}, Ryan and {Fetherolf}, Tara and {Azadi}, Mojegan and {Coil}, Alison L. and {Freeman}, William R. and {de Groot}, Laura and {Leung}, Gene and {Price}, Sedona H. and {Siana}, Brian and {Zick}, Tom},
        title = "{The MOSDEF Survey: The Variation of the Dust Attenuation Curve with Metallicity}",
      journal = {\apj},
         year = 2020,
        month = aug,
       volume = {899},
       number = {2},
          eid = {117},
        pages = {117},
          doi = {10.3847/1538-4357/aba35e},
archivePrefix = {arXiv},
       eprint = {2005.01742},
 primaryClass = {astro-ph.GA},
       adsurl = {https://ui.adsabs.harvard.edu/abs/2020ApJ...899..117S}
}

@ARTICLE{shivaei2025,
       author = {{Shivaei}, Irene and {Naidu}, Rohan P. and {Rodr{\'\i}guez Montero}, Francisco and {Matsumoto}, Kosei and {Leja}, Joel and {Matthee}, Jorryt and {Johnson}, Benjamin D. and {Oesch}, Pascal A. and {Chevallard}, Jacopo and {Adamo}, Angela and {Bodansky}, Sarah and {Bunker}, Andrew J. and {Covelo Paz}, Alba and {Di Cesare}, Claudia and {Egami}, Eiichi and {Furtak}, Lukas J. and {Heintz}, Kasper E. and {Kramarenko}, Ivan and {Meyer}, Romain A. and {Reddy}, Naveen A. and {Rinaldi}, Pierluigi and {Tacchella}, Sandro and {Torralba}, Alberto and {Witstok}, Joris and {Wozniak}, Michael A. and {Xiao}, Mengyuan},
        title = "{The Diversity and Evolution of Dust Attenuation Curves from Redshift z \raisebox{-0.5ex}\textasciitilde 1 to 9}",
      journal = {arXiv e-prints},
         year = 2025,
        month = sep,
          eid = {arXiv:2509.01795},
        pages = {arXiv:2509.01795},
          doi = {10.48550/arXiv.2509.01795},
archivePrefix = {arXiv},
       eprint = {2509.01795},
 primaryClass = {astro-ph.GA},
       adsurl = {https://ui.adsabs.harvard.edu/abs/2025arXiv250901795S}
}

@ARTICLE{Goicoechea2015,
       author = {{Goicoechea}, Javier R. and {Teyssier}, D. and {Etxaluze}, M. and {Goldsmith}, P.~F. and {Ossenkopf}, V. and {Gerin}, M. and {Bergin}, E.~A. and {Black}, J.~H. and {Cernicharo}, J. and {Cuadrado}, S. and {Encrenaz}, P. and {Falgarone}, E. and {Fuente}, A. and {Hacar}, A. and {Lis}, D.~C. and {Marcelino}, N. and {Melnick}, G.~J. and {M{\"u}ller}, H.~S.~P. and {Persson}, C. and {Pety}, J. and {R{\"o}llig}, M. and {Schilke}, P. and {Simon}, R. and {Snell}, R.~L. and {Stutzki}, J.},
        title = "{Velocity-resolved [CII] Emission and [CII]/FIR Mapping along Orion with Herschel}",
      journal = {\apj},
         year = 2015,
        month = oct,
       volume = {812},
       number = {1},
          eid = {75},
        pages = {75},
          doi = {10.1088/0004-637X/812/1/75},
archivePrefix = {arXiv},
       eprint = {1508.03801},
 primaryClass = {astro-ph.GA},
       adsurl = {https://ui.adsabs.harvard.edu/abs/2015ApJ...812...75G}
}

@ARTICLE{casey2014,
       author = {{Casey}, Caitlin M. and {Narayanan}, Desika and {Cooray}, Asantha},
        title = "{Dusty star-forming galaxies at high redshift}",
      journal = {\physrep},
         year = 2014,
        month = aug,
       volume = {541},
       number = {2},
        pages = {45-161},
          doi = {10.1016/j.physrep.2014.02.009},
archivePrefix = {arXiv},
       eprint = {1402.1456},
 primaryClass = {astro-ph.CO},
       adsurl = {https://ui.adsabs.harvard.edu/abs/2014PhR...541...45C}
}

@ARTICLE{byun2025NCdust,
       author = {{Byun}, Gyeong-Hwan and {Jang}, J.~K. and {Scofield}, Zachary P. and {Ahn}, Eunmo and {Baes}, Maarten and {Dubois}, Yohan and {Han}, San and {Jeon}, Seyoung and {Kim}, Juhan and {Pichon}, Christophe and {Rhee}, Jinsu and {Rodr{\'\i}guez Montero}, Francisco and {Yi}, Sukyoung K.},
        title = "{How Dust Models Shape High-z Galaxy Morphology: Insights from the NewCluster Simulation}",
      journal = {\apj},
         year = 2025,
        month = oct,
       volume = {992},
       number = {1},
          eid = {92},
        pages = {92},
          doi = {10.3847/1538-4357/adfed9},
archivePrefix = {arXiv},
       eprint = {2508.18374},
 primaryClass = {astro-ph.GA},
       adsurl = {https://ui.adsabs.harvard.edu/abs/2025ApJ...992...92B}
}

@ARTICLE{murphy2011,
       author = {{Murphy}, E.~J. and {Condon}, J.~J. and {Schinnerer}, E. and {Kennicutt}, R.~C. and {Calzetti}, D. and {Armus}, L. and {Helou}, G. and {Turner}, J.~L. and {Aniano}, G. and {Beir{\~a}o}, P. and {Bolatto}, A.~D. and {Brandl}, B.~R. and {Croxall}, K.~V. and {Dale}, D.~A. and {Donovan Meyer}, J.~L. and {Draine}, B.~T. and {Engelbracht}, C. and {Hunt}, L.~K. and {Hao}, C.-N. and {Koda}, J. and {Roussel}, H. and {Skibba}, R. and {Smith}, J.-D.~T.},
        title = "{Calibrating Extinction-free Star Formation Rate Diagnostics with 33 GHz Free-free Emission in NGC 6946}",
      journal = {\apj},
         year = 2011,
        month = aug,
       volume = {737},
       number = {2},
          eid = {67},
        pages = {67},
          doi = {10.1088/0004-637X/737/2/67},
archivePrefix = {arXiv},
       eprint = {1105.4877},
 primaryClass = {astro-ph.CO},
       adsurl = {https://ui.adsabs.harvard.edu/abs/2011ApJ...737...67M}
}

@ARTICLE{algera2025b,
       author = {{Algera}, H.~S.~B. and {Herrera-Camus}, R. and {Aravena}, M. and {Assef}, R. and {Bakx}, T.~L.~J.~C. and {Bolatto}, A. and {Cescon}, K. and {Chen}, C.-C. and {da Cunha}, E. and {Dayal}, P. and {De Looze}, I. and {Diaz-Santos}, T. and {Faisst}, A. and {Ferrara}, A. and {F{\"o}rster Schreiber}, N. and {Hathi}, N. and {Ikeda}, R. and {Inami}, H. and {Jones}, G.~C. and {Koekemoer}, A. and {Lutz}, D. and {Rela{\~n}o}, M. and {Romano}, M. and {Rowland}, L. and {Sommovigo}, L. and {Vallini}, L. and {Vijayan}, A. and {Villanueva}, V. and {van der Werf}, P.},
        title = "{How much gas and dust is in the $z=5.7$ Lyman Break Galaxy HZ10? An ALMA Band 10 to 4 and JWST/NIRSpec study of its interstellar medium}",
      journal = {arXiv e-prints},
         year = 2025,
        month = dec,
          eid = {arXiv:2512.02320},
        pages = {arXiv:2512.02320},
          doi = {10.48550/arXiv.2512.02320},
archivePrefix = {arXiv},
       eprint = {2512.02320},
 primaryClass = {astro-ph.GA},
       adsurl = {https://ui.adsabs.harvard.edu/abs/2025arXiv251202320A}
}

@ARTICLE{zavala2021,
       author = {{Zavala}, J.~A. and {Casey}, C.~M. and {Manning}, S.~M. and {Aravena}, M. and {Bethermin}, M. and {Caputi}, K.~I. and {Clements}, D.~L. and {Cunha}, E. da and {Drew}, P. and {Finkelstein}, S.~L. and {Fujimoto}, S. and {Hayward}, C. and {Hodge}, J. and {Kartaltepe}, J.~S. and {Knudsen}, K. and {Koekemoer}, A.~M. and {Long}, A.~S. and {Magdis}, G.~E. and {Man}, A.~W.~S. and {Popping}, G. and {Sanders}, D. and {Scoville}, N. and {Sheth}, K. and {Staguhn}, J. and {Toft}, S. and {Treister}, E. and {Vieira}, J.~D. and {Yun}, M.~S.},
        title = "{The Evolution of the IR Luminosity Function and Dust-obscured Star Formation over the Past 13 Billion Years}",
      journal = {\apj},
         year = 2021,
        month = mar,
       volume = {909},
       number = {2},
          eid = {165},
        pages = {165},
          doi = {10.3847/1538-4357/abdb27},
archivePrefix = {arXiv},
       eprint = {2101.04734},
 primaryClass = {astro-ph.GA},
       adsurl = {https://ui.adsabs.harvard.edu/abs/2021ApJ...909..165Z}
}

@ARTICLE{freundlich2019,
       author = {{Freundlich}, J. and {Combes}, F. and {Tacconi}, L.~J. and {Genzel}, R. and {Garcia-Burillo}, S. and {Neri}, R. and {Contini}, T. and {Bolatto}, A. and {Lilly}, S. and {Salom{\'e}}, P. and {Bicalho}, I.~C. and {Boissier}, J. and {Boone}, F. and {Bouch{\'e}}, N. and {Bournaud}, F. and {Burkert}, A. and {Carollo}, M. and {Cooper}, M.~C. and {Cox}, P. and {Feruglio}, C. and {F{\"o}rster Schreiber}, N.~M. and {Juneau}, S. and {Lippa}, M. and {Lutz}, D. and {Naab}, T. and {Renzini}, A. and {Saintonge}, A. and {Sternberg}, A. and {Walter}, F. and {Weiner}, B. and {Wei{\ss}}, A. and {Wuyts}, S.},
        title = "{PHIBSS2: survey design and z = 0.5 - 0.8 results. Molecular gas reservoirs during the winding-down of star formation}",
      journal = {\aap},
         year = 2019,
        month = feb,
       volume = {622},
          eid = {A105},
        pages = {A105},
          doi = {10.1051/0004-6361/201732223},
archivePrefix = {arXiv},
       eprint = {1812.08180},
 primaryClass = {astro-ph.GA},
       adsurl = {https://ui.adsabs.harvard.edu/abs/2019A&A...622A.105F}
}

@ARTICLE{accard2025,
       author = {{Accard}, C. and {B{\'e}thermin}, M. and {Boquien}, M. and {Buat}, V. and {Vallini}, L. and {Renaud}, F. and {Kraljic}, K. and {Aravena}, M. and {Cassata}, P. and {da Cunha}, E. and {Dam}, P. and {de Looze}, I. and {Dessauges-Zavadsky}, M. and {Dubois}, Y. and {Faisst}, A. and {Fudamoto}, Y. and {Ginolfi}, M. and {Gruppioni}, C. and {Han}, S. and {Herrera-Camus}, R. and {Inami}, H. and {Koekemoer}, A.~M. and {Lemaux}, B.~C. and {Li}, J. and {Li}, Y. and {Mobasher}, B. and {Molina}, J. and {Nanni}, A. and {Palla}, M. and {Pozzi}, F. and {Rela{\~n}o}, M. and {Romano}, M. and {Sawant}, P. and {Spilker}, J. and {Tsujita}, A. and {Veraldi}, E. and {Villanueva}, V. and {Wang}, W. and {Yi}, S.~K. and {Zamorani}, G.},
        title = "{The ALPINE-CRISTAL-JWST survey: Spatially resolved star formation relations at z {\ensuremath{\sim}} 5}",
      journal = {\aap},
         year = 2025,
        month = oct,
       volume = {702},
          eid = {A206},
        pages = {A206},
          doi = {10.1051/0004-6361/202556140},
archivePrefix = {arXiv},
       eprint = {2508.13136},
 primaryClass = {astro-ph.GA},
       adsurl = {https://ui.adsabs.harvard.edu/abs/2025A&A...702A.206A}
}

\begin{appendix}
\section{Choice of number of photon packets}\label{apdx:photon_packets}
\begin{figure}[h!]
    \centering
    \includegraphics[width=\linewidth]{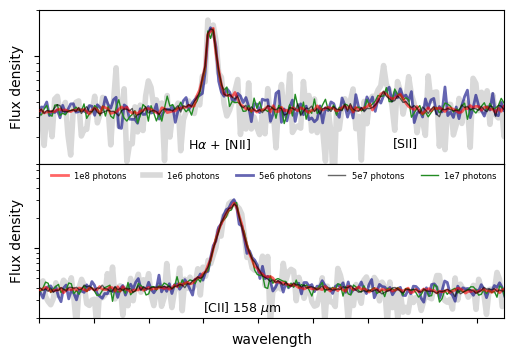}
    \caption{Zoomed-in view of a galaxy SED showing optical (top) and far-IR (bottom) spectral lines computed using different numbers of photon packets.}
    \label{fig:nphotons}
\end{figure}

To determine the number of photon packets required for an optimal balance between the signal and the Poisson noise, we perform a test on the most massive galaxy from NewCluster ($M_* = 1.1 \times 10^{10} \, \rm M_{\odot}$) by varying the number of photon packets between $10^6$ and $10^8$ photons. As shown in Fig. \ref{fig:nphotons}, the Poisson noise decreases significantly for $5 \times 10^7$ photon packets for emission lines in the optical and far-IR with different spectral resolutions. However, there is no significant change when using a larger amount of photons. 

\section{SFR variation with inclination}
\begin{figure*}[t]
    \includegraphics[width=\textwidth]{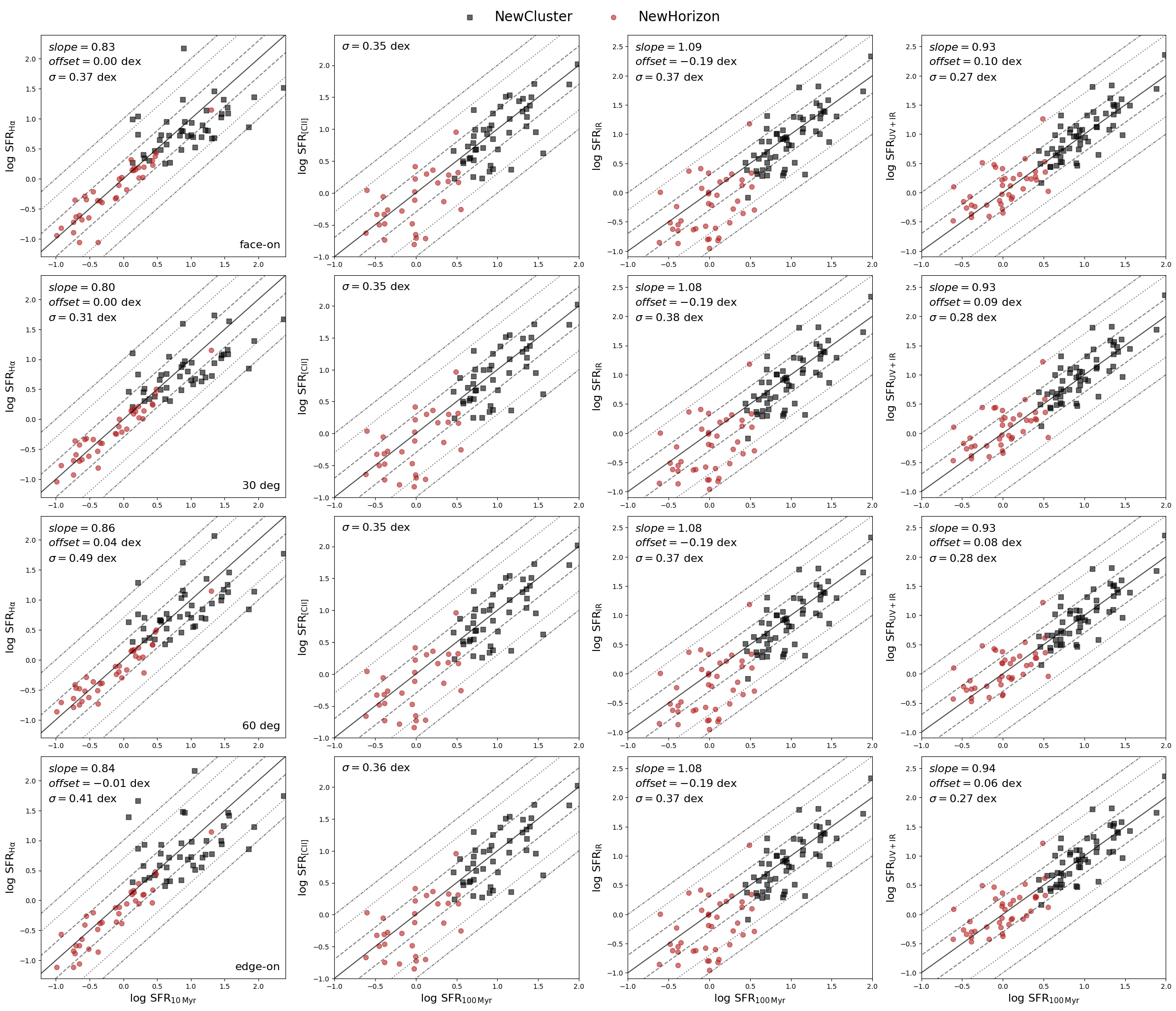}
    \caption{Left to right: SFR comparisons of H$\alpha$ versus SFR$_{\rm 10 \, Myr}$, [\textsc{Cii}] at $158 \, \mu$m versus SFR$_{\rm 100 \, Myr}$, IR versus SFR$_{\rm \, 100 Myr}$, and (UV + IR) versus SFR$_{\rm \, 100 Myr}$ for galaxies in the \textsc{NewCluster} (black squares) and \textsc{NewHorizon} (red circles) simulations. Each row corresponds to the different viewing angles: face-on, $30\degree$, $60\degree$, and edge-on (\textit{top to bottom}). The solid grey line is the identity line, while the dashed, dotted and dash-dot lines indicate offsets of $\pm 0.3$, $\pm 0.7$, and $\pm 1$ dex, respectively. On the top left corner of each plot, the dispersion of the entire sample is shown.}
    \label{fig:sfr-tracers}
\end{figure*}

We investigate the impact of the galactic viewing angle on the derivation of SFRs across different tracers. Figure \ref{fig:sfr-tracers} compares the SFRs inferred from H$\alpha$, [\textsc{Cii}], total IR luminosity, and hybrid (UV + IR) luminosities against the intrinsic SFRs averaged over 10 and 100 Myr. By varying the inclination from face-on to edge-on, we observe how dust attenuation and geometry affect the scatter and bias of these indicators which is further discussed in Sect. \ref{section:results}. Overall, the H$\alpha$-inferred SFR is the most sensitive to dust attenuation, showing an increasing scatter especially at $60\degree$ and edge-on configurations. In contrast, the other tracers show little to no dependence on viewing angle, highlighting their robustness against geometric effects and dust optical depths. 

\section{Leakage of UV photons: Effect on SFR$_{\rm IR}$ scatter}
\begin{figure}[h!]
    \centering
    \includegraphics[width=\linewidth]{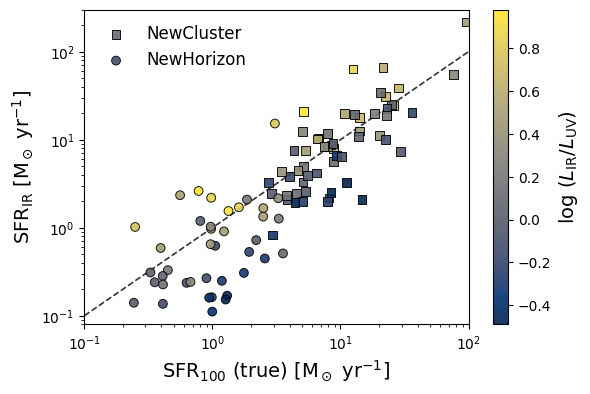}
    \caption{IR inferred SFRs versus time-averaged SFR over 100 Myr for both \textsc{NewCluster} (squares) and \textsc{NewHorizon} (circles) simulated galaxies. We color code the data points based on the ratio of the total IR ($8-1000 \, \mu$m) and far-UV ($1500 \AA$) luminosity (see Sect. \ref{section:ir-sfr} for details).}
    \label{fig:uv-lir-scatter}
\end{figure}

The relationship between $L_{\rm IR}$ and the SFR relies on the assumption that dust effectively absorbs and reprocesses UV radiation from young stars. To understand whether the scatter discussed in Sect. \ref{section:ir-sfr} is dominated by UV photon leakage, we explore the $\rm SFR_{IR} - SFR_{100 \, Myr}$ as a function of $L_{\rm IR}/L_{\rm FUV}$ ratio. Figure \ref{fig:uv-lir-scatter} shows that IR SFRs are systematically underestimated as $L_{\rm IR}/L_{\rm FUV}$ ratio becomes smaller, and it is more pronounced for low-mass galaxies (\textsc{NewHorizon}). Conversely, sources with IR-to-UV ratios $\sim 1$ are closer to the identity line. This highlights that UV-photon leakage contributes significantly to the scatter observed, particularly for underestimated SFRs. On the other hand, $L_{\rm IR}/L_{\rm FUV} > 1$ do not show a systematic offset and likely reflect effects such as stochastic star formation histories and stochastic heating of small dust grains, rather than UV leakage. 

\section{Star Formation History: effect on $L_{\rm IR}$} \label{apdx:sfh}
\begin{figure}[h!]
    \centering
    \includegraphics[width=\linewidth]{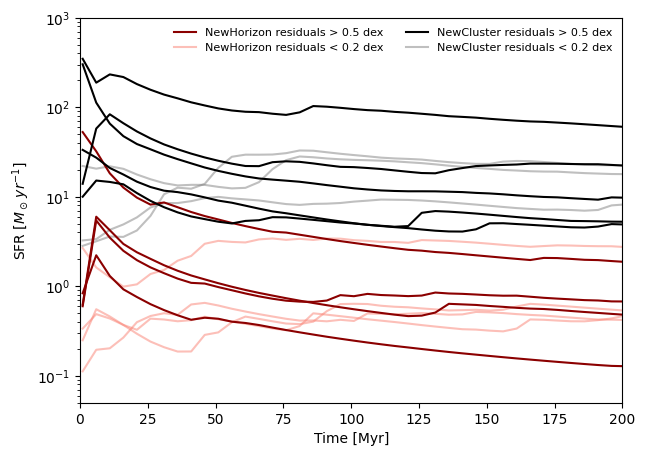}
    \caption{SFR as a function of lookback time relative to $z=5$, shown over 200 Myr interval. The black and dark red lines corresponds to the \textsc{NewCluster} and \textsc{NewHorizon} sources, respectively, with $\rm SFR_{\rm IR}$ residuals $> 0.5 \rm \, dex$. The grey and light red lines corresponds to the \textsc{NewCluster} and \textsc{NewHorizon} sources, respectively, with $\rm SFR_{\rm IR}$ residuals $< 0.2 \rm \, dex$. (see Sect. \ref{section:ir-sfr} for details).}
    \label{fig:sfr-scattered}
\end{figure}

We select sources from both simulations with the largest scatter in the $\rm SFR_{IR} - SFR_{100 \, Myr}$ plane (Fig. \ref{fig:sfr-ir}) with residuals larger than 0.5 dex, and derive the SFH over 200 Myr to inspect their burstiness at the galaxies' young age ($\rm < 20 \, Myr$). In Fig. \ref{fig:sfr-scattered}, we show in black and red the SFH of \textsc{NewCluster} and \textsc{NewHorizon} sources, respectively, which clearly display rising peaks of star formation rates at $t \sim 10 \, \rm  Myr$. The burstiness is seen to be stronger in lower mass galaxies from \textsc{NewHorizon} with $M_* < 10^9 \, M_\odot$, and a continuous decreases before stabilizing the rate of star formation in comparison to higher mass ones from \textsc{NewCluster}. For comparison, we also plot the SFH of sources from both simulations with residuals $< 0.2$ dex in grey and light red, where the SFRs show a more stable evolution with time.

\section{SFR with updated attenuation curves}
\begin{figure}[h!]
    \centering
    \includegraphics[width=\linewidth]{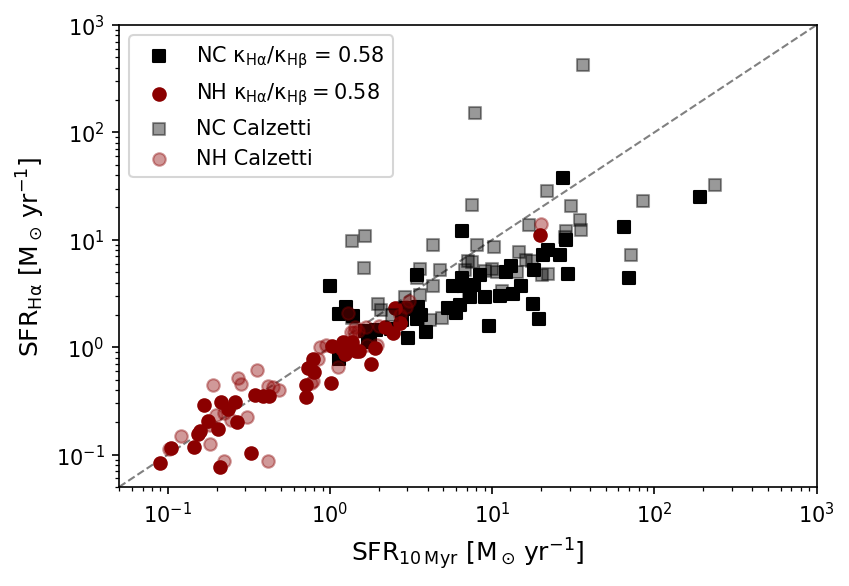}
    \caption{H$\alpha$-inferred SFRs versus 10 Myr time-averaged SFR. \textsc{NewHorizon} sources are shown in red circles, while \textsc{NewCluster} ones are shown in black squares. The solid colors indicate the SFRs inferred using the attenuation curves derived in Sect. \ref{section:attenuation} with $k_{\rm H\alpha} / k_{\rm H\beta} =0.58$, while transparent colors denote SFRs inferred with \citet{calzetti2000} attenuation law. The dashed line represents the identity line.}
    \label{fig:new-attenuation-sfr}
\end{figure}

To evaluate the improvement provided by attenuation modeling, we compare the H$\alpha$-inferred SFRs using the dust attenuation ratio derived in Sect. \ref{section:attenuation}. Figure \ref{fig:new-attenuation-sfr} displays the results using both the standard \citet{calzetti2000} law and our updated curves, which displays a tighter correlation between H$\alpha$-inferred SFRs and the intrinsic $\rm SFR_{10 \, Myr}$ and a significant decrease in the scatter compared to the canonical attenuation curve. Nevertheless, a systematic offset is observed for intrinsically higher SFRs, leading to their underestimation by a factor of $\sim 2$. 

\end{appendix}
\end{document}